\documentclass[10pt,twocolumn,letterpaper]{article}

\usepackage[pagenumbers]{cvpr} 
\usepackage{xcolor,colortbl}
\usepackage{booktabs}   
\usepackage{multirow}   
\usepackage{siunitx}    
\usepackage{xcolor}     
\usepackage{graphicx}   
\usepackage[skip=2pt]{caption}
\usepackage{makecell}

\definecolor{cvprblue}{rgb}{0.21,0.49,0.74}
\usepackage[pagebackref,breaklinks,colorlinks,allcolors=cvprblue]{hyperref}
\usepackage{multirow}
\usepackage{amssymb} 
\usepackage{pifont}  
\newcommand{\xmark}{\ding{55}}   
\newcommand{\cmark}{\checkmark}  
\newcommand{\pmark}{\ding{108}}
\usepackage{soul}

\def\paperID{23198} 
\def\confName{CVPR}
\def\confYear{2026}

\title{BackFed: A Standardized and Efficient Benchmark Framework for Evaluating Backdoor Attacks in Federated Learning}

\author{%
  Thinh Dao\textsuperscript{1,2}, Dung Thuy Nguyen\textsuperscript{3}, Khoa D Doan\textsuperscript{1,2}, Kok-Seng Wong\textsuperscript{1,2} \\[0.5em]
  \textsuperscript{1}VinUni-Illinois Smart Health Center, VinUniversity, Hanoi, Vietnam \\
  \textsuperscript{2}College of Engineering \& Computer Science, VinUniversity, Hanoi, Vietnam \\
  \textsuperscript{3}Vanderbilt University, Nashville, TN 37235 \\[0.5em]
  {\tt\small \{21thinh.dd, khoa.dd, wong.ks\}@vinuni.edu.vn} \\
  {\tt\small dung.t.nguyen@vanderbilt.edu}
}

\begin{document}
\maketitle
\begin{abstract}
Research on backdoor attacks in Federated Learning (FL) has accelerated in recent years, with new attacks and defenses continually proposed in an escalating arms race. However, the evaluation of these methods remains neither standardized nor reliable. First, there are severe inconsistencies in the evaluation settings across studies, and many rely on unrealistic threat models. Second, our code review uncovers semantic bugs in the official codebases of several attacks that artificially inflate their reported performance. These issues raise fundamental questions about whether current methods are truly effective or simply overfitted to narrow experimental setups. We introduce \textbf{BackFed}, a benchmark designed to standardize and stress-test FL backdoor evaluation by unifying attacks and defenses under a common evaluation framework that mirrors realistic FL deployments. Our benchmark on three representative datasets with three distinct architectures reveals critical limitations of existing methods. Malicious clients often require excessive training time and computation, making them vulnerable to server-enforced time constraints. Meanwhile, several defenses incur severe accuracy degradation or aggregation overhead. Popular defenses and attacks achieve limited performance in our benchmark, which challenges their previous efficacy claims. We establish BackFed as a rigorous and fair evaluation framework that enables more reliable progress in FL backdoor research.



\end{abstract}    
\section{Introduction}
Federated Learning (FL) has emerged as a promising framework for training machine learning models on decentralized datasets across multiple devices~\cite{konevcny2016federated, mcmahan2017communication}. This paradigm addresses increasing privacy concerns by enabling collaborative model training without sharing sensitive data, making it ideal for applications where data cannot be centralized due to privacy regulations or practical limitations. However, the distributed nature of FL introduces the risk of backdoor attacks \cite{howto}, in which malicious clients intentionally poison their local training processes using specially crafted triggers. These triggers cause the resulting global model to misclassify inputs containing the trigger during inference, while maintaining normal behavior on clean inputs. Such backdoors allow attackers to, for example, cause autonomous vehicles to misinterpret traffic signs, enable financial fraud through misclassified transactions, or bypass content filters in healthcare and social media systems \cite{chen2017targeted,anonymous2025cleanlabel}. In FL, backdoor attacks pose a major security concern because clients train privately on local data and share only model updates, making it hard to detect malicious behavior. This allows attackers to stealthily poison the global model without accessing the full dataset.

As FL gains adoption in critical domains \cite{backdoor_survey,fedunlearn}, the study of backdoor attacks and defenses in FL has evolved into a fast-moving arms race. However, the reliability and comparability of existing evaluations remain deeply flawed. \textbf{We identify major inconsistencies in threat models and evaluation settings among attacks and defenses.} For example, attacks such as 3DFed~\cite{3dfed} and DBA~\cite{xie2020dba} assume an unrealistic setting where malicious clients are continuously selected to poison the global model; defenses such as Indicator~\cite{li2024backdoorindicator} and Snowball~\cite{snowball} require server-side training that departs from the standard FL protocol. Furthermore, our review of public codebases uncovers fidelity bugs in trigger injection and backdoor evaluation that substantially inflate reported attack effectiveness. Together, these issues obscure a clear understanding of whether these methods genuinely work in practice.

To address these issues, we propose BackFed as a comprehensive benchmark that unifies backdoor attacks and defenses in FL under \textit{standardized} protocols. Compared to existing codebases, BackFed offers significant advantages in speed, scope, and extensibility. With a multi-processing backend, BackFed significantly accelerates experimentation runtime via parallel client training. The benchmark currently supports eight datasets across both CV and NLP domains, covering diverse architectures such as CNNs, Transformers, and LSTMs. Following the LEAF benchmark \cite{leaf}, BackFed supports the evaluation of three naturally non-IID datasets: FEMNIST, Sentiment140, and Reddit.


Using BackFed, we conduct an extensive benchmark of \textbf{10 attacks} and \textbf{21 defenses} across \textbf{CIFAR-10, FEMNIST, and Tiny-ImageNet}. Our study uncovers critical limitations in existing methods. For example, malicious clients often require substantial training time and computational resources, making them susceptible to server-enforced round-time constraints. Likewise, some defenses suffer from severe accuracy degradation and aggregation overhead compared to FedAvg. Notably, several popular attacks and defenses exhibit limited performance in our benchmark, challenging their previously reported efficacy. Two practical guidelines are derived from our results to enhance the robustness of FL systems against poisoning attacks: (1) adopt a low server learning rate, and
(2) maintain a large pool of sampling clients with a low selection threshold, which can prevent the influence of malicious updates. To establish BackFed as an evaluation framework for future research, we release the codebase with the recommended evaluation protocol so that researchers can efficiently and rigorously evaluate their methods against the baselines.

\section{Background and Related Work}

\noindent\textbf{Federated Learning.}
In FL, each client \( i \) holds a local dataset \( D_i \) and collaboratively trains a shared global model \( \theta \) without sharing raw data. The goal is to minimize the total loss \( \sum_{i=1}^{N} \mathcal{L}(D_i; \theta) \). 
At each round \( t \), the server broadcasts the current global model \( \theta^t \) to \( K \) selected clients, who train it on their local datasets and return the updates \( \Delta_i^t = \theta_i^t - \theta^t \) to the server. The server then aggregates these updates as:
\begin{align}
    \theta^{t+1} = \theta^{t} - \eta \cdot \text{AGR}(\Delta_1^t, \dots, \Delta_K^t),
\end{align}
where \( \eta \) is the server learning rate and \( \text{AGR}(\cdot) \) denotes the aggregation rule. 
For FedAvg, it becomes:
\begin{align}
    \text{AGR}(\Delta_1^t, \dots, \Delta_K^t) = \tfrac{1}{K}\sum_{i=1}^{K} \Delta_i^t.
    \label{eq:fedavg}
\end{align}
\textbf{Note.}
Standard FedAvg \citep{mcmahan2017communication} weights client updates by dataset size, but this strategy can be exploited by malicious clients to gain disproportionate influence by exaggerating their dataset sizes. Following \cite{howto}, we use equal weighting \( \tfrac{1}{K} \) for all clients as in \cref{eq:fedavg}.

\subsection{Taxonomy of Backdoor Attacks in FL} 
Backdoor attacks  in FL can be grouped into two categories:
\begin{itemize}
    \item \textbf{Data-poisoning attacks} differ in the use of \emph{triggers}: \emph{Static trigger} remains the same across malicious clients and training rounds. \citet{howto} uses a white square patch on bottom left corner as the trigger. Other types of static triggers proposed in centralized settings \cite{blended,gu2019badnets,sig} also belong to this category. \emph{Distributed trigger} (DBA, FCBA) \cite{fcba,xie2020dba} splits a global trigger into local pieces held by different malicious clients. Each malicious client trains with the local trigger but the server will be evaluated against the aggregated global trigger. \emph{Dynamic trigger} \cite{nguyen2024iba,zhang2024a3fl,ceberus,flat} is dynamically optimized against the current global model at each round to maximize attack success. IBA \cite{nguyen2024iba}, for instance, uses a trigger generator (U-Net or Autoencoder) that is retrained each round based on feedback from the global model. \emph{Edge-case triggers} \cite{edge_case} exploit out-of-distribution samples paired with target labels as natural backdoor triggers.
    
    \item \textbf{Model-poisoning attacks} directly manipulate the local model's parameter or the training process to strengthen the attack. These attacks can be categorized by purpose of manipulation: (1) \emph{Model-replacement} attack \cite{howto} scales malicious updates to dominate other benign updates during aggregation; (2) \emph{Durability-enhanced} techniques aim to extend the durability of attacks so the attack remains relevant even after malicious clients stop poisoning. Examples include Chameleon \cite{dai2023chameleon}, Neurotoxin \cite{zhang2022neurotoxin}, and Anticipate \cite{anticipate}. (3) \emph{Defense-evasion} approaches \cite{sun2019can,backdoor_critical,3dfed} are designed to bypass server-side defenses. For example, \citet{sun2019can} uses Projected Gradient Descent (PGD) to constrain the divergence of malicious updates. \citet{backdoor_critical} selectively poisons a minimal subset of layers to reduce attack traceability.
\end{itemize}

\noindent\textbf{Note.} Model-poisoning and data-poisoning attacks are orthogonal and thus can be combined to create a more potent attack. BackFed supports such composition by decoupling the trigger design from the model-poisoning mechanism, enabling flexible combinations of attack strategies.

\subsection{Taxonomy of Backdoor Defenses in FL} 
Backdoor defenses in FL can be grouped into 3 categories: 
\begin{itemize}
    \item \textbf{Robust Aggregation defenses} change the aggregation rule $\text{AGGR(.)}$ (\cref{eq:fedavg}) to improve the robustness of the aggregated model. (1) \emph{Robust Statistical Aggregation} replaces averaging with more robust statistics, such as coordinate median, trimmed mean of client updates at each dimension of model parameters \cite{pmlr-v80-yin18a}, and geometric median - the point minimizing the sum of distances to all client updates \cite{pillutla2022robust}. (2) \emph{Client Update Reweighting} methods assign different weights to client updates based on their trustworthiness, such as FLTrust \cite{cao2020fltrust} which uses a bootstrap dataset to evaluate client reliability. (3) \emph{DP-based Aggregation} methods clip the model's weight norm and inject Gaussian noise as a form of regularization that improves model robustness \cite{sun2019can,xie2021crfl}. (4) \emph{Distance-based Aggregation} methods such as Krum \cite{blanchard2017machine} selects the updates that are closest to the majority for aggregation, filtering out outliers. (5) \emph{Robust Learning Rate (RLR)} \cite{ozdayi2021defending} adjusts the sign of the learning rate for each parameter according to the sum of signs across all updates.

    \item \textbf{Anomaly Detection defenses} focus on detecting suspicious updates~\cite{rieger2022deepsight,nguyen2022flame,li2024backdoorindicator} or supicious clients~\cite{zhang2022fldetector,xie2024fedredefense}. (1) \emph{Update Filtering} methods often utilize clustering-based algorithm (HDBSCAN, KMeans) to isolate the cluster of malicious updates \cite{rieger2022deepsight,nguyen2022flame,wang2022rflbat}, or construct an outlier scoring method to quantify the suspiciousness of an update \cite{huang2023multi,aligninns,li2024backdoorindicator}. On top of anomaly detection, some methods also apply techniques from Robust Aggregation such as norm clipping to further defend the global model \cite{rieger2022deepsight,nguyen2022flame,li2024backdoorindicator}. (2) \emph{Client Filtering} defenses detect malicious clients based on their updates \cite{zhang2022fldetector,xie2024fedredefense}. Once marked as malicious, the client is permanently excluded from participating in subsequent rounds.

    \item \textbf{Client-side defenses} aim to enhance the robustness of local training to mitigate the impact of backdoor attacks before aggregation at the server. These approaches include perturbation-based methods which add controlled noise to model updates or parameters to mask potential malicious manipulations~\cite{flwbc,localdp,huang2023lockdown}. Another strategy is trigger inversion \cite{zhang2023flip}, which seeks to reconstruct potential trigger patterns to perform adversarial training. 
\end{itemize}

\begin{table}[t]
\centering
\caption{Summary of differences between BackFed and other two frameworks. \pmark\; means partially satisfying.}
\label{tab:benchmark_comparison}
\resizebox{0.9\linewidth}{!}{%
\begin{tabular}{lccc}
\toprule
Feature & FLPoison & Backdoors101 & \textbf{BackFed} \\
\midrule
Focus on backdoor attacks & \xmark & \cmark & \cmark \\
Modular architecture & \cmark & \pmark & \cmark \\
Multi-processing backend & \xmark & \xmark & \cmark \\
Natural non-IID datasets & \xmark & \xmark & \cmark \\
Support NLP evaluation & \xmark & \xmark & \cmark \\
\bottomrule
\end{tabular}%
}
\end{table}

\subsection{Existing Benchmarks and Their Limitations}

While several comprehensive backdoor benchmarks exist for in the centralized setting~\cite{karra2020trojai,wu2022backdoorbench, trojanzoo}, FL has limited options. The most comparable benchmark to ours is FLPoison~\cite{zhang2025sok}, which evaluates data poisoning attacks. Additionally, \textit{Backdoors101}~\cite{backdoors101} is a popular framework that researchers use to develop new attacks and defenses. However, However, as summarized in \cref{tab:benchmark_comparison}, they suffer from two main limitations:
(1) Performance bottlenecks – sequential execution restricts scalability, whereas BackFed’s multi-processing design accelerates experimentation;
(2) Narrow scope – prior frameworks focus mainly on CV tasks with synthetic data, whereas BackFed supports both CV and NLP domains with diverse, naturally non-IID datasets.

\section{Why Do We Need a Benchmark?}
Current research on backdoor attacks in FL suffers from significant evaluation inconsistencies, unrealistic threat models, and implementation errors. These factors collectively compromise reliable evaluation in the field.

\begin{itemize}
\item \textbf{Disparate FL simulation settings.} Non-IID data partitioning is typically simulated with a Dirichlet distribution, but the heterogeneity parameter $\alpha$ is often chosen arbitrarily (e.g., $\alpha{=}0.2$~\cite{li2024backdoorindicator}, $0.5$~\cite{xie2020dba,edge_case,nguyen2024iba}, $0.9$~\cite{3dfed,howto,zhang2024a3fl}). Additionally, relatively few works evaluate naturally heterogeneous datasets such as FEMNIST. \emph{Server-side hyperparameters.} Differences exist in the \underline{client sampling ratios} (e.g., 3DFed \cite{3dfed} selects all clients; others use 10\%), the \underline{FedAvg variant} (unweighted \cite{howto,xie2020dba} vs.\ weighted \cite{zhang2024a3fl,rlr}), and the \underline{server learning rates} ($\eta=0.1$ \cite{howto,xie2020dba}, $\eta=0.5$ \cite{dai2023chameleon,zhang2022neurotoxin}). RLR~\cite{rlr} evaluates with $\eta{=}0.001$ while comparing against a FedAvg baseline with $\eta{=}1$, which is misleading since $\eta$ influences attack convergence and durability (\cref{sec:server_side}).

\item \textbf{Disparate attack settings.} Attack evaluations also vary widely. \emph{Poison scheduling.} Two common strategies are \underline{random-sampling} (malicious clients are selected randomly)~\cite{zhang2024a3fl,howto,nguyen2024iba} and \underline{fixed-frequency} (malicious clients are selected every $f$ rounds)~\cite{nguyen2024iba,edge_case,xie2020dba,zhang2022neurotoxin,3dfed}. \emph{Poisoning hyperparameters.} Attacks vary in the number of compromised clients, poison ratios, and local learning rates, all of which affect attack performance. One key difference that is often overlooked is whether malicious clients \underline{share the same poisoned data}. Shared poisons yield correlated gradient directions that amplify the backdoor signal, whereas independent poisons produce more diverse and stealthier gradient directions. It is worth noting that attacks such as \cite{zhang2022neurotoxin,dai2023chameleon,howto} assume malicious clients share up to 200 batches of data for poisoning in CIFAR10, which far exceeds benign client data sizes and introdces unrealistic training bottlenecks. 

\item \textbf{Unrealistic threat models.}
Several methods rely on assumptions that are unlikely to hold in practical FL systems. \emph{Communication complexity.} Dynamic attacks such as IBA~\cite{nguyen2024iba}, A3FL~\cite{zhang2024a3fl}, and Cerberus~\cite{ceberus} require coordinated optimization and synchronization of triggers among colluding malicious clients across poisoning rounds. Such coordination introduces considerable communication and computation overhead, prolonging local training and making these clients susceptible to timeout constraints. \emph{Impractical defense assumptions.}
Defenses such as FLARE~\cite{flare} and FLTrust~\cite{cao2020fltrust} assume access to a task-representative dataset on the server side, which is often unavailable—especially in privacy-sensitive domains such as healthcare. Other defenses requires heavy computation on the server that deviates from standard FL protocol. For instance, BackdoorIndicator~\cite{li2024backdoorindicator} requires training the global model on out-of-distribution (OOD) data, while Snowball~\cite{snowball} fits a variational autoencoder (VAE) on high-dimensional model updates. Both methods introduce substantial overhead and aggregation bottlenecks.

\item \textbf{Implementation errors.}
We discovered fidelity bugs in multiple attack codebases that substantially inflate reported attack success rates. \emph{Trigger injection space.} Many implementations apply the trigger \underline{after} input normalization, which is unrealistic given that the attacker can only inject the trigger in the raw input domain at inference time. This bug is found in \href{https://github.com/ybdai7/Chameleon-durable-backdoor/blob/main/image_helper.py}{Chameleon}~\cite{dai2023chameleon} codebase, the pixel value for the injected trigger is $255$ while the inputs have been scaled to $[0,1]$. We also found the same bug in \href{https://github.com/qzzqzzb/A3FL/blob/main/fl_utils/attacker.py}{A3FL}~\cite{zhang2024a3fl} codebase, where each pixel in the trigger is clipped to $[-2,2]$ range during optimization, which is larger than the expected $[0,1]$ range of the pixel space. Such mismatches artificially amplify the trigger’s gradient signal, leading to exaggerated ASR. \emph{Evaluation leakage.} We found that Neurotoxin~\cite{zhang2022neurotoxin} and Chameleon~\cite{dai2023chameleon} use evaluation settings that leak information between backdoor training and testing. Neurotoxin \underline{trains and tests on the same split}, reusing samples from a single class (class 2) for both stages (\href{https://github.com/jhcknzzm/Federated-Learning-Backdoor/blob/master/FL_Backdoor_CV/image_helper.py#L71-L72}{Lines 71–72, 295–304}). Chameleon constructs its backdoor test set from the first \underline{1,000 training samples}, potentially overlapping with the malicious clients’ dataset (\href{https://github.com/ybdai7/Chameleon-durable-backdoor/blob/main/image_helper.py#L445-L453}{Lines 445–453}). In both cases, the backdoor test set overlaps with the poisoned training set, enabling easy backdoor memorization that inflates ASR. In BackFed, attacks are evaluated on a separate held-out test set used for accuracy evaluation.


\end{itemize}
\begin{table*}[!htbp]
\centering
\caption{Configurations in BackFed benchmark. Common settings: Server learning rate $\eta$=0.5; all clients train with batch size 64 and SGD optimizer. Benign clients: \textit{lr}=0.1 for 2 epochs. Malicious clients (10\%): \textit{lr}=0.05 for 6 epochs, poisoning 20/64 samples per batch.} 
\resizebox{0.9\textwidth}{!}{%
\begin{tabular}{lcccccc}
\toprule
\textbf{Dataset} & \textbf{Data distribution}  & \textbf{Trigger size} & \textbf{Model} & \textbf{\# samples / \# classes} & \textbf{\# clients per round} & \textbf{ACC @ 2000} \\ \midrule
CIFAR-10      & Dirichlet $\alpha=0.9$       &$5\times5$& ResNet18 & 50,000 / 10   & 10 / 100  & 92.50\% \\
Tiny-ImageNet & Dirichlet $\alpha=0.5$       &$10\times10$& VGG11 & 200,000 / 200 & 20 / 200  & 55.06\% \\
FEMNIST       & Natural non-IID (by writer)  &$4\times4$& 2 conv + 2 linear    & 817,851 / 62  & 30 / 3596 & 87.15\% \\
\bottomrule
\end{tabular}%
}
\label{tab:data_model}
\end{table*}

\section{BackFed: A Standardized and Efficient Benchmark for Backdoor Attacks in FL}

To address the identified evaluation inconsistencies and implementation errors, we introduce \textbf{BackFed}, a evaluation framework designed with three core principles:

\begin{itemize}
    \item \textbf{Standardized Evaluation Protocol}: Backfed reimplements disparate attacks and defenses with unified interfaces and evaluate them with common and realistic experimental settings to ensure a fair comparison.

    \item \textbf{Efficient Multi-Processing Backend}: A key bottleneck of FL is client model training that prevents scalable studies with high number of clients and task complexity. We address this with a highly-optimised multi-processing backend via Ray~\cite{ray} that parallelizes client model training. Our runtime benchmark shows that parallel training can enable up to $7\times$ speedup over sequential training. We note that BackFed is the \textit{first} FL security framework to have a multi-processing pipeline.

    \item \textbf{Extensible Modular Architecture}: BackFed decouples attack mechanisms (data poisoning vs. model poisoning) and defense strategies (robust aggregation vs. anomaly detection). This design allows researchers to easily compose new attacks and defenses and extend the framework with minimal code changes.
\end{itemize}


\subsection{Framework Structure}
\begin{figure}
    \centering
    \includegraphics[width=\linewidth]{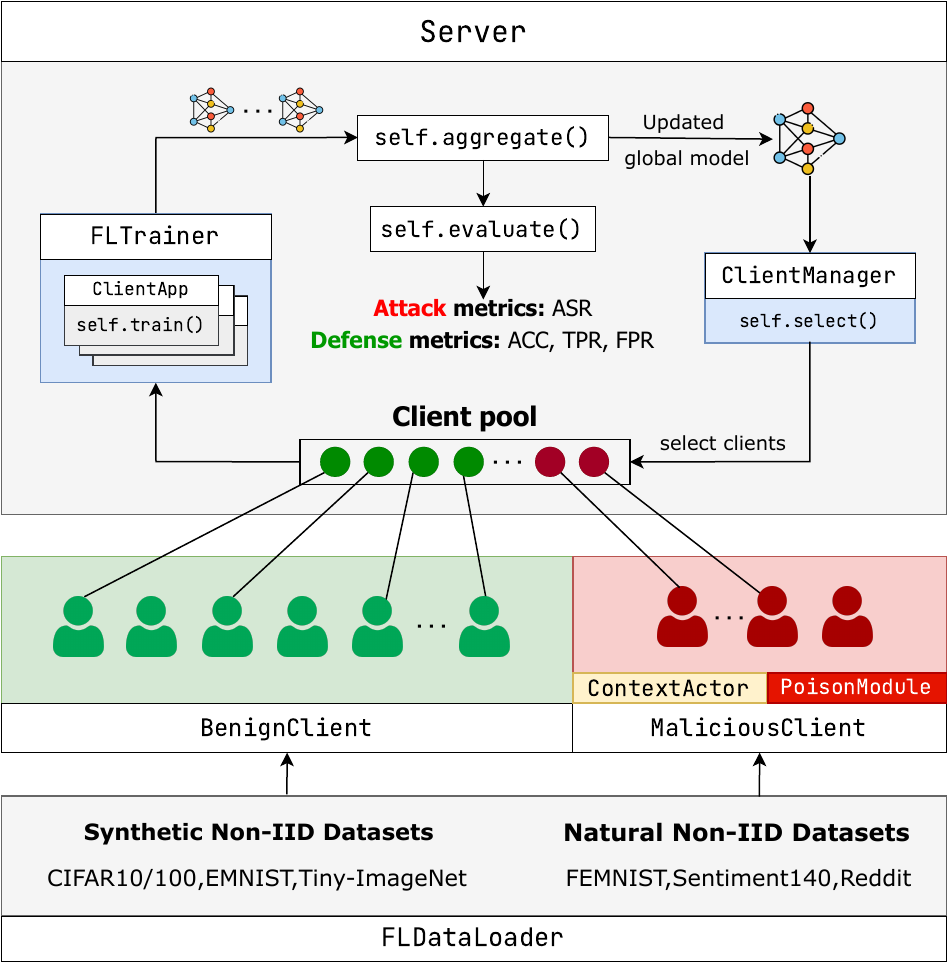}
    \caption{The overall structure and pipeline of BackFed.}
    \label{fig:framework_structure}
\end{figure}

\cref{fig:framework_structure} shows the overall structure of BackFed. We summarize key components below (details in Appendix~\ref{sec:framework_struc}): 
\begin{itemize}
    \item \texttt{FLDataLoader} partitions datasets following either a Dirichlet-based distribution (e.g., CIFAR10) or natural non-iid splits (e.g., FEMNIST).
    
    \item \texttt{Server} orchestrates training through two components: \texttt{ClientManager}, which handles client selection and attack scheduling, and \texttt{FLTrainer}, which manages parallel client training (via Ray package \cite{ray}) and submission of model updates to the server. Server-side defenses are implemented as subclasses of \texttt{Server} that mostly differ in the \texttt{aggregate()} function.

    \item \texttt{Client} includes both benign and malicious participants. \texttt{MaliciousClient}  uses \texttt{PoisonModule} to handle data-poisoning operations such as trigger injection and label poisoning. \texttt{ContextActor} is a communication route that enables malicious clients to synchronize attack resources, particularly for dynamic triggers. Model-poisoning attacks are implemented as \texttt{MaliciousClient} subclasses while Client-side defenses are implemented as \texttt{BenignClient} subclasses.
    
\end{itemize}

\subsection{Implementation Scope}
At the time of this writing, BackFed has incorporated \textbf{18} attacks (11 data poisoning + 7 model poisoning) and \textbf{25} defenses (13 RA defenses + 10 AD defenses + 2 client-side defenses). BackFed supports the evaluation of \textbf{8} datasets (5 CV + 2 NLP) with diverse model families (CNN for CV, Transformer + LSTM for NLP). The framework provides preprocessed datasets for non-IID datasets (FEMNIST, Reddit, Sentiment140) and pretrained models so that researchers can conduct experiments directly without the complexity of dataset preparation and clean model training procedures. While we have not included \textit{all} attacks and defenses, our library has (1) all popular methods that are widely used as baselines and published in top-tier conferences/journals and (2) most state-of-the-art methods that have been recently published, such as \cite{li2024backdoorindicator,aligninns,feddlad}. We specifically choose methods that have open-sourced codebases to ensure the validity of our implementation. 

\section{Evaluation Methodology}
With wide variations in evaluation settings and mentioned semantic bugs, it remains uncertain whether the proposed attacks and defenses are truly effective and robust across diverse scenarios. To answer this question, we aim to design a standardized evaluation protocol that \textit{mirrors} realistic FL systems and spans representative evaluation scenarios. 


\subsection{Threat Model}
To cover a wide range of attacks and defenses, we assume a general and relaxed threat model. On the attack side, an attacker controls a small fraction of clients (\textit{malicious} clients) that collaboratively attempt to poison the global model. Each malicious client has full control over its local training process, enabling online poisoning and arbitrary adjustments to local hyperparameters (e.g., learning rate, number of epochs, objective). However, the attacker has no knowledge of other clients’ data, the client-sampling process, or the server’s aggregation rule. On the defense side, we assume a trustworthy server with full visibility into unencoded client updates. The server can deploy protective mechanisms by modifying the aggregation rule or filtering suspicious updates. We evaluate only a \textit{realistic} \textbf{random-sampling} server that selects $K$ clients each round, so malicious clients are not guaranteed repeated participation as assumed by several attacks. Following \cite{howto,xie2020dba}, attacks start near global model convergence, where we set at round 2001 across datasets. The attack phase spans for 200 rounds, during which any malicious clients that are selected will submit backdoored model updates. For dynamic attacks~\cite{nguyen2024iba,ceberus,zhang2024a3fl}, the trigger is first optimized by a random malicious client and then shared to synchronize poisoning.

\subsection{Evaluation Setup (\texorpdfstring{\cref{tab:data_model}}{Table~1}).}
\noindent\textbf{Standardized evaluation settings.}
For both attacks and defenses, we closely follow method-specific hyperpameters from the original papers, but standardize certain settings to ensure a fair comparison across methods. For attacks, we standardize the trigger size, number of malicious clients, poison ratio, and training settings. For defenses, we fix the server learning rate and enforce unweighted aggregation for all anomaly detection methods (\cref{eq:fedavg}). Although lower values of $\eta$ would improve robustness, we set $\eta=0.5$ to stress-test the performance of defenses against attacks.

\noindent\textbf{Datasets \& Models.} We evaluate attacks and defenses on three standard benchmarks—CIFAR-10, FEMNIST, and Tiny-ImageNet—covering varying levels of heterogeneity and learning difficulty. FEMNIST exhibits strong client-level heterogeneity, CIFAR-10 (Dirichlet $\alpha{=}0.9$) approximates an IID setup, and Tiny-ImageNet introduces a substantial domain shift. ViTs are excluded as their computing resource and data requirements are not suitable for FL.


\noindent\textbf{Evaluation Metrics.} Two standard metrics are used: Attack Success Rate (ASR) measures the percentage of \textit{non-target} samples with the trigger that are misclassified as the target class, while Accuracy (ACC) measures the classification accuracy on normal inputs. For Anomaly Detection defenses, we evaluate detection performance with True Positive Rate (TPR) and True Negative Rate (TNR), which measure the proportion of correctly identified \textit{malicious} and \textit{benign} updates respectively. A good detector should achieve a high TPR without sacrificing TNR.


\section{Benchmarking Backdoor Attacks in FL}
Our benchmark addresses the following research questions:

\noindent\textbf{RQ1:} How do server-side hyperparameters influence the success of backdoor attacks?

\noindent\textbf{RQ2:} Under realistic and standardized settings, how effective are \textit{data-poisoning} and \textit{model-poisoning} attacks?

\noindent\textbf{RQ3:} Are these attacks applicable to real-world FL systems in terms of training time and computational resources?

\noindent\textbf{RQ4:} How effective are server-side defenses (\textit{Robust Aggregation} and \textit{Anomaly Detection}) in mitigating backdoor attacks, and what are the trade-offs of these defenses?


\subsection{Impact of Server-side Hyperparameters (RQ1).}
\label{sec:server_side}
\begin{figure}[h]
    \centering
    \includegraphics[width=0.975\linewidth]{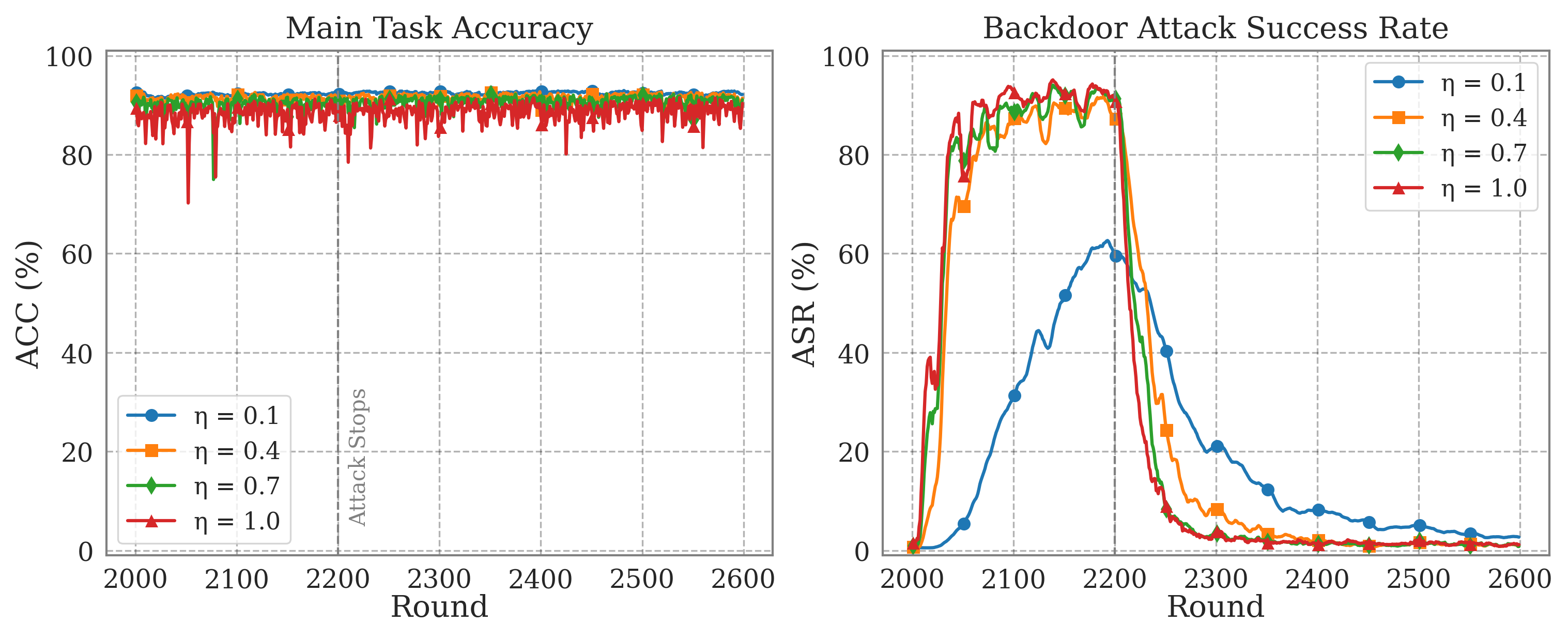}
    \caption{Impact of server learning rate.}
    \label{fig:server_lr}
\end{figure}

We study two server-side hyperparameters that directly affect backdoor attacks: the server's  learning rate ($\eta$) and the number of sampled clients per round ($K$). $\eta$ controls how strongly poisoned updates could impact the global model, while $K$ governs attack participation frequency. While varying $K$ has minimal effect on ACC and ASR, we find that $\eta$ critically affects both attack \textit{convergence} and \textit{durability} (\cref{fig:server_lr}). A higher $\eta$ makes the model more vulnerable to attacks and accuracy degradation. However, it facilitates quicker \textit{backdoor forgetting} after the attack stops. On the other hand, a lower $\eta$ slows convergence but makes the backdoor more persistent after the attack stops. Our results suggest that \textbf{lower $\eta$ improves the robustness of FL training} in most cases. Future defenses could develop an \textit{adaptive} $\eta$ based on the \textit{trustworthiness} of client updates, balancing attack durability and global model stability.

\begin{figure*}[ht]
    \centering
    \resizebox{\textwidth}{!}{%
    \begin{minipage}[c]{0.7\textwidth} 
        \centering
        \begin{subfigure}[t]{0.333\textwidth}
            \centering
            \includegraphics[width=\linewidth]{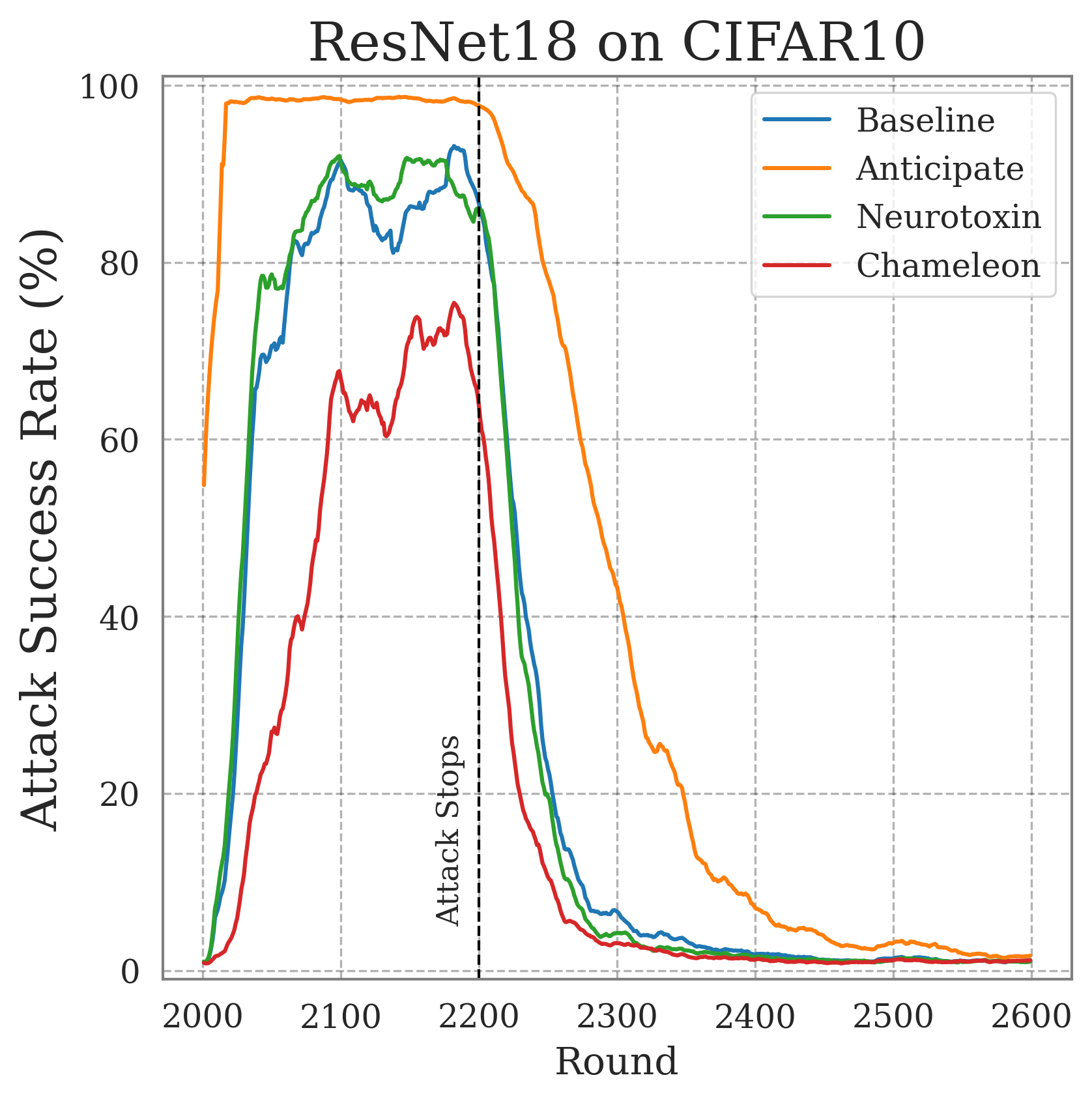}
            \caption{CIFAR10}
            \label{fig:cifar10_durability}
        \end{subfigure}%
        \begin{subfigure}[t]{0.333\textwidth}
            \centering
            \includegraphics[width=\linewidth]{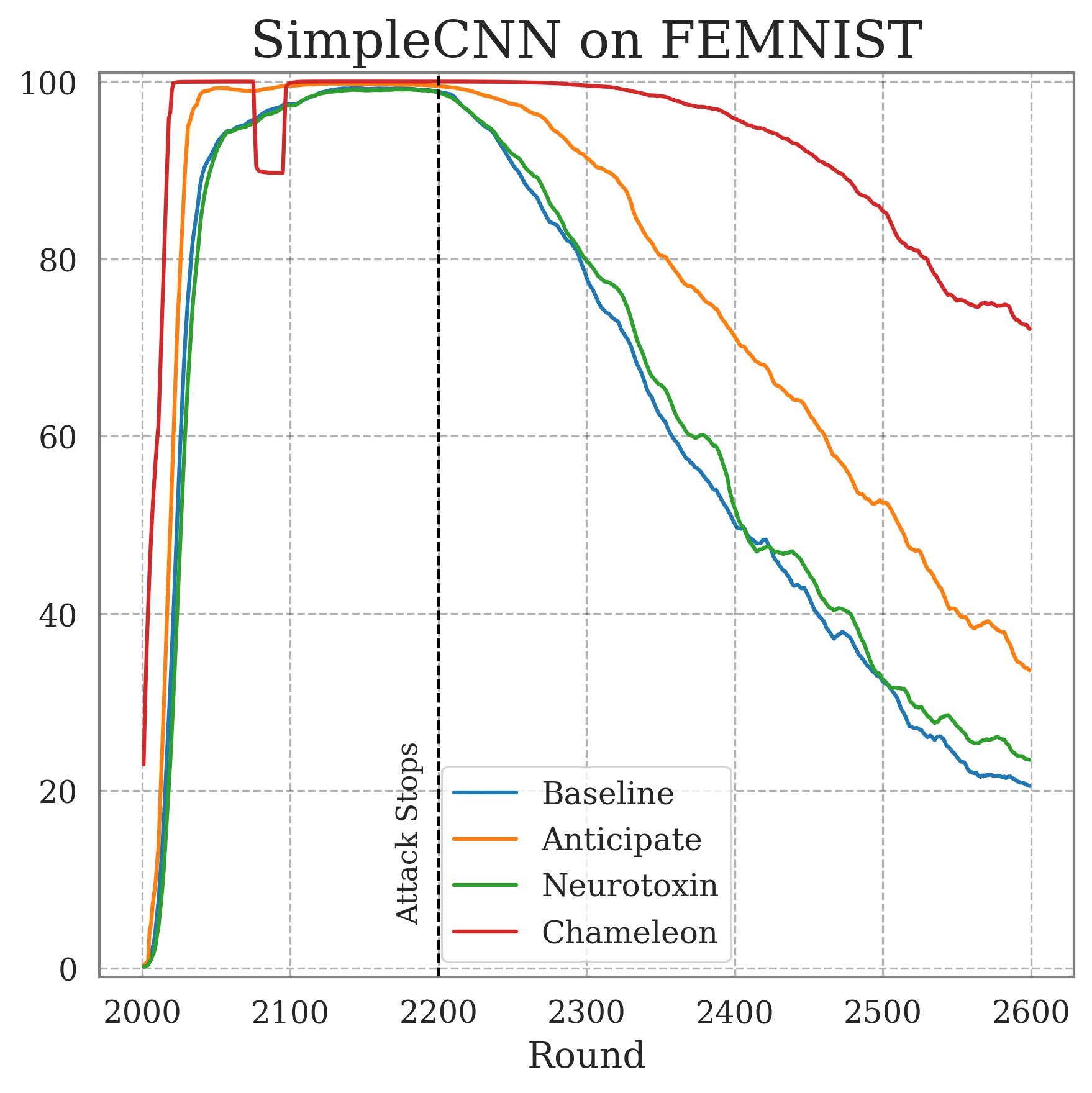}
            \caption{FEMNIST}
            \label{fig:femnist_durability}
        \end{subfigure}%
        \begin{subfigure}[t]{0.333\textwidth}
            \centering
            \includegraphics[width=\linewidth]{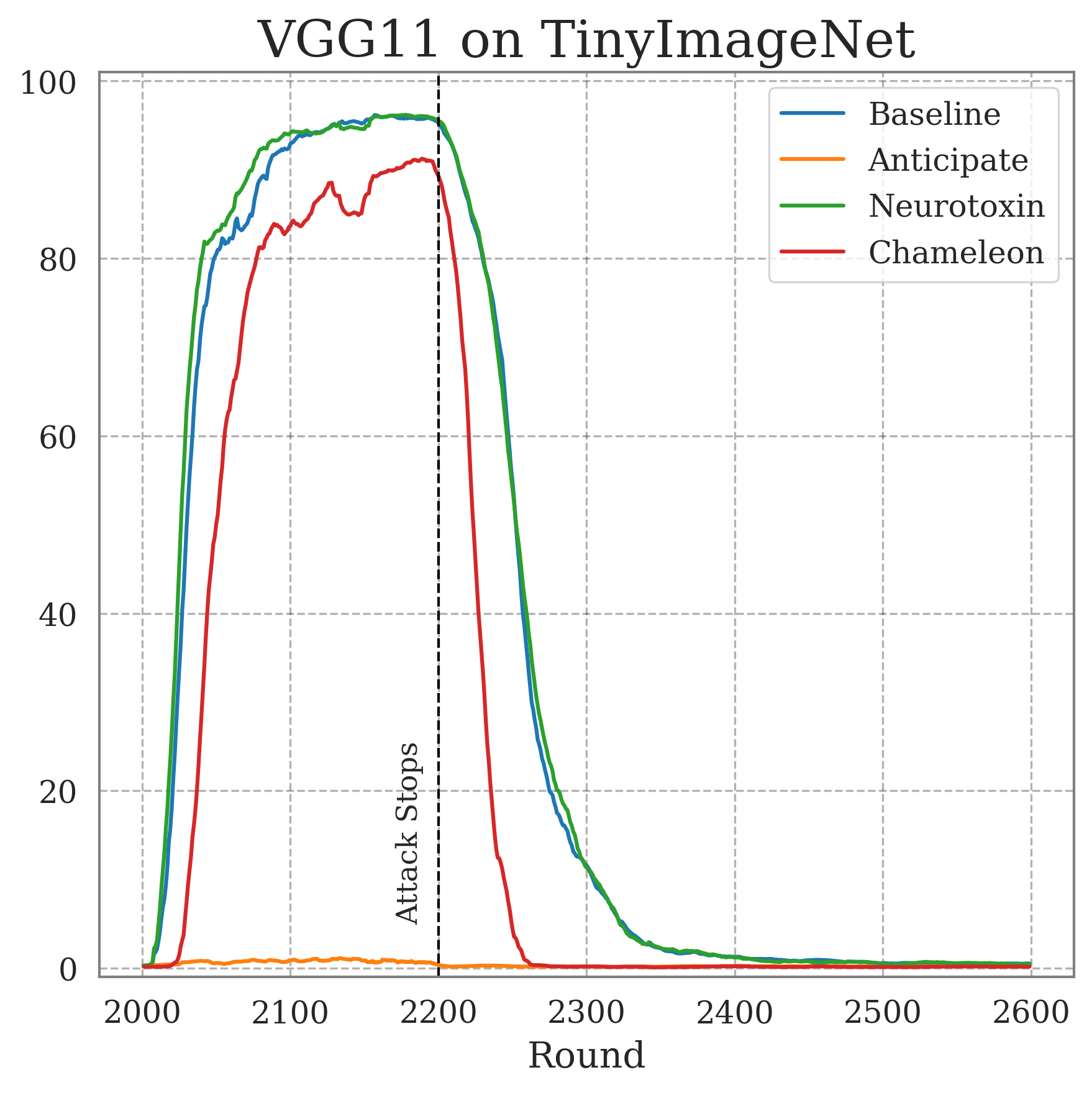}
            \caption{TinyImageNet}
            \label{fig:tiny_durability}
        \end{subfigure}
    \end{minipage}%
    \hspace{2mm}
    \begin{minipage}[c]{0.3\textwidth} 
        \centering
        \captionsetup{skip=2pt} 
        \captionof{table}{Summary metrics. Lowest ACC is \textbf{bolded}, highest ASR is \textcolor{red}{highlighted}.}
        \label{tab:model_poisoning}
        \setlength{\tabcolsep}{5pt}
        \resizebox{\textwidth}{!}{%
        \begin{tabular}{@{}l|l|cccc@{}}
        \toprule
        & Metric & Baseline & NTX & CML & ANT \\ \midrule
        \multirow{3}{*}{(a)} 
            & ACC$_\text{min}$ & 83.5 & 82.1 & 81.3 & \textbf{32.2} \\
            & ASR$_\text{during}$ & 72.8 & 75.9 & 48.0 & \textcolor{red}{96.2} \\
            & ASR$_\text{post}$ & 30.6 & 30.9 & 19.5 & \textcolor{red}{48.7} \\ \midrule
        \multirow{3}{*}{(b)}
            & ACC$_\text{min}$ & 85.6 & 85.6 & \textbf{79.3} & 85.4 \\
            & ASR$_\text{during}$ & 86.4 & 84.9 & \textcolor{red}{95.1} & 90.5 \\
            & ASR$_\text{post}$ & 65.3 & 66.3 & \textcolor{red}{92.8} & 77.0 \\ \midrule
        \multirow{3}{*}{(c)}
            & ACC$_\text{min}$ & 50.8 & \textbf{50.5} & 51.2 & 50.9 \\
            & ASR$_\text{during}$ & 78.9 & \textcolor{red}{81.2} & 66.0 & 0.7 \\
            & ASR$_\text{post}$ & 35.8 & \textcolor{red}{36.7} & 25.9 & 0.4 \\ 
        \bottomrule
        \multicolumn{6}{@{}l@{}}{\footnotesize NTX=Neurotoxin, ANT=Anticipate, CML=Chameleon} \\
        \end{tabular}}
    \end{minipage}
    }
    \caption{Evaluation of Durability-Enhanced attacks (Model-Poisoning). Figures (left) visualize ASR evolution; Table (right) summarizes metrics: ACC$_\text{min}$ records minimum clean accuracy in a round; ASR$_\text{during}$ and ASR$_\text{post}$ record average ASR during and after the attack phase.}
    \label{fig:model_poison}
\end{figure*}

\begin{figure*}[ht]
    \centering
    \resizebox{\textwidth}{!}{%
    \begin{minipage}[c]{0.7\textwidth}
        \centering
        \begin{subfigure}[t]{0.333\textwidth}
            \centering
            \includegraphics[width=\linewidth]{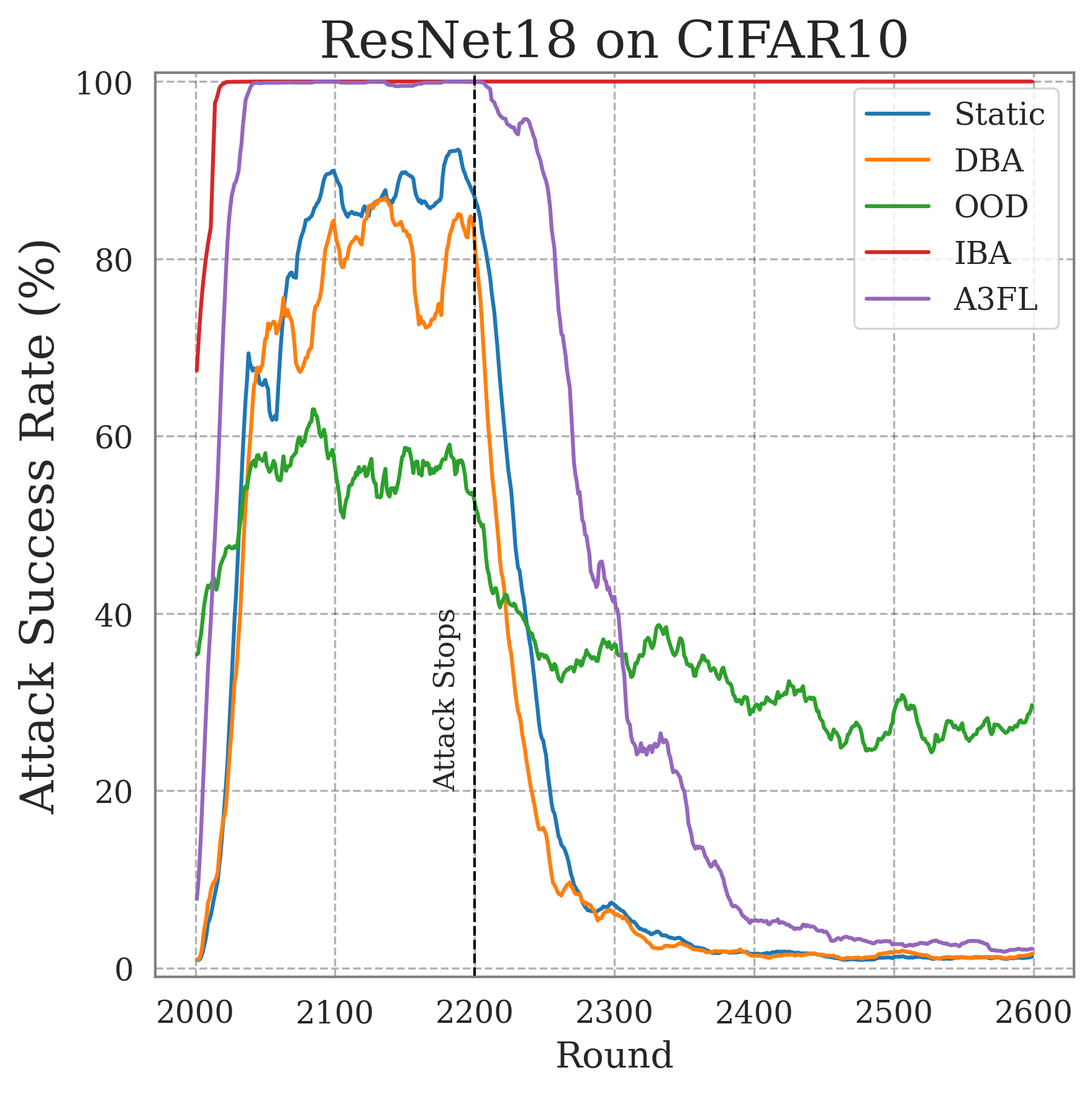}
            \caption{CIFAR10}
            \label{fig:cifar10_data_poison}
        \end{subfigure}%
        \begin{subfigure}[t]{0.333\textwidth}
            \centering
            \includegraphics[width=\linewidth]{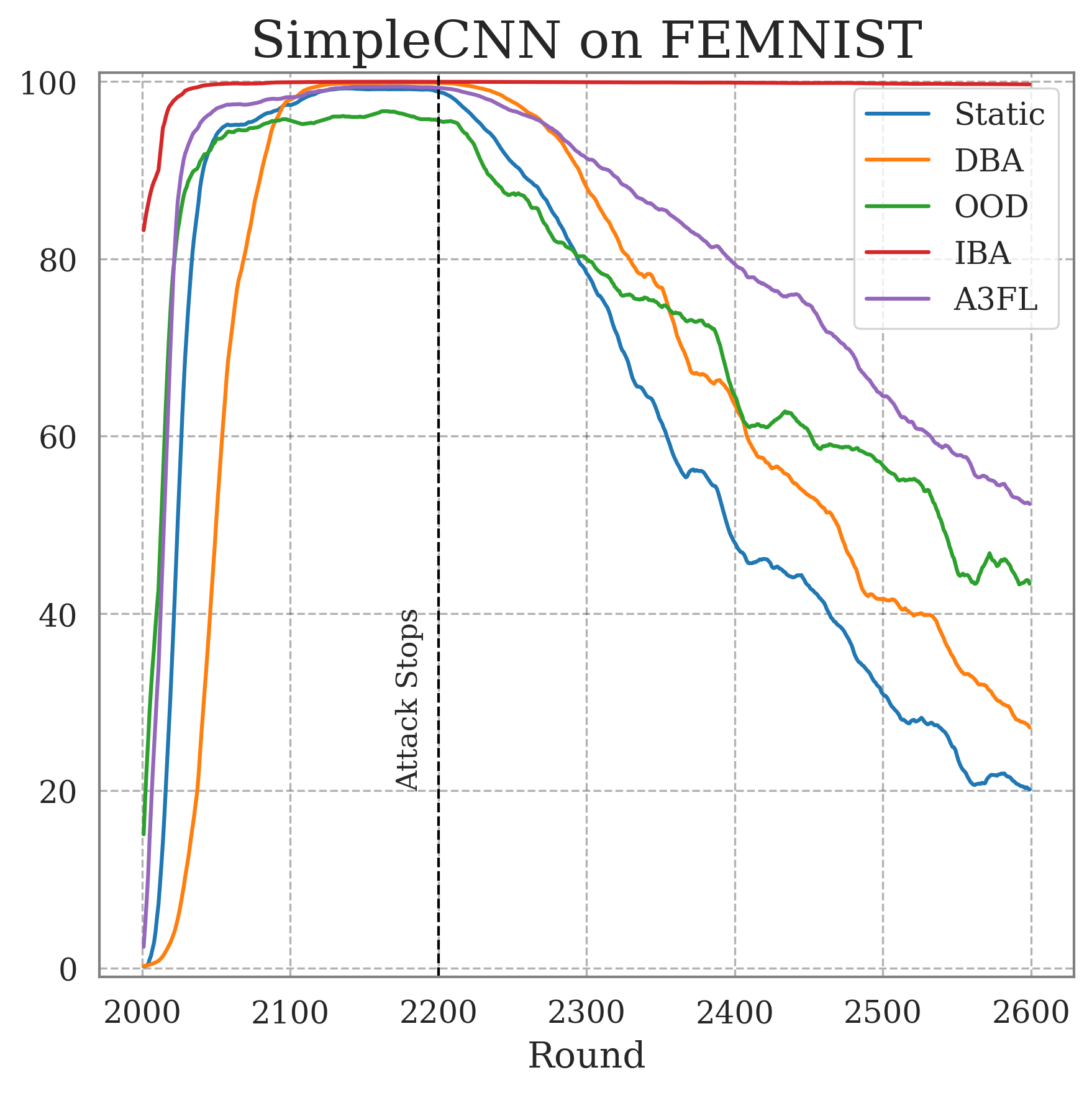}
            \caption{FEMNIST}
            \label{fig:femnist_data_poison}
        \end{subfigure}%
        \begin{subfigure}[t]{0.333\textwidth}
            \centering
            \includegraphics[width=\linewidth]{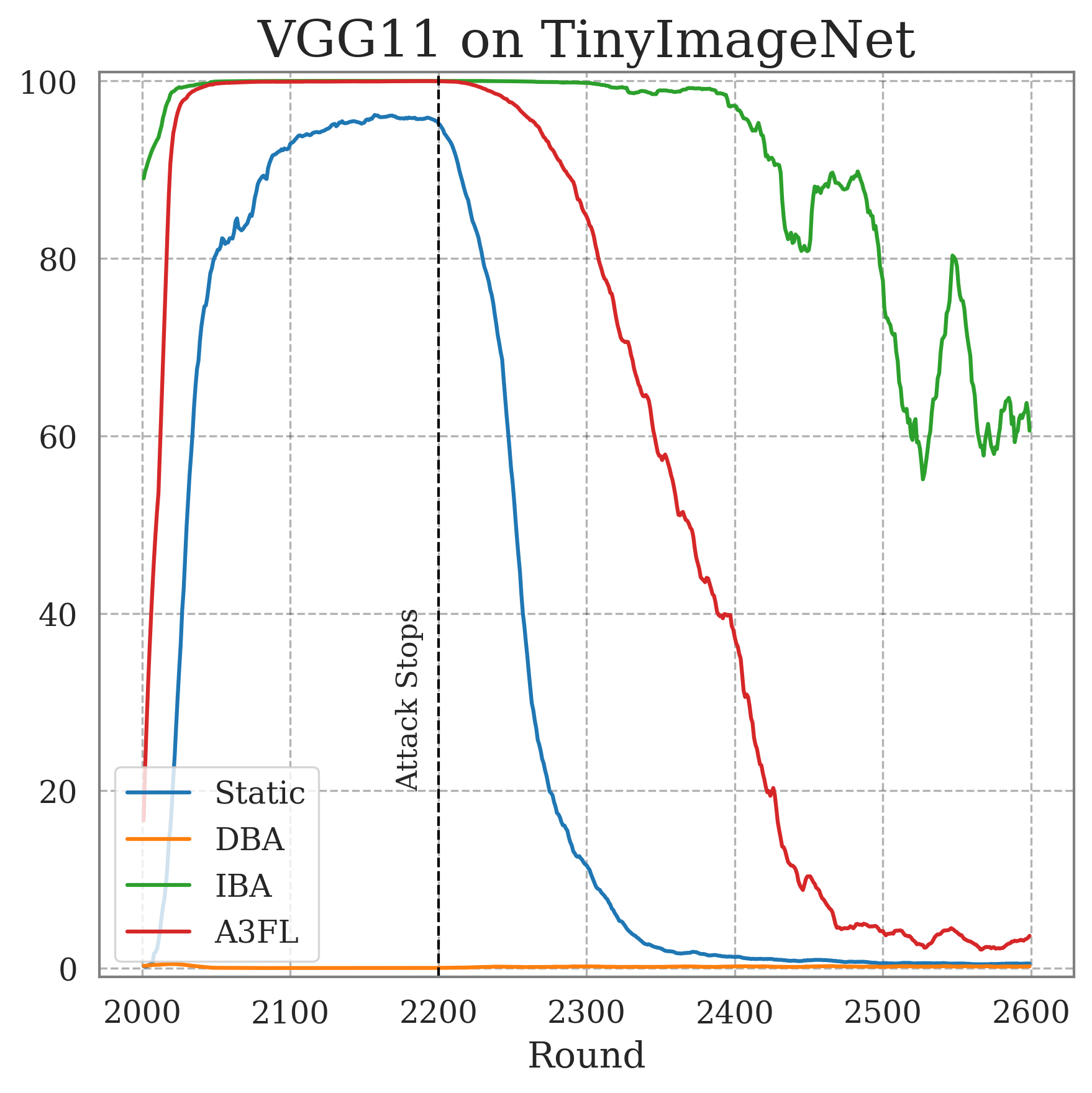}
            \caption{TinyImageNet}
            \label{fig:tiny_data_poison}
        \end{subfigure}
    \end{minipage}%
    \hspace{2mm}
    \begin{minipage}[c]{0.3\textwidth}
        \centering
        \captionsetup{skip=2pt}
        \captionof{table}{Summary metrics. Lowest ACC is \textbf{bolded}, highest ASR is \textcolor{red}{highlighted}.}
        \label{tab:data_poisoning}
        \setlength{\tabcolsep}{3pt}
        \resizebox{\textwidth}{!}{%
            \small
            \begin{tabular}{l|l|ccccc}
            \toprule
            & Metric & Static & DBA & OOD & IBA & A3FL \\ \midrule
            \multirow{3}{*}{(a)}
                & ACC$_\text{min}$ & \textbf{70.0} & 75.7& 79.3 & 71.1 & 84.7 \\
                & ASR$_\text{during}$ & 72.5 & 67.4& 54.7 & \textcolor{red}{98.4} & 92.8 \\
                & ASR$_\text{post}$ & 9.6& 7.2& 31.9& \textcolor{red}{100.0}& 25.4\\
            \midrule
            \multirow{3}{*}{(b)}
                & ACC$_\text{min}$ & 85.6 & 85.9 & 85.9 & \textbf{85.6} & 85.9 \\
                & ASR$_\text{during}$ & 86.3 & 74.2& 89.4 & \textcolor{red}{98.8} & 91.5 \\
                & ASR$_\text{post}$ & 54.5& 64.6& 67.4& \textcolor{red}{99.9}& 78.2\\
            \midrule
            \multirow{3}{*}{(c)}
                & ACC$_\text{min}$ & 50.8 & \textbf{50.0} & --- & 51.6 & \textbf{50.0} \\
                & ASR$_\text{during}$ & 78.9 & 0.1 & --- & \textcolor{red}{99.3} & 95.1 \\
                & ASR$_\text{post}$ & 14.2& 0.2& --- & \textcolor{red}{88.0}& 42.5\\
            \bottomrule
            \end{tabular}
        }
    \end{minipage}
    }
    \caption{Evaluation of Data-Poisoning attacks with 5 trigger types. Figures (left) visualize ASR evolution; Table (right) summarizes metrics: ACC$_\text{min}$ records minimum clean accuracy in a round; ASR$_\text{during}$ and ASR$_\text{post}$ record average ASR during and after the attack phase.}
    \label{fig:data-poison-asr-combined}
\end{figure*}

\subsection{Data and Model-Poisoning Attacks (RQ2).}
While model-poisoning attacks can serve other goals such as defense evasion \cite{layer-aware,3dfed,f3ba}, we focus on \textit{durability-enhancement} attacks \cite{dai2023chameleon,zhang2022neurotoxin,anticipate}, which pose a serious threat for FL systems since the attacker can only poison in a limited number of rounds but leave a long-lasting vulnerability on the global model. For data-poisoning attacks, we compare the performance between different trigger types based on our taxonomy: a \textit{Static} white-patch trigger \cite{howto}, an \textit{OOD} trigger using edge-case samples \cite{edge_case}, the \textit{DBA} attack where a global trigger is divided among malicious clients \cite{xie2020dba}, and dynamic attacks with \textit{A3FL} and \textit{IBA}, which adversarially optimize the triggers with the global model at each round to maximize attack success rate. We note that OOD attack is not evaluated on TinyImageNet since edge-case samples for this dataset are not available.

\noindent\textbf{Model-Poisoning Attacks (\cref{fig:model_poison}).} After fixing the bugs in trigger injection space and backdoor evaluation dataset, we find that durability-enhanced attacks \textbf{show inconsistent performance} across datasets and often \textbf{underperform Baseline attack}. \textit{Neurotoxin} shows only marginal improvements over the baseline, with post-attack ASR increasing by less than 1\% across all datasets. \textit{Anticipate} achieves strong ASR gains on CIFAR-10 and FEMNIST but completely fails on TinyImageNet, where ASR remains below 1\%. It is also important to note that Anticipate could disrupt normal training pattern and degrade ACC, as noted by the low ACC of 32.2\% in CIFAR10. \textit{Chameleon} achieves the strongest ASR gain on CIFAR-10 but performs worse than the Baseline on the other two datasets.

\noindent\textbf{Data-Poisoning Attacks (\cref{fig:data-poison-asr-combined}).} 
Dynamic attacks, particularly IBA, achieve significantly higher ASRs compared to other trigger types. Even after the attack phase, IBA shows no signs of ASR degradation. Another key observation is that \textbf{DBA is not inherently stronger than the centralized (static) attack}. While DBA is more persistent than static attack in FEMNIST, it perform significantly worse in CIFAR10 and TinyImageNet, with ASR only fluctuating around $0.1\%$ for the latter. We believe that as task complexity increases, it becomes harder for local trigger to exert a meaningful signal for backdoor learning, causing the distributed learning of global trigger to fail. While our results contradict the main efficacy claim in \cite{xie2020dba}, an additional study suggests that the effectiveness of DBA \textbf{relies on Model-Replacement strategy} which scales the model update to override aggregation. The performance of the OOD attack, on the other hand, heavily depends on edge-case samples and the target class. Compared to pattern-based attacks, it is generally harder for the model to learn OOD backdoor samples due to the required generalization; however, the implanted backdoor tends to be more persistent.

\begin{table}[h]
\centering
\caption{Average ASR (\%) (round 2000 to 2200) of dynamic attacks with/without training the local models on poisoned samples.}
\label{tab:asr_noattack}
\begin{tabular}{@{}lcccc@{}}
\toprule
            & \textbf{Static} & \textbf{A3FL} & \textbf{CerP} & \textbf{IBA} \\ \midrule
No Attack   & 0.8           & 2.4           & 1.7               & 98.1         \\
With Attack & 72.5          & 92.8          & 85.3              & 98.4         \\ \bottomrule
\end{tabular}%
\end{table}

Given the strong performance of dynamic attacks, we further investigate \textit{whether they are essentially adversarial attacks}. In other words, \textit{can these attacks remain effective even if malicious clients do not train on poisoned samples?} As shown in \cref{tab:asr_noattack}, Static, A3FL, and CerP attacks yield low ASR in the no-attack scenario, as expected. In contrast, IBA maintains nearly identical ASR in both attack and no-attack settings. This indicates that \textbf{IBA behaves fundamentally as an adversarial attack}. Unlike conventional backdoor attacks, IBA does not rely on model poisoning—it can be executed purely as inference-time adversarial examples. Consequently, existing backdoor defenses cannot mitigate IBA as they do not inherently improve global model robustness. We believe that effective defenses against IBA must bridge the gap between backdoor mitigation and adversarial robustness. Potential directions include adapting \textit{adversarial training} to FL by locally generating adversarial examples for robustness improvement \cite{madry2018towards,zhang2019theoretically,lin2020dual}, applying \textit{adversarial purification} at inference to remove trigger-like perturbations \cite{song2017pixeldefend,nie2022diffpure}, and designing \textit{test-time firewall detectors} to identify and reject anomalous inputs \cite{pang2022defense,ma2021effective}.

\subsection{Resource Overhead of Malicious Clients (RQ3).} \cref{tab:attack_resource_overhead} shows that malicious clients often incur significantly higher training costs than benign ones. Attacks such as A3FL and Anticipate prolong local training by up to 574$\times$–108$\times$ and consume over 2$\times$ more GPU memory, making them impractical in resource-constrained FL settings. This increased cost is due to attack optimization, trigger coordination, and higher local epochs for backdoor convergence. This result suggests that a robust FL design should impose \textit{training timeouts} or \textit{update-selection thresholds} to avoid the influence of straggling malicious clients.

\begin{table}[h]
\centering
\caption{Resource overhead and training time comparison between malicious and benign clients on ResNet18 + CIFAR-10.}
\label{tab:attack_resource_overhead}
\resizebox{\linewidth}{!}{%
  \begin{tabular}{@{}lccccccc@{}}
    \toprule
    \textbf{Metric} & \textbf{Benign} & \textbf{Static} & \textbf{A3FL} & \textbf{IBA} & \textbf{NTX} & \textbf{ANT} & \textbf{CML} \\
    \midrule
    Time (s)   & 0.61 & 1.87 & 350.95 & 4.74 & 2.11 & 66.59 & 6.19 \\
    VRAM (GB)  & 0.278 & 0.221 & 0.377 & 0.621 & 0.366 & 6.623 & 0.237 \\
    \bottomrule
    \multicolumn{8}{@{}l@{}}{\footnotesize NTX=Neurotoxin, ANT=Anticipate, CML=Chameleon} \\
  \end{tabular}%
}
\end{table}

\begin{table*}[t]
\centering
\caption{Representative results of RA defenses from round 1001 to 1200. Mean ASR and ACC are reported. We mark \textcolor{red}{ASR $>$ 50\%} and \textbf{ACC $<$ 75\%}. (1) -- Robust Statistics, (2) -- Client-Reweighting, (3) -- DP-based, (4) -- Distance-based.}
\label{tab:robust_aggregation_defenses}
\resizebox{0.95\textwidth}{!}{%
\begin{tabular}{@{}lc|cccccccccc|cccccccccc@{}}
\toprule
\makebox[0pt][l]{Dataset+Model $\rightarrow$} & &
\multicolumn{10}{c|}{CIFAR10 + ResNet18} &
\multicolumn{10}{c}{FEMNIST + SimpleCNN} \\
\midrule
\multirow{2}{*}{Defense} &
\multirow{2}{*}{Metric} & \multicolumn{4}{c}{Baseline} & \multicolumn{3}{c}{PGD} & \multicolumn{3}{c|}{Model-Replacement} &
\multicolumn{4}{c}{Baseline} & \multicolumn{3}{c}{PGD} & \multicolumn{3}{c}{Model-Replacement} \\
\cmidrule(lr){3-6} \cmidrule(lr){7-9} \cmidrule(lr){10-12}
\cmidrule(lr){13-16} \cmidrule(lr){17-19} \cmidrule(lr){20-22}
 &  & \textbf{Static} & \textbf{OOD} & \textbf{A3FL} & \textbf{CerP} & \textbf{Static} & \textbf{OOD} & \textbf{A3FL} & \textbf{Static} & \textbf{OOD} & \textbf{DBA} & \textbf{Static} & \textbf{OOD} & \textbf{A3FL} & \textbf{CerP} & \textbf{Static} & \textbf{OOD} & \textbf{A3FL} & \textbf{Static} & \textbf{OOD} & \textbf{DBA} \\
\midrule
\multirow{2}{*}{\textbf{FedAvg}} & ASR & \textcolor{red}{73.6} & 30.3 & \textcolor{red}{93.2} & \textcolor{red}{86.7} & \textcolor{red}{76.3} & 22.9 & \textcolor{red}{93.8} & \textcolor{red}{96.1} & \textcolor{red}{80.1} & \textcolor{red}{94.6} & \textcolor{red}{86.8} & \textcolor{red}{91.2} & \textcolor{red}{88.1} & \textcolor{red}{82.9} & \textcolor{red}{70.7} & \textcolor{red}{82.7} & \textcolor{red}{74.8} & \textcolor{red}{97.9} & \textcolor{red}{98.4} & \textcolor{red}{97.1} \\
 & ACC & 90.1 & 89.9 & 90.2 & 90.1 & 89.8 & 90.0 & 90.0 & 83.9 & 80.3 & 82.2 & 83.3 & 83.3 & 83.3 & 83.3 & 83.2 & 83.0 & 83.3 & 83.1 & 83.1 & 82.9 \\
\midrule
\multirow{2}{*}{\textbf{CMed (1)}} & ASR & 24.9 & 12.4 & 48.4 & 24.2 & 40.3 & 14.1 & \textcolor{red}{69.7} & 36.9 & 10.5 & 29.0 & 0.4 & \textcolor{red}{56.8} & 3.2 & 1.0 & 0.5 & \textcolor{red}{54.2} & 2.5 & 0.4 & \textcolor{red}{55.7} & 0.4 \\
 & ACC & \textbf{55.7} & \textbf{58.5} & \textbf{56.3} & \textbf{54.1} & \textbf{51.6} & \textbf{56.7} & \textbf{54.3} & \textbf{54.0} & \textbf{55.0} & \textbf{52.4} & 82.3 & 82.2 & 82.2 & 82.3 & 82.2 & 82.2 & 82.2 & 82.2 & 82.2 & 82.3 \\
\midrule
\multirow{2}{*}{\textbf{FoolsGold (2)}} & ASR & \textcolor{red}{72.7} & 25.8 & \textcolor{red}{94.5} & \textcolor{red}{86.3} & \textcolor{red}{74.7} & 20.8 & \textcolor{red}{94.4} & \textcolor{red}{95.9} & \textcolor{red}{78.6} & \textcolor{red}{94.5} & \textcolor{red}{86.7} & \textcolor{red}{91.4} & \textcolor{red}{92.3} & \textcolor{red}{84.5} & \textcolor{red}{70.0} & \textcolor{red}{84.9} & \textcolor{red}{82.5} & \textcolor{red}{98.0} & \textcolor{red}{97.6} & \textcolor{red}{97.6} \\
 & ACC & 89.8 & 90.1 & 89.9 & 89.8 & 89.8 & 90.0 & 89.7 & 83.5 & 80.2 & 82.3 & 86.2 & 86.2 & 86.2 & 86.2 & 86.2 & 86.2 & 86.2 & 85.6 & 85.6 & 85.6 \\
\midrule
\multirow{2}{*}{\textbf{FLTrust (2)}} & ASR & 10.9 & 11.6 & \textcolor{red}{61.3} & \textcolor{red}{56.9} & 6.4 & 8.3 & \textcolor{red}{53.4} & 7.3 & 23.2 & 2.4 & 0.5 & 0.7 & 1.4 & 5.2 & 38.3 & 1.4 & 44.8 & 0.5 & 1.6 & 0.3 \\
 & ACC & \textbf{67.3} & \textbf{67.6} & \textbf{67.2} & \textbf{67.7} & \textbf{68.2} & \textbf{68.1} & \textbf{66.2} & \textbf{68.2} & \textbf{67.3} & \textbf{67.4} & 82.2 & 82.2 & 82.1 & 82.2 & 82.2 & 82.1 & 82.2 & 82.2 & 82.1 & 82.2 \\
\midrule
\multirow{2}{*}{\textbf{WeakDP (3)}} & ASR & 25.2 & 18.3 & \textcolor{red}{85.6} & \textcolor{red}{61.8} & 27.3 & 17.8 & \textcolor{red}{89.6} & 43.5 & 24.0 & \textcolor{red}{58.9} & \textcolor{red}{60.1} & \textcolor{red}{89.4} & \textcolor{red}{75.1} & \textcolor{red}{68.7} & 27.8 & \textcolor{red}{76.8} & \textcolor{red}{51.9} & \textcolor{red}{64.3} & \textcolor{red}{88.7} & \textcolor{red}{74.0} \\
 & ACC & \textbf{73.7} & \textbf{73.7} & \textbf{73.8} & \textbf{73.8} & \textbf{73.4} & \textbf{73.4} & \textbf{73.4} & \textbf{73.6} & \textbf{73.6} & \textbf{73.5} & \textbf{71.1} & \textbf{71.2} & \textbf{71.2} & \textbf{71.0} & \textbf{70.8} & \textbf{70.9} & \textbf{71.0} & \textbf{71.0} & \textbf{71.4} & \textbf{71.2} \\
\midrule
\multirow{2}{*}{\textbf{Krum (4)}} & ASR & 4.0 & 6.5 & 27.1 & 22.2 & 3.7 & 5.5 & 14.7 & 3.8 & 4.8 & 3.8 & 0.3 & 4.0 & 0.3 & \textcolor{red}{93.9} & \textcolor{red}{96.2} & \textcolor{red}{98.1} & \textcolor{red}{97.2} & 0.4 & 4.0 & 0.5 \\
 & ACC & \textbf{74.7} & \textbf{74.7} & \textbf{74.5} & \textbf{75.0} & \textbf{75.0} & \textbf{74.7} & \textbf{74.9} & 75.3 & 75.1 & \textbf{74.8} & \textbf{70.3} & \textbf{70.3} & \textbf{70.5} & \textbf{74.1} & \textbf{74.8} & \textbf{74.3} & \textbf{74.9} & \textbf{71.2} & \textbf{71.0} & \textbf{71.8} \\
\midrule
\multirow{2}{*}{\textbf{RLR (5)}} & ASR & \textcolor{red}{67.7} & 44.5 & \textcolor{red}{91.0} & \textcolor{red}{89.4} & \textcolor{red}{72.4} & 40.4 & \textcolor{red}{90.2} & 1.1 & 1.8 & 1.5 & \textcolor{red}{69.4} & \textcolor{red}{88.3} & \textcolor{red}{81.3} & \textcolor{red}{70.1} & 47.2 & \textcolor{red}{79.7} & \textcolor{red}{66.2} & \textcolor{red}{97.0} & \textcolor{red}{97.5} & 42.1 \\
 & ACC & 84.2 & 83.7 & 84.7 & 84.4 & 83.6 & 83.5 & 84.1 & 82.5 & 80.7 & 82.3 & 84.4 & 84.4 & 84.4 & 84.4 & 84.4 & 84.3 & 84.4 & 84.2 & 84.1 & 77.9 \\
\bottomrule
\end{tabular}%
}
\end{table*}

\begin{table*}[t]
\centering
\caption{Evaluation of AD defenses in round 2001-2200. Mean ASR and ACC are recorded. We highlight \textcolor{red}{ASR $>$ 50\%} and \textbf{ACC $<$ 75\%}.}
\label{tab:ad_defense_results_combined}
\resizebox{0.95\textwidth}{!}{%
\begin{tabular}{@{}lc|cccccccccc|cccccccccc@{}}
\toprule
\makebox[0pt][l]{Dataset+Model $\rightarrow$} & &
\multicolumn{10}{c|}{CIFAR10 + ResNet18} &
\multicolumn{10}{c}{FEMNIST + SimpleCNN} \\
\midrule
\multirow{2}{*}{Defense} &
\multirow{2}{*}{Metric} & \multicolumn{4}{c}{Baseline} & \multicolumn{3}{c}{PGD} & \multicolumn{3}{c|}{Model-Replacement} &
\multicolumn{4}{c}{Baseline} & \multicolumn{3}{c}{PGD} & \multicolumn{3}{c}{Model-Replacement} \\
\cmidrule(lr){3-6} \cmidrule(lr){7-9} \cmidrule(lr){10-12}
\cmidrule(lr){13-16} \cmidrule(lr){17-19} \cmidrule(lr){20-22}
 &  & \textbf{Static} & \textbf{OOD} & \textbf{A3FL} & \textbf{CerP} & \textbf{Static} & \textbf{OOD} & \textbf{A3FL} & \textbf{Static} & \textbf{OOD} & \textbf{DBA} & \textbf{Static} & \textbf{OOD} & \textbf{A3FL} & \textbf{CerP} & \textbf{Static} & \textbf{OOD} & \textbf{A3FL} & \textbf{Static} & \textbf{OOD} & \textbf{DBA} \\
\midrule
\multirow{2}{*}{\textbf{FedAvg}} & ASR & \textcolor{red}{73.2} & 39.3 & \textcolor{red}{93.1} & \textcolor{red}{85.3} & \textcolor{red}{75.0} & 29.8 & \textcolor{red}{93.2} & \textcolor{red}{95.7} & \textcolor{red}{82.0} & \textcolor{red}{75.9} & \textcolor{red}{86.6} & \textcolor{red}{91.1} & \textcolor{red}{92.2} & \textcolor{red}{84.3} & \textcolor{red}{69.8} & \textcolor{red}{84.4} & \textcolor{red}{83.3} & \textcolor{red}{97.6} & \textcolor{red}{97.9} & 45.9 \\
 & ACC & 91.0 & 90.9 & 91.0 & 91.0 & 90.9 & 90.9 & 91.1 & 85.8 & 82.6 & 84.1 & 86.8 & 86.8 & 86.9 & 86.9 & 86.9 & 86.8 & 86.9 & 86.2 & 86.2 & 86.2 \\
\midrule
\multirow{2}{*}{\textbf{MKrum}} & ASR & \textcolor{red}{76.6} & 46.3 & \textcolor{red}{94.1} & \textcolor{red}{87.3} & \textcolor{red}{79.3} & 36.8 & \textcolor{red}{94.1} & 0.9 & 5.8 & 0.8 & \textcolor{red}{77.7} & \textcolor{red}{91.1} & \textcolor{red}{88.6} & \textcolor{red}{86.1} & \textcolor{red}{74.5} & \textcolor{red}{86.8} & \textcolor{red}{85.7} & 0.1 & 4.6 & 0.1 \\
 & ACC & 90.8 & 90.8 & 90.9 & 91.1 & 91.0 & 90.4 & 90.9 & 91.0 & 90.3 & 91.0 & 86.8 & 86.8 & 86.8 & 86.8 & 86.8 & 86.8 & 86.8 & 86.8 & 86.8 & 86.8 \\
\midrule
\multirow{2}{*}{\textbf{AlignIns}} & ASR & 1.0 & 4.1 & \textcolor{red}{88.8} & 4.6 & 1.0 & 4.1 & 3.6 & 1.1 & 9.1 & 0.9 & 0.1 & \textcolor{red}{66.4} & \textcolor{red}{51.4} & 0.1 & 0.1 & 48.1 & 0.1 & \textcolor{red}{89.5} & 46.3 & \textcolor{red}{88.7} \\
 & ACC & 90.2 & 89.8 & 89.7 & 90.0 & 90.4 & 90.2 & 90.1 & 89.8 & 90.0 & 89.8 & 86.5 & 86.3 & 86.4 & 86.5 & 86.6 & 86.3 & 86.5 & \textbf{6.5} & \textbf{7.9} & \textbf{6.5} \\
\midrule
\multirow{2}{*}{\textbf{DeepSight}} & ASR & 0.9 & 4.5 & 2.3 & 1.6 & 0.8 & 4.9 & 2.9 & 0.8 & 18.2 & 0.7 & \textcolor{red}{76.6} & \textcolor{red}{90.0} & \textcolor{red}{89.5} & \textcolor{red}{81.1} & \textcolor{red}{67.2} & \textcolor{red}{83.5} & \textcolor{red}{80.5} & \textcolor{red}{78.5} & \textcolor{red}{91.2} & 21.2 \\
 & ACC & 89.3 & 90.6 & 90.5 & 90.7 & 90.7 & 90.3 & 91.1 & 91.2 & 90.0 & 90.3 & 86.9 & 86.8 & 86.9 & 86.9 & 86.9 & 86.8 & 86.9 & 86.8 & 86.8 & 86.8 \\
\midrule
\multirow{2}{*}{\textbf{FLAME}} & ASR & \textcolor{red}{56.2} & 29.1 & \textcolor{red}{88.8} & \textcolor{red}{82.6} & 49.5 & 20.7 & \textcolor{red}{90.3} & \textcolor{red}{78.4} & \textcolor{red}{55.0} & 28.9 & \textcolor{red}{72.8} & \textcolor{red}{85.6} & \textcolor{red}{82.5} & \textcolor{red}{81.3} & 45.2 & \textcolor{red}{83.5} & \textcolor{red}{70.4} & \textcolor{red}{80.2} & \textcolor{red}{90.7} & 23.2 \\
 & ACC & 86.0 & 86.0 & 86.3 & 86.1 & 85.9 & 86.1 & 85.5 & 83.5 & 82.8 & 84.0 & 86.3 & 86.2 & 86.3 & 86.3 & 86.3 & 86.2 & 86.3 & 86.2 & 86.1 & 86.1 \\
\midrule
\multirow{2}{*}{\textbf{MultiM}} & ASR & \textcolor{red}{59.4} & 45.8 & \textcolor{red}{91.1} & \textcolor{red}{76.8} & \textcolor{red}{55.2} & 11.3 & \textcolor{red}{86.6} & \textcolor{red}{92.4} & \textcolor{red}{68.8} & 48.8 & \textcolor{red}{77.1} & \textcolor{red}{93.2} & \textcolor{red}{92.0} & \textcolor{red}{81.1} & \textcolor{red}{68.6} & \textcolor{red}{92.0} & \textcolor{red}{83.4} & 0.2 & 23.3 & 0.1 \\
 & ACC & 89.3 & 87.2 & 89.6 & 89.8 & 89.1 & 89.6 & 89.2 & 80.7 & 80.5 & 80.8 & 83.0 & 83.2 & 83.2 & 83.2 & 82.9 & 83.5 & 83.4 & 84.1 & 84.6 & 84.1 \\
\midrule
\multirow{2}{*}{\textbf{Indicator}} & ASR & 1.2 & 7.7 & \textcolor{red}{86.9} & 2.7 & 1.1 & 32.3 & 23.5 & 1.1 & 4.3 & 1.1 & 5.1 & 39.1 & \textcolor{red}{50.2} & 0.0 & 3.9 & \textcolor{red}{96.5} & 5.8 & \textcolor{red}{58.9} & 44.3 & 34.5 \\
 & ACC & 89.0 & 88.8 & 89.3 & 88.7 & 88.4 & 89.3 & 88.7 & 89.0 & 88.4 & 89.2 & \textbf{37.1} & \textbf{45.7} & \textbf{38.7} & \textbf{5.4} & \textbf{14.6} & \textbf{5.5} & \textbf{34.5} & \textbf{47.3} & \textbf{45.3} & \textbf{5.3} \\
\midrule
\multirow{2}{*}{\textbf{FedDLAD}} & ASR & 1.5 & 9.6 & 3.3 & 2.6 & 3.0 & 9.6 & 8.6 & \textcolor{red}{52.7} & 45.9 & 9.9 & 0.1 & 10.9 & 0.2 & 0.1 & 0.0 & 0.0 & 0.0 & 0.2 & 46.9 & 0.3 \\
 & ACC & 89.2 & 88.3 & 89.6 & 89.3 & 89.3 & 89.0 & 89.6 & 90.8 & 90.4 & 90.9 & 80.8 & 81.3 & 81.4 & 81.7 & 85.6 & 85.3 & 83.2 & 80.9 & 81.3 & 80.7 \\
\bottomrule
\end{tabular}%
}
\end{table*}

\subsection{Evaluation of Server-Side Defenses (RQ4).}

\textbf{Robust Aggregation (RA) Defenses.} We evaluate 13 RA defenses against 11 attacks created by combining Model-Poisoning (second row) and Data-Poisoning (third row) strategies. This combination is carefully designed to guarantee the diversity of attack mechanisms. Baseline attacks apply no modifications to client training; PGD projects malicious updates within an $\ell_\infty$ bound of the global model after each gradient step; Model-Replacement scales up poisoned updates to disproportionately influence the global aggregation. Since RA defenses modify aggregation rule, we train each of these defense from scratch until round 1000, when attacks happen in the following 200 rounds. \cref{tab:robust_aggregation_defenses} shows representative results of best-performing defense for each sub-category. We have the following observations:
\begin{enumerate}
    \item The general conclusion is that no RA defenses can effectively counter all attacks. Dynamic attacks (A3FL, CerP) are particularly strong and can bypass almost all defenses. \textit{Krum} displays the best overall trade-off, but it remains vulnerable to PGD attacks in FEMNIST.
    \item RA defenses with stronger robustness (\textit{Krum}, \textit{CMed}, \textit{Krum}) suffer from \textit{greater} ACC degradation. Besides, we observe that DP-based defenses and RLR exhibit unstable training dynamics with potential loss explosions if the hyperparameters are not carefully configured, making them undesirable for real-world deployment.
    \item Model-Replacement attacks are particularly effective against FedAvg but vulnerable to RA defenses.
\end{enumerate}

\noindent\textbf{Anomaly Detection (AD) Defenses.} We further assess seven state-of-the-art AD defenses under identical attack configurations. All experiments initialize from the same global model (FedAvg at round 2000). \textit{MKrum} is adapted for anomaly detection by applying a selection threshold of $0.5K$ per round (we inspect that the number of malicious clients never surpasses $0.5K$ in a round). \cref{tab:ad_defense_results_combined} reports only ASR and ACC, while detection-specific metrics are deferred to Appendix~\ref{sec:full_result}. Our findings are as follows:

\begin{enumerate}
\item \textit{FedDLAD} \cite{feddlad} demonstrates the strongest overall performance in ASR and ACC across both datasets. However, it is susceptible to Model-Replacement attacks.

\item \textbf{Inconsistent performance.} \textit{DeepSight} and \textit{Indicator} perform well on CIFAR10 but fail on FEMNIST. \textit{MKrum} is effective only against Model-Replacement attacks and fails against Baseline and PGD attacks.

\item Similar to RA defenses, AD defenses are also more vulnerable to dynamic attacks. PGD strategy does not offer better resistance to AD defenses than Baseline strategy. 

\item \textbf{Defense trade-offs.} Most AD defenses preserve accuracy, except \textit{Indicator} and \textit{AlignIns} on FEMNIST. Furthermore, \cref{tab:aggregation-time} shows that all methods (except \textit{AlignIns}) incur at least $10\times$ longer aggregation time than FedAvg. Notably, the \textit{DeepSight} server requires 25.2s on average for aggregation, compared to 0.06s for FedAvg. 
\end{enumerate}

\noindent\textbf{Recommendation.} Our study suggests that future defenses should (1) focus on mitigating dynamic attacks such as A3FL and CerP, (2) be efficient enough to leave only minimal aggregation overhead, (3) minimize accuracy reduction. The inconsistent performance across datasets and architectures suggests that defenses should conduct a more holistic evaluation to ensure generalization.

\vspace{0.3em}

\noindent\textbf{Additional Studies.} In the main paper, we fix several key hyperparameters, including the server learning rate and the number of malicious clients. To provide a more comprehensive evaluation, we conduct an extended set of experiments that vary these hyperparameters, including a detailed sensitivity analysis of attack settings, in Appendix~\ref{sec:full_result}.

\section{Conclusion}

This paper introduces BackFed, a standardized and efficient framework for benchmarking backdoor attacks in FL. We uncover key evaluation inconsistencies and fidelity bugs in prior work, and conduct a comprehensive benchmark that reveals the limitations of existing methods. We also identify two simple design choices to improve the robustness of FL systems: using a lower server learning rate and increasing the client sampling ratio with a low enforced selection threshold. We hope BackFed will provide a reliable evaluation framework for future research on FL backdoor attacks.

{
  \small
  \bibliographystyle{ieeenat_fullname}

}

\clearpage
\setcounter{page}{1}
\maketitlesupplementary


\noindent The structure of the Appendix:
\begin{itemize}
    \item \cref{sec:framework_struc} describes the structure of BackFed in detail and compares it with FLPoison and Backdoors101.
    \item \cref{sec:atk_des} and \cref{sec:def_des} describe the evaluated attacks and defenses and their settings in BackFed benchmark.
    \item \cref{sec:full_result} shows the full results of defense experiments together with a sensitivity analysis of attacks.
\end{itemize}

\section{Dataset Partitioning}
We show the data distribution of Dirichlet-based partitioning for CIFAR10 and Tiny-ImageNet that we use in our experimentts in the two figures below.

\begin{figure}[h]
    \centering
    \includegraphics[width=\linewidth]{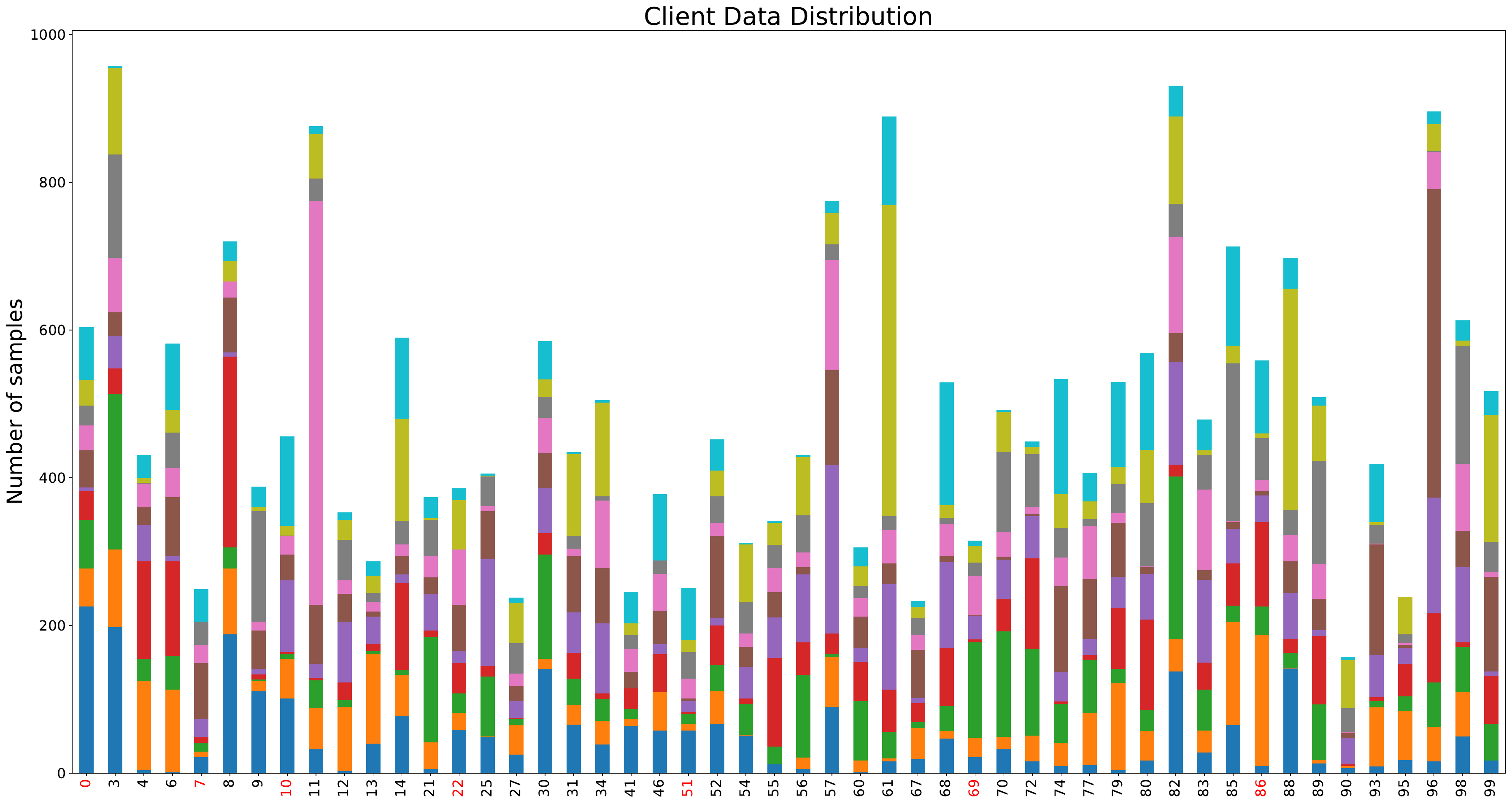}
    \caption{Data distribution of 50 clients with Dirichlet $\alpha=0.9$ on CIFAR10. Red ticks are datasets of malicious clients.}
    \label{fig:cifar_dis}
\end{figure}

\begin{figure}[h]
    \centering
    \includegraphics[width=\linewidth]{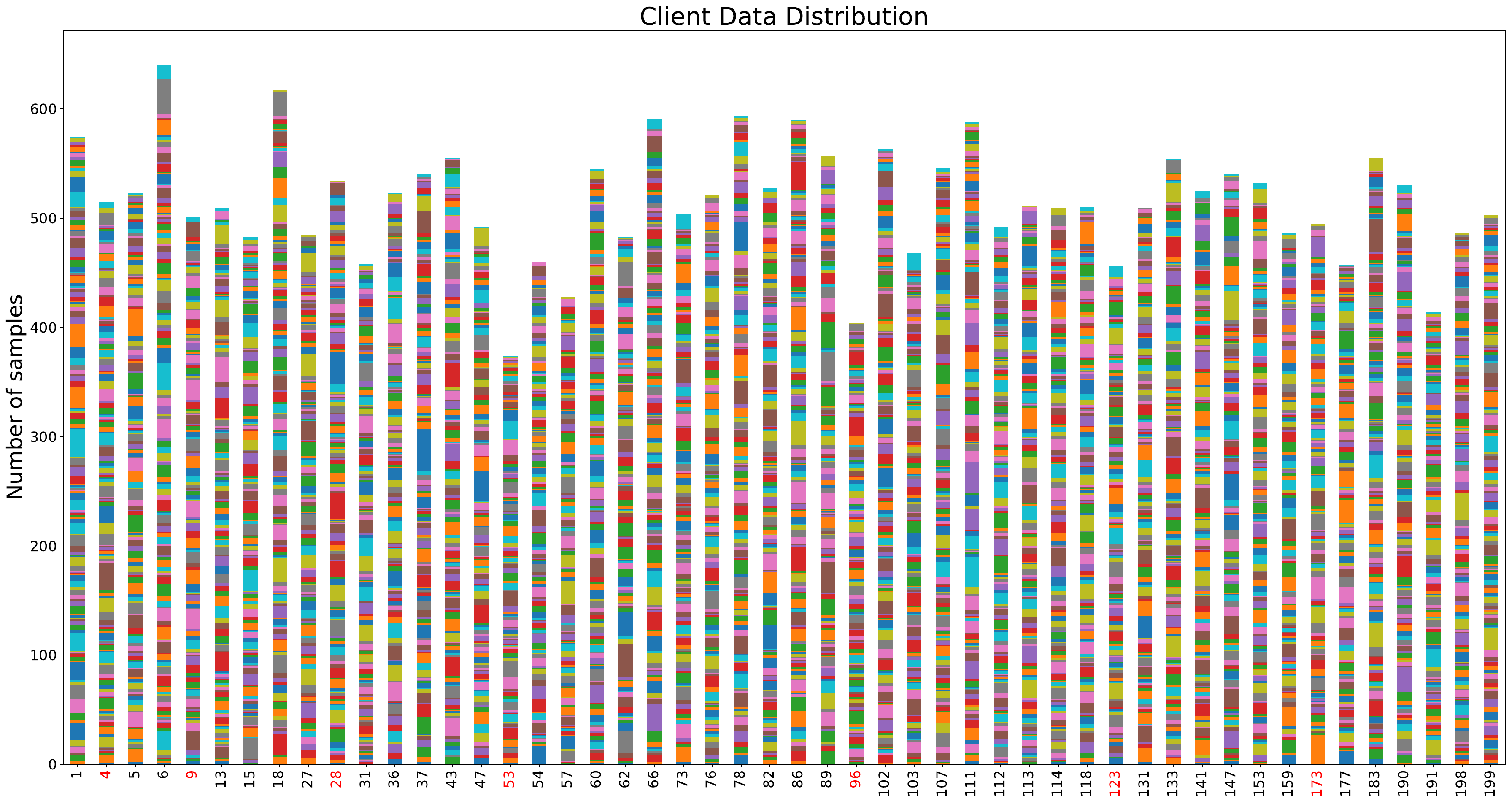}
    \caption{Data distribution of 50 clients with Dirichlet $\alpha=0.5$ on Tiny-ImageNet. Red ticks are datasets of malicious clients.}
    \label{fig:tiny_dist}
\end{figure}

\section{BackFed Framework Structure}
\label{sec:framework_struc}

\begin{figure}
    \centering
    \includegraphics[width=\linewidth]{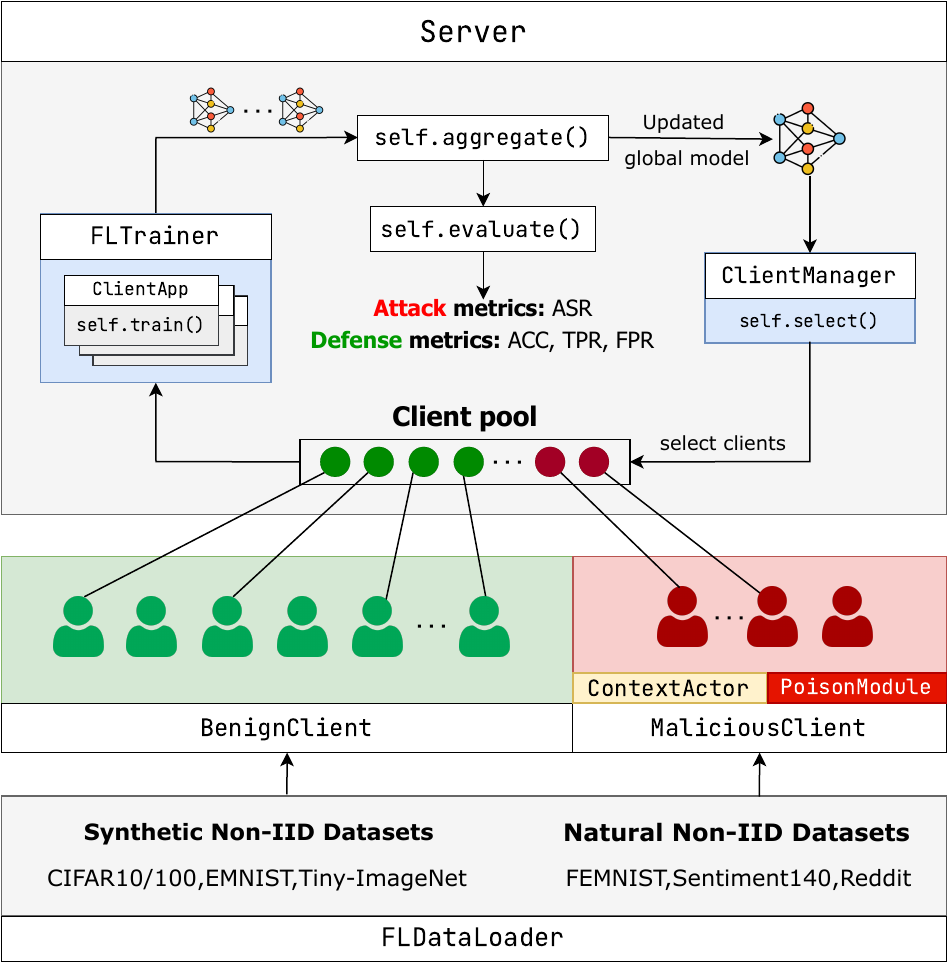}
    \caption{The overall structure and pipeline of BackFed.}
    \label{fig:framework_structure_appendix}
\end{figure}

\noindent\textbf{FLDataLoader.}
The \textbf{\texttt{FLDataLoader}} module prepares dataset partitions among clients. It supports two partitioning strategies: (1) \textit{Dirichlet-based partitioning} using parameter $\alpha$ to control the degree of non-IIDness—lower $\alpha$ values create stronger heterogeneity across clients; (2) \textit{Natural splits} that preserve natural data distributions in datasets like FEMNIST (partitioned by writer), Sentiment140 (by user), and Reddit (by subreddit). All partitions are cached to disk to avoid redundant preprocessing across experiments.

\noindent\textbf{Server Module.}
The \textbf{\texttt{Server}} module orchestrates the entire FL training process through two key components:
\begin{itemize}[leftmargin=*]
    \item \textbf{\texttt{ClientManager}} handles client selection and attack scheduling. It implements multiple selection strategies. The two examples include \textit{random sampling} where clients are selected uniformly at random each round and \textit{fixed-frequency poisoning} where malicious clients are scheduled every $f$ rounds.
    
    \item \textbf{\texttt{FLTrainer}} manages all client-server interactions through a Ray ActorPool—a fixed-size pool of worker processes that execute client training in parallel. In each round, \texttt{FLTrainer}: (1) broadcasts the global model and training configuration to selected clients, (2) collects model updates and evaluation metrics asynchronously as clients complete training, (3) enforces optional timeout constraints or selection thresholds to prevent stragglers, and (4) submits collected updates to the server for aggregation. Beyond standard FL operations, \texttt{FLTrainer} supports custom client-server interactions through function references, enabling flexible protocol modifications for specialized defenses.
\end{itemize}
Server-side defenses are implemented as \texttt{Server} subclasses. Robust aggregation defenses (e.g., Krum, Trimmed-Mean, FLTrust) override the \texttt{aggregate()} method to modify the aggregation rule. Anomaly detection defenses (e.g., FLAME, DeepSight) perform \texttt{detect\_anomalies()} to filter out malicious updates before the aggregation. Our implementation tracks aggregation time for each defense to assess deployability.

\noindent\textbf{Client Module.}
BackFed maintains a pool of persistent client actors to avoid repeated initialization overhead. The number of actors in the pool determines the extent of parallel training. Two types of clients in BackFed are:

\begin{itemize}[leftmargin=*]
    \item \texttt{BenignClient} represents honest participants who train normally with standard hyperparameters (learning rate, batch size, number of local epochs). These clients simply minimize the cross-entropy loss on their local data and return model updates $\Delta_i = \theta_i - \theta^t$ to the server.
    
    \item \texttt{MaliciousClient} extends benign behavior with backdoor attack capabilities through two mechanisms:
    \begin{itemize}[leftmargin=*]
        \item \textit{Data poisoning} via \texttt{PoisonModule}, which injects triggers into training batches and flips labels to the target class. The module supports multiple trigger types: static patterns (fixed across rounds), distributed triggers (DBA/FCBA where each malicious client holds a fragment), dynamic triggers (IBA/A3FL optimized each round), and edge-case triggers (OOD samples).
        \item \textit{Model poisoning} through subclass-specific methods that extend the base \texttt{MaliciousClient} to manipulate the training process and/or the final update. For example, model-replacement scales the update by a factor $\lambda$ to dominate aggregation; Neurotoxin projects updates into dormant parameter subspace to maximize backdoor retention.
    \end{itemize}
    
    \item \texttt{BenignClient} with client-side defenses implements protective mechanisms that can be applied locally, such as norm clipping (FL-WBC), differential privacy noise injection (LDP), or trigger inversion (FLIP). These defenses are applied before submitting updates to the server.
\end{itemize}
All malicious clients share access to a \texttt{ContextActor}—a Ray Actor that serves as a centralized communication hub for synchronizing attack resources across parallel processes. This is crucial for dynamic attacks where one malicious client optimizes the trigger each round and shares it with others for coordinated poisoning.

\noindent\textbf{PoisonModule.}
The \texttt{PoisonModule} encapsulates all backdoor-specific operations to enable modular composition of attacks. Its core functionalities include:

\begin{itemize}[leftmargin=*]
    \item \textit{Trigger management}: Generates and maintains triggers according to the attack type. For static triggers, this is a fixed pattern (e.g., 5×5 white square). For dynamic triggers, it maintains a generator network (U-Net for IBA) or optimized trigger pattern (A3FL, CerP) that are updated in each poisoning round based on feedback from the global model.
    
    \item \textit{Batch poisoning}: Injects triggers into training batches during local training. For a batch of size $B$ with poison ratio $\rho$, the module: (1) selects the first $\lfloor B \cdot \rho \rfloor$ samples, (2) applies the trigger transformation, (3) flips labels to the target class, (4) returns the poisoned batch for training. The injection occurs \textit{before} input normalization to ensure triggers are realistic at inference time.
    
    \item \textit{Attack evaluation}: Computes Attack Success Rate (ASR) on a held-out test set by: (1) applying triggers to all non-target-class samples, (2) feeding them through the global model, (3) measuring the fraction classified as the target class. This evaluation is performed on the server side to track attack progress.
    
    \item \textit{Trigger synchronization}: For dynamic attacks requiring coordination, \texttt{PoisonModule} communicates with \texttt{ContextActor} to retrieve or update the latest optimized trigger. This ensures consistent triggers among malicious clients within each round.
\end{itemize}

By decoupling trigger types from model-poisoning strategies, \texttt{PoisonModule} enables flexible attack composition. For example, researchers can easily test IBA (dynamic trigger) combined with model-replacement (update scaling) by simply configuring both components without modifying the attack logic.

\noindent\textbf{Parallel Execution Pipeline.}
BackFed's efficiency stems from its Ray-based execution model. At the start of each round: (1) \texttt{FLTrainer} submits client training tasks to the ActorPool, (2) Ray schedules tasks across available CPU cores or GPUs based on resource availability, (3) Clients execute training independently and asynchronously, (4) \texttt{FLTrainer} collects updates as they complete (with optional timeout), (5) The server aggregates collected updates and broadcasts the new global model. This pipeline enables true parallelism—when 10 clients are selected and 8 CPU cores are available, 8 clients train simultaneously while 2 wait in the queue, dramatically reducing wall-clock time compared to sequential execution.

\begin{figure}[!h]
    \centering
    \includegraphics[width=\linewidth]{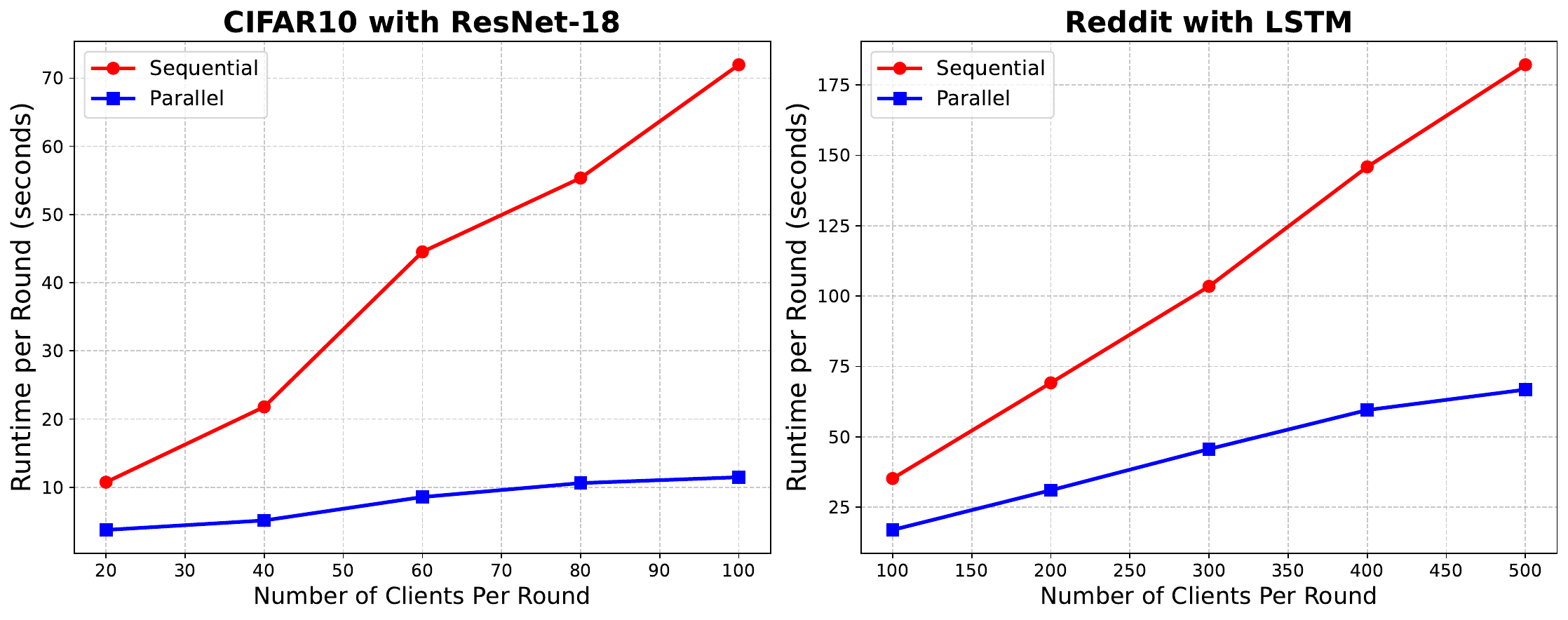}
    \caption{Average runtime of a communication round with increasing number of clients per round on CIFAR10 and Reddit.}
    \label{fig:scalability}
\end{figure}

\subsection{Efficiency of BackFed}
We evaluate the efficiency of BackFed's multi-processing pipeline by recording the average runtime of a communication round as the number of sampled clients per round increases. All experiments are performed on 4 A5000 GPUS. As shown in \cref{fig:scalability}, BackFed's parallel training delivers consistent speedups in both CIFAR10 and Reddit, especially as the number of selected clients increases. For instance, when training with 100 clients per round on CIFAR-10, BackFed achieves over a 7$\times$ reduction in runtime compared to sequential training.

\begin{table}[!ht]
\centering
\small
\caption{Taxonomy of backdoor attack in FL.}
\resizebox{\columnwidth}{!}{%
\begin{tabular}{@{}lll@{}}
\toprule
\textbf{Category} & \textbf{Sub-category} & \textbf{Literature} \\
\midrule
\multirow{4}{2.5cm}{\textbf{Data Poisoning}} 
& Static trigger& \cite{howto,blended,gu2019badnets,sig} \\
& Distributed trigger & \cite{fcba,xie2020dba} \\
& Edge-case trigger & \cite{edge_case} \\
& Dynamic trigger & \cite{nguyen2024iba,zhang2024a3fl,f3ba,ceberus} \\
\midrule
\multirow{3}{2.5cm}{\textbf{Model Poisoning}} 
& Durability-enhancement & \cite{zhang2022neurotoxin,anticipate,dai2023chameleon} \\
& Defense-evasion & \cite{3dfed,layer-aware,howto,edge_case} \\
& Model-replacement & \cite{howto} \\
\bottomrule
\end{tabular}%
}
\label{tab:attack-taxonomy}
\end{table}

\begin{figure*}
    \centering
    \includegraphics[width=\linewidth]{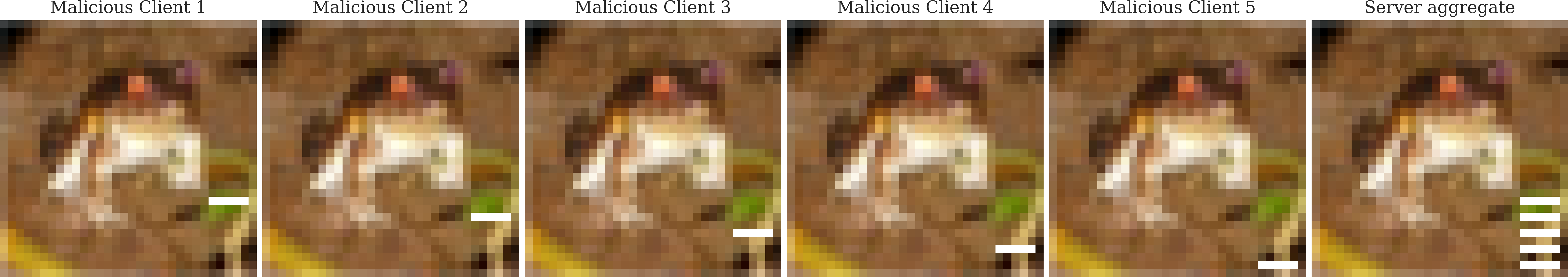}
    \caption{Visualization of poisoned images of DBA attack in CIFAR10.}
    \label{fig:dba_visualize}
\end{figure*}

\begin{figure*}
    \centering
    \includegraphics[width=\linewidth]{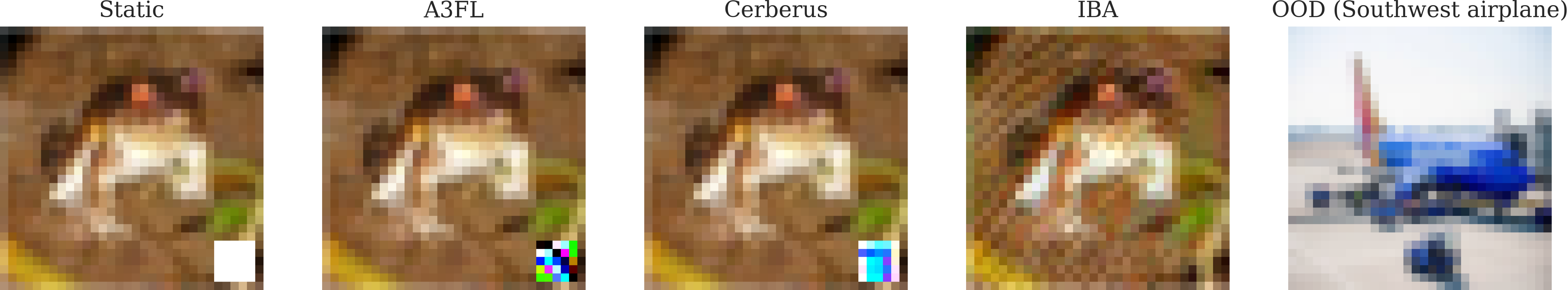}
    \caption{Visualization of poisoned images of 5 data-poisoning attacks in CIFAR10.}
    \label{fig:data_poison_visual}
\end{figure*}

\section{Description of Backdoor Attacks}
\label{sec:atk_des}

\subsection{Data Poisoning Attacks}


\noindent\textbf{Static.} Static attack uses a fixed trigger pattern that is static across rounds and shared by all malicious clients. We follow \cite{howto} to use a white square patch.

\noindent\textbf{DBA.} DBA~\cite{xie2020dba} attack splits a global trigger into multiple local patterns, with each malicious client contributing only a partial trigger.

\noindent\textbf{OOD.} Edge-case attack chooses samples that lie on the tail of the data distribution~\cite{edge_case} as triggers. Rather than adding artificial triggers such as a square pattern, the adversary selects semantically natural but rarely occurring samples such as ``Southwest airplane images'' for the class `airplane'. 

\noindent\textbf{A3FL.} A3FL~\cite{zhang2024a3fl} improves backdoor persistence by anticipating how the global model will attempt to unlearn the trigger during training. 
Instead of optimizing the trigger only for the current global model $\theta_t$, A3FL also constructs an adversarially updated model $\theta'_t$ that simulates the worst-case ``unlearning'' behavior. 
The trigger is then jointly optimized to remain effective on both $\theta_t$ and $\theta'_t$, effectively bridging the local-to-global gap and making the backdoor more resistant to server-side aggregation and defenses.

Formally, A3FL solves the following bilevel optimization problem:
\begin{equation*}
\arg\min_{\delta} \;
\mathbb{E}_{(x,y)\sim D_i}
\left[
\mathcal{L}(x \oplus \delta, \tilde{y}; \theta_t)
+
\lambda \, \mathcal{L}(x \oplus \delta, \tilde{y}; \theta'_t)
\right],
\label{eq:a3fl_outer}
\end{equation*}

\noindent
where the adversarial global model $\theta'_t$ is defined as
\begin{equation*}
\theta'_t
= \arg\min_{\theta} \;
\mathbb{E}_{(x,y)\sim D_i}
\left[
\mathcal{L}(x \oplus \delta, y;\theta)
\right],
\qquad
\theta \leftarrow \theta_t.
\label{eq:a3fl_inner}
\end{equation*}

\noindent
Here, $\delta$ is the backdoor trigger, $\tilde{y}$ is the target label, and $\lambda$ balances the backdoor effectiveness between the current global model and the adversarially adapted one. 
This formulation explicitly bridges the local–global gap by ensuring the optimized trigger remains effective even under simulated worst-case unlearning.

\noindent\textbf{IBA.} IBA~\cite{nguyen2024iba} learns an instance-specific trigger generator $T_{\xi}$ that transforms any benign input $x$ into a poisoned version whose prediction is pushed toward an attacker-specified target label $y_T$. The generator parameters $\xi$ are optimized so that the triggered samples reliably induce the target behavior while remaining visually unobtrusive. Formally, the trigger generator is trained by solving:
\begin{equation*}
\label{eq:iba_argmin}
\xi^{*}=\arg\min_{\xi} \;
\mathbb{E}_{x \sim \mathcal{D}}
\Big[
\mathcal{L}\big( f_{w_k}( T_{\xi}(x) ), \, y_T \big)
\Big],
\end{equation*}
where $f_{w_k}$ is the local model, $T_{\xi}$ is the learnable trigger generator, and $y_T$ is the attacker's target label.

\subsection{Model Poisoning Attacks}

\noindent\textbf{PGD. }~We adopt a strong model-poisoning baseline based on projected gradient descent (PGD) optimization~\cite{edge_case,sun2019can}. 
Instead of relying solely on poisoned data, the adversary directly optimizes its local update in parameter space to maximize an attack objective (e.g., increasing the global loss or enforcing a targeted backdoor), while enforcing a norm constraint so that the malicious update remains within the server’s clipping or detection thresholds.

\noindent\textbf{Model Replacement. }~Model replacement~\cite{howto} is a canonical, highly effective model-poisoning attack in FL. 
A malicious client locally trains a backdoored model and then scales its update so that, after aggregation, the global model parameters are approximately replaced by the adversary’s model. 
This enables the attacker to implant a backdoor in a single or a few rounds while preserving high accuracy on benign inputs.

\noindent\textbf{Neurotoxin. }~Neurotoxin~\cite{zhang2022neurotoxin} is a durability-enhanced backdoor attack that focuses on the manipulation of parameters whose magnitudes change the least during benign training, where they call bottom neurons. 
By leveraging these rarely updated neurons, the attack substantially increases the lifespan of the implanted backdoor across subsequent federated rounds, even after the adversary stops participating.

\noindent\textbf{Chameleon. }~Chameleon~\cite{dai2023chameleon} further improves backdoor durability by adapting poisoned samples to their benign ``peer’’ images using contrastive learning. 
By explicitly shaping the relationship between benign and poisoned examples with the original and target labels, Chameleon plants backdoors that survive many more FL rounds and remain effective under a wide range of datasets, trigger types, and model architectures.

\subsection{Attack Settings in our benchmark}
We standardize common attack configurations to ensure a fair comparison among attacks. For each dataset, we use a square trigger with a consistent size and location across data-poisoning attacks: FEMNIST (4×4), CIFAR-10 (5×5), and TinyImageNet (10×10). For DBA, the global trigger has the same pixel size as centralized attacks and is split horizontally among malicious clients (\cref{fig:dba_visualize}). 

Poisoning is performed online, modifying 31.25\% of each training batch. Malicious clients train for 6 local epochs with SGD (lr=0.05, momentum=0.9, weight decay=0.005). We assume 10\% of clients are adversarial under random sampling, and poisoning occurs in multi-shot rounds (2001–2200) whenever an adversarial client is selected. Each malicious client holds a distinct local subset, avoiding unrealistic assumptions that require excessive shared data as in previous works \cite{howto,zhang2022neurotoxin,dai2023chameleon}.

Dataset-specific parameters are listed in \cref{tab:backfed-atk-config}. We set other method-specific hyperparameters based on the official codebases and the original papers. Since these hyperparameters are dataset-dependent, we refer readers to BackFed codebase for more details.

\begin{table*}[tbh]
\centering
\caption{Backdoor attack configurations across datasets and poisoning regimes.}
\label{tab:backfed-atk-config}
\begin{tabular}{llcccccccc}
\toprule
\multicolumn{1}{l}{\textbf{Dataset}} & \multicolumn{1}{c}{\textbf{Setting}} &
\makecell[c]{\textbf{Aversary}\\ \textbf{Fracion}}
 &
\makecell[c]{\textbf{Attack}\\ \textbf{Type}} &
\textbf{Target} &
\makecell[c]{\textbf{Poisoning}\\ \textbf{Rounds}}  &
\makecell[c]{\textbf{Poison}\\ \textbf{Mode}}  &
\makecell[c]{\textbf{Poison}\\ \textbf{Ratio}}  &
\makecell[c]{\textbf{Poison}\\ \textbf{Lr}}  &
\makecell[c]{\textbf{Poison}\\ \textbf{Epochs}} \\
\midrule
\multirow{2}{*}{CIFAR-10}
  & single-shot & 0.1 & all2one & 2 & 2001--2001 & online & 0.3125 & 0.05 & 6 \\
  & multi-shot  & 0.1 & all2one & 2 & 2001--2200 & online & 0.3125 & 0.05 & 6 \\
\midrule
\multirow{2}{*}{FEMNIST}
  & single-shot & 0.1 & all2one & 7 & 2001--2001 & online & 0.3125 & 0.05 & 6 \\
  & multi-shot  & 0.1 & all2one & 7 & 2001--2200 & online & 0.3125 & 0.05 & 6 \\
\midrule
\multirow{2}{*}{Tiny-ImageNet}
  & single-shot & 0.1 & all2one & 2 & 2001--2001 & online & 0.3125 & 0.05 & 6 \\
  & multi-shot  & 0.1 & all2one & 2 & 2001--2200 & online & 0.3125 & 0.05 & 6 \\
\bottomrule
\end{tabular}
\end{table*}

\begin{table}[!ht]
\centering
\small
\caption{Taxonomy of backdoor defense in FL.}
\resizebox{\columnwidth}{!}{%
\begin{tabular}{@{}lll@{}}
\toprule
\textbf{Category} & \textbf{Sub-category} & \textbf{Literature} \\
\midrule
\multirow{5}{3cm}{\textbf{Robust Aggregation}}
& Robust Statistics & \cite{pmlr-v80-yin18a,pillutla2022robust} \\
& Client Update Reweighting & \cite{cao2020fltrust,flare,fung2018foolsgold} \\
& DP-based & \cite{sun2019can,xie2021crfl} \\
& Distance-based & \cite{blanchard2017machine,bulyan} \\
& Robust Learning Rate & \cite{rlr} \\
\midrule
\multirow{3}{3cm}{\textbf{Anomaly Detection}}
& \multirow{2}{*}{Update Filtering} & \cite{nguyen2022flame,rieger2022deepsight,nguyen2022flame} \\ 
& & \cite{li2024backdoorindicator,aligninns,feddlad,wang2022rflbat} \\ 
& Client Filtering & \cite{zhang2022fldetector,xie2024fedredefense} \\
\bottomrule
\end{tabular}%
}
\label{tab:defense-taxonomy}
\end{table}
\section{Description of Backdoor Defenses}
\label{sec:def_des}

\subsection{Robust Aggregation Defenses}

\noindent\textbf{Coordinate Median~\cite{yin2018byzantine}}. Coordinate-wise median (CMed) aggregator replaces the arithmetic mean with the element-wise median of client updates, providing robustness to outliers and Byzantine deviations. Because the median has a breakdown point of 50\%, CMed tolerates a substantial fraction of adversarial clients as long as the majority behave honestly.
\noindent\textbf{Geometric median~\cite{rfa}}. The geometric median aggregator finds the update vector that minimizes the sum of distances to all client updates, yielding a central, distance-robust estimate that down-weights adversarial outliers without explicitly removing any clients. 

\noindent\textbf{Trimmed-Mean~\cite{yin2018byzantine}}. Trimmed-Mean aggregates client updates by sorting each parameter dimension and removing the largest and smallest values before averaging the remaining middle portion. By discarding extreme coordinate-wise outliers—which are typical signatures of adversarial or corrupted updates—Trimmed-Mean provides strong robustness under Byzantine settings.

\noindent\textbf{FoolsGold~\cite{fung2018foolsgold}. }FoolsGold mitigates Sybil-based poisoning attacks by assigning adaptive weights to client updates based on their historical pairwise cosine similarity. Clients with highly similar gradient directions—which is characteristic of Sybil or colluding attackers—receive exponentially smaller weights, effectively suppressing coordinated malicious behavior while preserving diverse honest updates.

\noindent\textbf{FLTrust~\cite{cao2020fltrust}. }FLTrust establishes trust by leveraging a small, server-held clean dataset to compute a trusted reference gradient. Client updates are first filtered by enforcing positive cosine similarity with the reference update, and then scaled by their alignment to the reference before aggregation. This design ensures that only updates consistent with clean data behavior are aggregated into the global model.

\noindent\textbf{Flare~\cite{flare}}. Flare detects model-poisoning attacks by analyzing penultimate-layer representations (PLRs) instead of raw model parameters. Flare computes pairwise PLR discrepancies among all client updates to estimate a trust score for each client. Updates that are far from the benign cluster get low trust. The server then performs trust-weighted aggregation, which suppresses adversarial influence while preserving accuracy. 

\noindent\textbf{WeakDP~\cite{sun2019can}. }Weak Differential Privacy (WeakDP) injects carefully calibrated Gaussian noise into the global aggregation to provide a weaker but utility-preserving privacy guarantee. WeakDP achieves a tunable trade-off between robustness and utility by controlling the noise variance.

\noindent\textbf{Krum~\cite{blanchard2017krum}}. Krum selects a single client update that is closest to the majority by computing, for each update, the sum of distances to its $n - f -2$ nearest neighbors, and choosing the update with the smallest score. This guarantees resilience against up to $f$ Byzantine clients by ensuring that only an update consistent with most clients is applied.

\noindent\textbf{Multi-Krum~\cite{blanchard2017krum}}. Multi-Krum extends the original Krum rule by selecting multiple candidate updates instead of a single one. Specifically, it chooses the top $m$ updates with the smallest Krum scores (distance-based robustness metric) and averages them. This allows the aggregator to remain Byzantine-resilient while still utilizing contributions from multiple honest clients, improving convergence compared to Krum’s single-update selection strategy.

\noindent\textbf{Bulyan~\cite{bulyan}}. Bulyan first uses Krum to select a set of candidate updates and then applies coordinate-wise trimmed-mean on this subset, combining selection and filtering to further suppress outlier influence and achieve stronger robustness than Multi-Krum alone. A key difference between Bulyan and Multi-Krum is that Bulyan applies Krum selection multiple times after the selected client is removed, while Multi-Krum applies Krum once to select the most consistent set of updates.  

\noindent\textbf{RLR~\cite{rlr}}. Robust Learning Rate (RLR) reduces the influence of anomalous or malicious updates by adjusting the learning rate applied to each client’s gradient based on its deviation from the majority update direction. Clients whose updates diverge significantly from the norm are assigned smaller effective learning rates, soft-limiting their impact without discarding them entirely.

\subsection{Anomaly Detection Defenses}

\noindent\textbf{AlignIns~\cite{aligninns}. }AlignIns identifies and filters suspicious client updates by aligning them with class-level feature centroids computed from server-side clean data. It compares the consistency of each client’s local representation statistics against trusted global statistics and removes updates whose internal feature alignment deviates beyond a threshold, effectively detecting and mitigating poisoned models with inconsistent representations.

\noindent\textbf{DeepSight~\cite{rieger2022deepsight}. }DeepSight detects malicious clients by analyzing the layer-wise sensitivity of client updates. It constructs a hierarchical influence map of how each client’s updates propagate through the network and identifies anomalous structural perturbations that differ from benign update patterns. DeepSight then prunes or down-weights suspicious clients, providing defense against backdoor and targeted manipulation attacks.

\noindent\textbf{FLAME~\cite{nguyen2022flame}. }FLAME defends against backdoor attacks in federated learning by first clustering similar model updates and clipping their weights to reduce variance, then injecting the minimum amount of noise necessary to eliminate backdoors while preserving benign model performance. This adaptive noise calibration approach, informed by the clustering and clipping steps, allows FLAME to effectively remove malicious backdoors without significantly degrading the accuracy of the global model on legitimate tasks.

\noindent\textbf{MultiM~\cite{huang2023multi}. }MultiM revisits distance-based methods and addresses their limitations—Euclidean distance becoming meaningless in high dimensions and single metrics failing to identify diverse malicious gradients—by employing multiple distance metrics with dynamic weighting to adaptively identify backdoors without relying on predefined assumptions about attack settings or data distributions. 

\noindent\textbf{Indicator~\cite{li2024backdoorindicator}. }Indicator uses out-of-distribution (OOD) data to inject indicator tasks into the global model, then detects poisoned client updates by exploiting the fact that backdoor samples are inherently OOD relative to benign data—allowing the server to identify backdoors without prior knowledge of the attack type or target label. 

\noindent\textbf{FedDLAD~\cite{feddlad}. }FedDLAD uses a dual-layer anomaly detection approach that first classifies clients using a Connectivity-Based Outlier Factor (COF) module to analyze structural data connectivity, then applies an Interquartile Range (IQR) module to remove residual malicious influences from the trusted group and a Pardon module to reintegrate misclassified benign clients from the untrusted group.

\subsection{Experiment Settings}
For all defenses, we standardize $\eta=0.5$ and use unweighted FedAvg \cite{howto}. We \textit{mostly} follow the method-specfic hyperparameters as in the papers and codebases, but adapt some key settings as follows:

\begin{itemize}
    \item \textbf{RLR.} We use a sign-agreement threshold $\texttt{robustLR\_threshold} = 4$ for all datasets. Higher thresholds enforce stricter sign agreements and, therefore, more aggressive suppression of inconsistent or potentially malicious updates. We observe that using $\texttt{robustLR\_threshold} = 8$ for CIFAR10, as recommended by the original paper, led to a very low ACC and occasion gradient explosions, so we fix the value to 4.
    \item \textbf{NormClipping and WeakDP.} Since the weight norm of models is highly dependent on architectures, we do hyperparameter tuning to find the lowest possible value of norm bound for NormClipping that does not compromise ACC below 80\% CIFAR10 and FEMNIST. Similarly for WeakDP, we fix the standard deviation of noise to be 0.025, and tune the norm bound so that ACC does not fall below 70\%. For Tiny-ImageNet, we heuristically set the norm bound of NormClipping to be approximately $90^\text{th}$ percentile of norms of benign client updates at round 1000. Finally, the norm bounds for NormClipping are 0.25, 1.0, and 10.0 for FEMNIST, CIFAR10, and Tiny-ImageNet; the norm bounds for WeakDP are 2.0, 3.0, and 15.0 for FEMNIST, CIFAR10, and Tiny-ImageNet.
    \item \textbf{Bulyan and Multi-Krum.} The two methods determine the number of updates to keep based on the estimated number of byzantine updates $f$. We set $f$ to 2, 4, and 6 for CIFAR10, Tiny-ImageNet, and FEMNIST. These values are within the range to preserve the theoretical robustness guarantee of both methods. 
    \item For some Anomaly Detection defenses that require a selection threshold such as \textbf{ADMultiKrum} \cite{blanchard2017krum} and \textbf{MultiM} \cite{huang2023multi}, we fix this threshold to 50\% of the number of sampled clients. In practice, the selected malicious clients never occupy over 50\% of the sampled clients in a single round.
    \item \textbf{Indicator.} Indicator trains the global model for 200 rounds on OOD data in each round, which causes a significant aggregation bottleneck. We observe that the training often converges after the first 10 epochs. Therefore, to improve the experiment speed, we set an early stopping for server-side training once the loss decrease does not pass a certain threshold after a number of epochs.
\end{itemize}

\section{Full Experiment Results}
\label{sec:full_result}

\subsection{Number of sampling clients}
\label{sec:server-side-results}

\begin{table}[!h]
\centering
\caption{Impact of different number of sampled clients per round. Results are averaged over 600 rounds (400 rounds post attack).}
\resizebox{\linewidth}{!}{%
\begin{tabular}{@{}lccccc@{}}
\toprule
Metric & \textbf{10 clients} & \textbf{20 clients} & \textbf{30 clients} & \textbf{40 clients} & \textbf{50 clients} \\ \midrule
ACC (\%) & 91.19 & 91.72 & 91.94 & 92.01 & 91.94 \\
ASR (\%) & 30.80 & 31.77 & 30.97 & 31.23 & 30.32 \\ 
\bottomrule
\end{tabular}%
}
\label{tab:client_scaling}
\end{table}

\cref{tab:client_scaling} shows that ACC and ASR remain largely unchanged across different numbers of sampled clients.

\begin{figure*}[t]
    \centering
    \includegraphics[width=\textwidth]{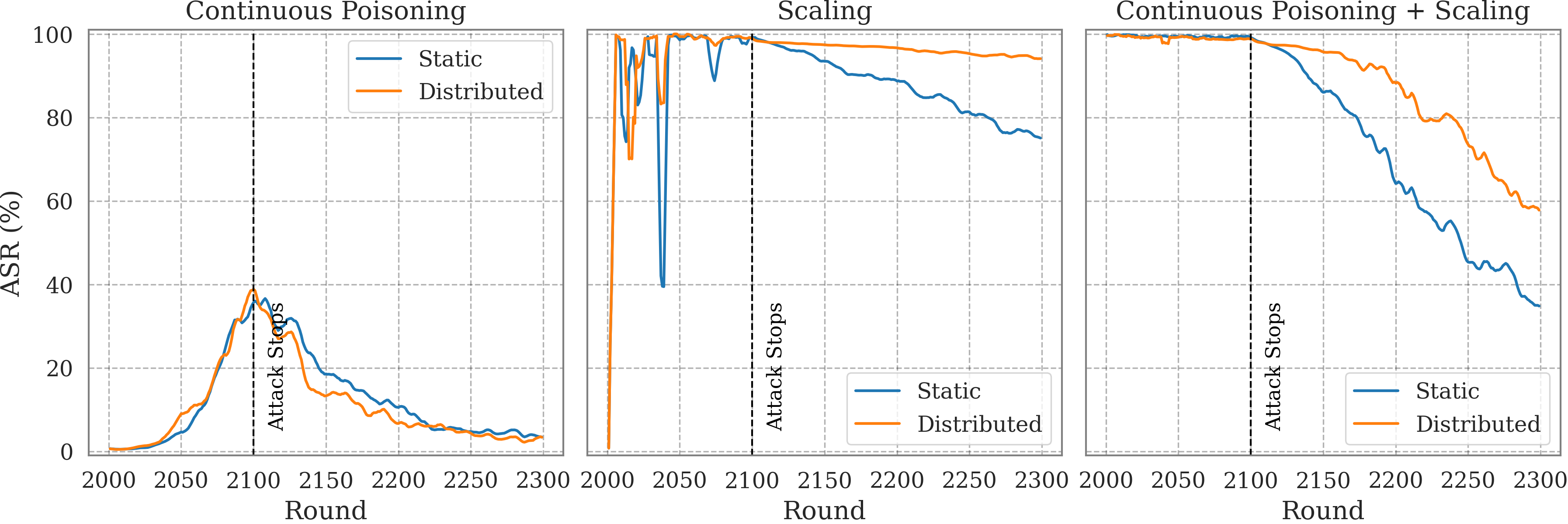}
    \caption{DBA vs Centralized attack on ResNet18 + CIFAR10 under three settings: Continuous Poisoning, Model-Replacement (Scaling), Continuous Poisoning + Model-Replacement. We set $\eta=0.1$.}
    \label{fig:dba_cifar10}
\end{figure*}

\begin{figure*}[t]
    \centering
    \includegraphics[width=\textwidth]{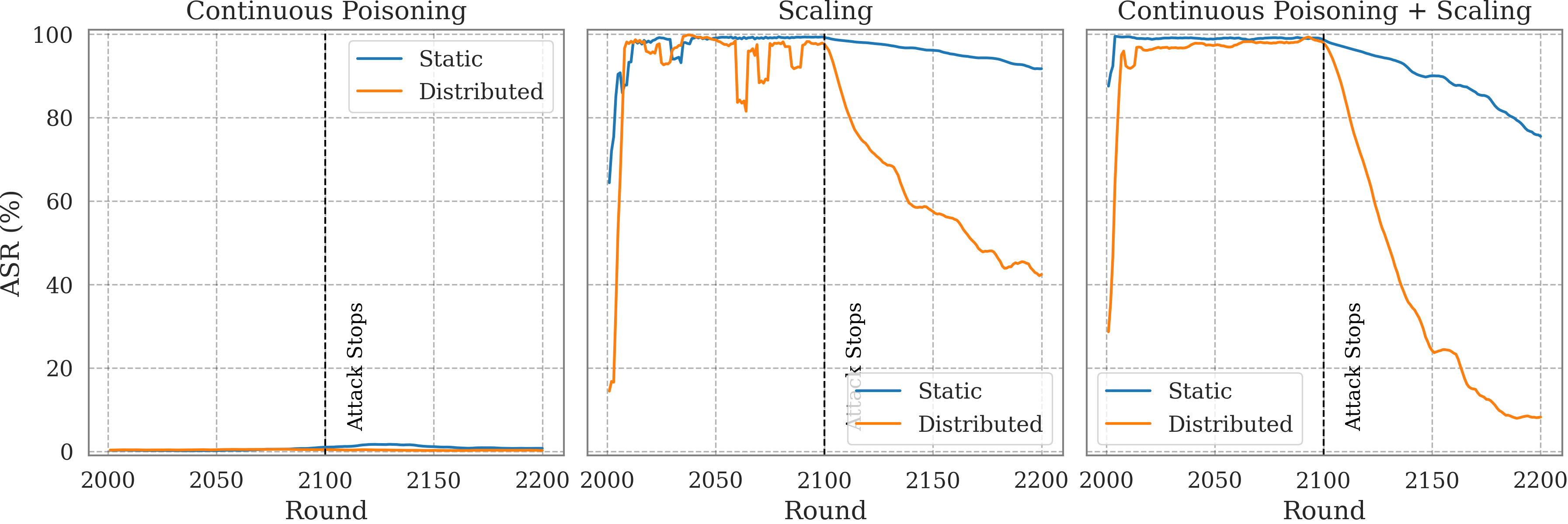}
    \caption{DBA vs Centralized attack on VGG11 + TinyImageNet under three settings: Continuous Poisoning, Model-Replacement (Scaling), Continuous Poisoning + Model-Replacement. We set $\eta=0.1$.}
    \label{fig:dba_tiny}
\end{figure*}

\subsection{Study of DBA.}
\textbf{Experiment Settings:} In order to reproduce the effects of DBA, we follow the original settings in the paper by comparing DBA to centralized attack under three settings: Continuous poisoning, where each malicious client is consecutively selected to poison the global model (malicious client 1 selected for round 1, malicious client 2 selected for round 2, etc.); Model-Replacement (Scaling) under Random-Sampling, where the malicious clients will scale the update by 100$\times$ for each round they are selected to poison the global model, and combine both settings. We note that DBA uses both Continuous Poisoning and Molde-Replacement in their evaluation settings. We also set the number of malicious clients to be equal to the number of local triggers and $\eta=0.1$ to align more closely with their settings. The only remaining difference is in the model-architecture for Tiny-ImageNet where they use ResNet-18 while we use VGG11. The trigger shape of DBA for Tiny-ImageNet resembles the shape for CIFAR10 in \cref{fig:dba_visualize}, but we set the trigger gap and trigger margin to be 2 pixels since the image size of Tiny-ImageNet doubles that of CIFAR10.

\noindent\textbf{Experiment Results:} The results in \cref{fig:dba_cifar10} shows that DBA is more persistent than Centralized Attack for CIFAR10 with ResNet-18, but \cref{fig:dba_tiny} shows the reverse where \textbf{DBA is less persistent than Centralized Attack}. It is also apparent that setting a low $\eta$ prevents both attacks from reaching high ASR, especially for VGG11 on Tiny-ImageNet.

\begin{table}[h]
\centering
\caption{Evaluation of 3 attacks under 25\% selection threshold with 20\% sampling ratio and 1.2 seconds training timeout. selections measure the total number of times malicious clients are selected. CIFAR10 + ResNet18 within round 2000-2200.}
\resizebox{\columnwidth}{!}{%
\begin{tabular}{@{}lc|ccc@{}}
\toprule
Attacks  & Metric & \multicolumn{1}{l}{Baseline} & \multicolumn{1}{l}{25\% selection threshold} & \multicolumn{1}{l}{1.2 seconds timeout} \\ \midrule
\multirow{2}{*}{\textbf{Static}}     & selections & 184 times & 0 times & 63 times \\
                            & ASR            & 72.8\%   & 1.1\%  & 24.4\%  \\
\multirow{2}{*}{\textbf{Neurotoxin}} & selections & 184 times & 0 times & 41 times \\
                            & ASR            & 75.9\%   & 1.1\%  & 12.2\%  \\
\multirow{2}{*}{\textbf{A3FL}}       & selections & 184 times & 0 times & 0 times  \\
                            & ASR            & 92.8\%   & 3.7\%  & 2.5\%   \\ \bottomrule
\end{tabular}%
}
\label{tab:runtime_improv}
\end{table}

\subsection{Evaluation of attacks under training timeouts and selection threshold}

To assess the impact of server-side runtime constraints, we evaluate three representative attacks—Static, Neurotoxin, and A3FL—under a 25\% selection threshold and a 1.2s training-timeout. As shown in \cref{tab:runtime_improv}, the selection threshold is highly effective: by always choosing the 25\% fastest clients, the server prevents all malicious clients from being selected over 200 poisoning rounds, driving ASR close to zero for every attack. The training-timeout strategy is weaker but still impactful: late-running malicious clients are frequently discarded, reducing their total selections by 66–100\% and lowering ASR by 50–80\% across attacks. Notably, A3FL, despite being a dynamic and adaptive attack, is neutralized completely by the selection threshold and strongly suppressed by the timeout. Overall, these results highlight the practicality of leveraging client-side computational heterogeneity as a defense—simple runtime-based filters substantially constrain malicious participation without modifying the FL training protocol.

\begin{table*}[h]
\centering
\caption{ACC(\%) at round 1000 of Robust Aggregation defenses on CIFAR10, FEMNIST, and Tiny-ImageNet. For CIFAR10, we highlight \textbf{ACC $<$ 80\%}; for FEMNIST, we highlight \textbf{ACC $<$ 75\%}; for Tiny-ImageNet, we highlight \textbf{ACC $<$ 40\%}.}
\resizebox{\textwidth}{!}{%
\begin{tabular}{lccccccccccccc}\toprule
 &
  \textbf{FedAvg} &
  \textbf{CMed} &
  \textbf{GMed} &
  \textbf{TM} &
  \textbf{Krum} &
  \multicolumn{1}{l}{\textbf{MKrum}} &
  \multicolumn{1}{l}{\textbf{Bulyan}} &
  \textbf{FG} &
  \textbf{RLR} &
  \textbf{FLTrust} &
  \multicolumn{1}{l}{\textbf{Flare}} &
  \textbf{NC} &
  \textbf{WDP} \\ \midrule
CIFAR10                & 90.7 & \textbf{59.9} & 88.8 & 89.6 & \textbf{75.6} & 88.9 & 87.6 & 90.9 & 80.7 & \textbf{64.2} & 87.2 & 90.1 & \textbf{72.6} \\
FEMNIST                & 86.2 & 82.4 & 85.3 & 84.1 & \textbf{71.2} & 86.2 & 79.4 & 86.1 & 84.4 & 82.0 & 78.6 & 80.6 & \textbf{72.1} \\
Tiny-ImageNet          & 48.37 & \textbf{19.51} & 50.5 & 46.77 & \textbf{20.90} & 49.72 & 48.09 & 50.4 & \textbf{39.12} & \textbf{6.34} & \textbf{39.86} & 49.60 & \textbf{31.45} \\

\bottomrule\end{tabular}%
}
\label{tab:robust_acc}
\end{table*}

\begin{table*}[ht]
\centering
\caption{Average Aggregation Time (seconds) for Anomaly Detection Defenses}
\label{tab:aggregation-time}
\begin{tabular}{lcccccccl}
\toprule
 & \textbf{FedAvg} & \textbf{Flame} & \textbf{DeepSight} & \textbf{MKrum} & \textbf{Indicator} & \textbf{AlignIns} & \textbf{MultiMetrics} & \textbf{FedDLAD} \\
\midrule
CIFAR10 & 0.0645 & 0.4673 & 25.1936 & 0.2364 & 1.0064 & 0.0909 & 1.1055 & 2.3618 \\
FEMNIST & 0.0673 & 0.2600 & 3.7536 & 1.8682 & 3.3364 & 0.0855 & 1.1136 & 1.2191 \\
\bottomrule
\end{tabular}%
\end{table*}

\subsection{Accuracy tradeoff of Robust Aggregation defenses.}
As shown in \cref{tab:robust_acc}, RA defenses that demonstrate better resistance against attacks (CMed, Krum, FLTrust) tend to have lower ACCs. We also note that the accuracy trade-off is worsened for complex models training on difficult tasks, as can be observed in the case of Tiny-ImageNet.

\subsection{Aggregation bottleneck of Anomaly Detection defenses.}
Compared to RA defenses, defenses introduce substantial aggregation overhead, as shown in \cref{tab:aggregation-time}. DeepSight is the most computationally expensive method, requiring \textasciitilde388$\times$ the aggregation time of FedAvg on CIFAR10, while FedDLAD is the next heaviest at \textasciitilde36$\times$. Other AD defenses such as MultiMetrics and Indicator also incur non-trivial slowdowns (15--50$\times$). In contrast, lighter-weight methods like AlignIns maintain near-FedAvg efficiency. These results reveal a clear scalability bottleneck: although AD defenses generally achieve stronger robustness than RA defenses, their prohibitive aggregation costs make them impractical for real-world FL deployments where servers must process updates from many clients under strict latency constraints.

\subsection{Full results of Defense Performance}
\begin{itemize}
    \item \cref{tab:ra_defense_results_tiny} and \cref{tab:ad_defense_results_tiny} show the results of RA and AD defenses on Tiny-ImageNet.
    \item \cref{tab:full_ra_defenses} and \cref{tab:full_ad_results} show full results of RA and AD defenses on CIFAR10 and FEMNIST.
\end{itemize}

\begin{table}[t]
\centering
\caption{Performance of RA defenses on TinyImageNet + VGG11. We highlight \textcolor{red}{ASR $>$ 50\%} and \textbf{ACC $<$ 40\%}.}
\label{tab:ra_defense_results_tiny}
\resizebox{\columnwidth}{!}{%
\begin{tabular}{@{}lc|ccc|cc|cc@{}}
\toprule
\multirow{2}{*}{Defense} &
\multirow{2}{*}{Metric} & \multicolumn{3}{c}{Baseline} & \multicolumn{2}{c}{PGD} & \multicolumn{2}{c}{ModelRep} \\
\cmidrule(lr){3-5} \cmidrule(lr){6-7} \cmidrule(lr){8-9}
 &  & \textbf{Static} & \textbf{A3FL} & \textbf{CerP} & \textbf{Static} & \textbf{A3FL} & \textbf{Static} & \textbf{DBA} \\
\midrule
\multirow{2}{*}{\textbf{FedAvg}} & ASR & \textcolor{red}{75.1} & \textcolor{red}{96.5} & \textcolor{red}{85.6} & \textcolor{red}{57.8} & \textcolor{red}{93.1} & \textcolor{red}{93.4} & \textcolor{red}{51.0} \\
 & ACC & \textbf{49.9} & 50.7 & 50.1 & \textbf{49.6} & 50.1 & \textbf{43.6} & \textbf{42.3} \\
\midrule
\multirow{2}{*}{\textbf{TM}} & ASR & \textcolor{red}{79.2} & \textcolor{red}{94.5} & \textcolor{red}{85.1} & \textcolor{red}{73.5} & \textcolor{red}{92.7} & \textcolor{red}{86.8} & 0.6 \\
 & ACC & \textbf{47.1} & \textbf{47.3} & \textbf{47.4} & \textbf{47.0} & \textbf{47.3} & \textbf{46.7} & \textbf{46.2} \\
\midrule
\multirow{2}{*}{\textbf{CMed}} & ASR & 25.3 & \textcolor{red}{61.0} & 28.8 & 27.0 & \textcolor{red}{56.2} & 25.1 & 1.2 \\
 & ACC & \textbf{13.0} & \textbf{10.3} & \textbf{11.9} & \textbf{12.6} & \textbf{12.6} & \textbf{12.9} & \textbf{12.3} \\
\midrule
\multirow{2}{*}{\textbf{GMed}} & ASR & \textcolor{red}{87.8} & \textcolor{red}{98.0} & \textcolor{red}{94.8} & \textcolor{red}{90.4} & \textcolor{red}{98.1} & \textcolor{red}{89.6} & 0.5 \\
 & ACC & 50.0 & 50.4 & 50.3 & 50.2 & 50.2 & \textbf{49.3} & \textbf{49.3} \\
\midrule
\multirow{2}{*}{\textbf{FG}} & ASR & \textcolor{red}{72.2} & \textcolor{red}{94.9} & \textcolor{red}{81.3} & \textcolor{red}{51.7} & \textcolor{red}{92.0} & \textcolor{red}{90.8} & 49.8 \\
 & ACC & \textbf{50.0} & 50.6 & 50.1 & \textbf{49.7} & 50.4 & \textbf{43.7} & \textbf{42.6} \\
\midrule
\multirow{2}{*}{\textbf{FLTrust}} & ASR & 2.3 & 28.9 & \textcolor{red}{62.9} & 2.1 & 5.0 & 4.7 & 3.9 \\
 & ACC & \textbf{6.6} & \textbf{5.9} & \textbf{6.0} & \textbf{6.3} & \textbf{6.3} & \textbf{6.4} & \textbf{6.6} \\
\midrule
\multirow{2}{*}{\textbf{Flare}} & ASR & \textcolor{red}{87.6} & \textcolor{red}{97.2} & \textcolor{red}{93.4} & \textcolor{red}{76.4} & \textcolor{red}{91.4} & 0.3 & 0.2 \\
 & ACC & \textbf{41.1} & \textbf{41.5} & \textbf{41.6} & \textbf{39.7} & \textbf{41.0} & \textbf{39.2} & \textbf{38.9} \\
\midrule
\multirow{2}{*}{\textbf{NC}} & ASR & \textcolor{red}{79.5} & \textcolor{red}{96.5} & \textcolor{red}{85.4} & \textcolor{red}{66.1} & \textcolor{red}{94.4} & \textcolor{red}{87.1} & 0.1 \\
 & ACC & \textbf{49.3} & \textbf{49.8} & \textbf{49.5} & \textbf{49.2} & \textbf{49.5} & \textbf{49.2} & \textbf{49.0} \\
\midrule
\multirow{2}{*}{\textbf{WDP}} & ASR & 22.7 & \textcolor{red}{93.9} & 46.1 & 0.6 & \textcolor{red}{88.9} & \textcolor{red}{81.9} & 0.7 \\
 & ACC & \textbf{32.6} & \textbf{32.7} & \textbf{32.6} & \textbf{32.7} & \textbf{32.3} & \textbf{32.4} & \textbf{32.0} \\
\midrule
\multirow{2}{*}{\textbf{Krum}} & ASR & 0.4 & 4.3 & 0.7 & 0.4 & 4.6 & 0.4 & 0.3 \\
 & ACC & \textbf{19.7} & \textbf{20.0} & \textbf{19.7} & \textbf{19.7} & \textbf{19.7} & \textbf{19.9} & \textbf{20.0} \\
\midrule
\multirow{2}{*}{\textbf{MKrum}} & ASR & 0.3 & 8.7 & 0.5 & 0.3 & 0.4 & 22.4 & 0.4 \\
 & ACC & \textbf{49.6} & 50.1 & \textbf{49.8} & \textbf{49.7} & \textbf{49.7} & \textbf{49.4} & \textbf{49.3} \\
\midrule
\multirow{2}{*}{\textbf{Bulyan}} & ASR & \textcolor{red}{82.0} & \textcolor{red}{96.4} & \textcolor{red}{86.8} & \textcolor{red}{74.9} & \textcolor{red}{93.7} & 0.3 & 0.2 \\
 & ACC & \textbf{48.5} & \textbf{47.9} & \textbf{48.7} & \textbf{48.1} & \textbf{48.7} & \textbf{48.2} & \textbf{48.3} \\
\midrule
\multirow{2}{*}{\textbf{RLR}} & ASR & \textcolor{red}{78.0} & \textcolor{red}{93.8} & \textcolor{red}{85.1} & \textcolor{red}{83.3} & \textcolor{red}{93.5} & \textcolor{red}{87.2} & 3.8 \\
 & ACC & \textbf{39.7} & \textbf{40.1} & \textbf{40.0} & \textbf{39.7} & \textbf{39.7} & \textbf{38.6} & \textbf{38.1} \\
\bottomrule
\end{tabular}%
}
\end{table}

\begin{table}[h]
\centering
\caption{Performance of AD defenses on TinyImageNet + VGG11. We highlight \textcolor{red}{ASR $>$ 50\%} and \textbf{ACC $<$ 40\%}.}
\label{tab:ad_defense_results_tiny}
\resizebox{\columnwidth}{!}{%
\begin{tabular}{@{}lc|ccc|cc|cc@{}}
\toprule
\multirow{2}{*}{Defense} &
\multirow{2}{*}{Metric} & \multicolumn{3}{c|}{Baseline} & \multicolumn{2}{c|}{PGD} & \multicolumn{2}{c}{ModelRep} \\
\cmidrule(lr){3-5} \cmidrule(lr){6-7} \cmidrule(lr){8-9}
 &  & \textbf{Static} & \textbf{A3FL} & \textbf{CerP} & \textbf{Static} & \textbf{A3FL} & \textbf{Static} & \textbf{DBA} \\
\midrule
\multirow{4}{*}{\textbf{FedAvg}} & ASR & \textcolor{red}{74.7} & \textcolor{red}{95.1} & \textcolor{red}{85.0} & 4.8 & \textcolor{red}{71.2} & \textcolor{red}{93.0} & \textcolor{red}{71.7} \\
 & ACC & 53.3 & 53.9 & 53.8 & 53.7 & 53.6 & 47.6 & 46.6 \\
\midrule
\multirow{4}{*}{\textbf{MKrum}} & ASR & \textcolor{red}{76.6} & \textcolor{red}{95.5} & \textcolor{red}{86.4} & 15.2 & \textcolor{red}{75.2} & 0.2 & 0.3 \\
 & ACC & 53.7 & 53.9 & 54.2 & 54.0 & 54.1 & 53.2 & 53.4 \\
 & TPR & 0.4 & 0.4 & 0.0 & 0.0 & 0.0 & 100.0 & 100.0 \\
 & TNR & 87.6 & 87.6 & 87.5 & 87.5 & 87.5 & 100.0 & 100.0 \\
\midrule
\multirow{4}{*}{\textbf{AligIns}} & ASR & 0.2 & 16.9 & 0.3 & 0.2 & 12.6 & 0.2 & 0.2 \\
 & ACC & 53.2 & 53.0 & 53.3 & 53.2 & 53.3 & 53.3 & 53.2 \\
 & TPR & 100.0 & 99.5 & 100.0 & 100.0 & 100.0 & 100.0 & 100.0 \\
 & TNR & 80.6 & 77.9 & 85.4 & 90.7 & 91.8 & 80.6 & 74.9 \\
\midrule
\multirow{4}{*}{\textbf{DeepSight}} & ASR & 2.0 & \textcolor{red}{64.7} & 30.2 & 0.2 & 13.9 & 25.0 & 0.2 \\
 & ACC & 53.6 & 53.9 & 53.7 & 53.7 & 53.5 & 53.6 & 53.6 \\
 & TPR & 89.2 & 49.2 & 89.6 & 90.1 & 91.2 & 88.7 & 90.5 \\
 & TNR & 97.9 & 97.7 & 97.9 & 98.1 & 97.8 & 98.0 & 98.1 \\
\midrule
\multirow{4}{*}{\textbf{FLAME}} & ASR & 14.7 & \textcolor{red}{87.4} & 0.4 & 0.6 & 1.1 & \textcolor{red}{52.8} & 1.6 \\
 & ACC & \textbf{16.2} & \textbf{15.8} & \textbf{16.2} & \textbf{16.7} & \textbf{16.7} & \textbf{16.1} & \textbf{16.4} \\
 & TPR & 24.4 & 10.4 & 26.4 & 75.3 & 84.1 & 24.9 & 23.8 \\
 & TNR & 55.3 & 53.7 & 56.9 & 63.6 & 64.0 & 54.9 & 56.0 \\
\midrule
\multirow{4}{*}{\textbf{RFLBAT}} & ASR & \textcolor{red}{77.6} & \textcolor{red}{95.5} & \textcolor{red}{86.5} & 12.9 & \textcolor{red}{75.8} & \textcolor{red}{71.5} & 17.1 \\
 & ACC & 53.7 & 54.1 & 54.1 & 54.1 & 53.9 & 52.6 & 51.9 \\
 & TPR & 0.1 & 0.4 & 0.0 & 0.0 & 0.0 & 92.8 & 92.0 \\
 & TNR & 87.9 & 87.6 & 87.2 & 87.9 & 87.9 & 97.0 & 96.9 \\
\midrule
\multirow{4}{*}{\textbf{MultiM}} & ASR & 39.3 & \textcolor{red}{86.6} & \textcolor{red}{78.9} & 0.5 & \textcolor{red}{71.3} & \textcolor{red}{85.0} & \textcolor{red}{58.8} \\
 & ACC & 51.5 & 51.5 & 51.5 & 51.8 & 51.9 & 44.9 & 44.3 \\
 & TPR & 85.7 & 81.7 & 74.7 & 67.3 & 57.5 & 71.8 & 73.6 \\
 & TNR & 53.9 & 53.5 & 52.4 & 51.5 & 50.1 & 52.1 & 52.2 \\
\midrule
\multirow{4}{*}{\textbf{Indicator}} & ASR & 4.0 & \textcolor{red}{97.0} & 8.6 & 0.2 & 14.1 & 0.2 & 0.3 \\
 & ACC & 45.9 & 47.4 & 46.4 & 46.7 & 45.9 & 45.8 & 50.9 \\
 & TPR & 92.5 & 21.1 & 94.1 & 100.0 & 100.0 & 100.0 & 100.0 \\
 & TNR & 73.7 & 73.9 & 74.7 & 74.7 & 74.1 & 75.1 & 82.1 \\
\midrule
\multirow{4}{*}{\textbf{FedDLAD}} & ASR & 9.8 & 11.0 & 7.5 & 0.2 & 0.0 & 20.2 & 11.7 \\
 & ACC & \textbf{38.2} & \textbf{39.7} & \textbf{36.2} & 46.8 & 46.0 & 42.7 & 42.3 \\
 & TPR & 76.1 & 74.1 & 64.1 & 26.5 & 28.1 & 86.0 & 89.5 \\
 & TNR & 24.8 & 24.4 & 23.2 & 19.0 & 17.6 & 27.2 & 27.3 \\
\bottomrule
\end{tabular}%
}
\end{table}

\begin{table*}[h]
\centering
\caption{Full results of RA defenses from round 1001 to 1200. Mean ASR and ACC are reported. CRFL* is evaluated only in the last round (with smoothing) We mark \textcolor{red}{ASR $>$ 50\%} and \textbf{ACC $<$ 75\%}. (1) -- Robust Statistics, (2) -- Client-Reweighting, (3) -- DP-based, (4) -- Distance-based.}
\label{tab:full_ra_defenses}
\resizebox{0.9\textwidth}{!}{%
\begin{tabular}{@{}lc|cccccccccc|cccccccccc@{}}
\toprule
\makebox[0pt][l]{Dataset+Model $\rightarrow$} & &
\multicolumn{10}{c|}{CIFAR10 + ResNet18} &
\multicolumn{10}{c}{FEMNIST + SimpleCNN} \\
\midrule
\multirow{2}{*}{Defense} &
\multirow{2}{*}{Metric} & \multicolumn{4}{c}{Baseline} & \multicolumn{3}{c}{PGD} & \multicolumn{3}{c|}{Model-Replacement} &
\multicolumn{4}{c}{Baseline} & \multicolumn{3}{c}{PGD} & \multicolumn{3}{c}{Model-Replacement} \\
\cmidrule(lr){3-6} \cmidrule(lr){7-9} \cmidrule(lr){10-12}
\cmidrule(lr){13-16} \cmidrule(lr){17-19} \cmidrule(lr){20-22}
 &  & \textbf{Static} & \textbf{OOD} & \textbf{A3FL} & \textbf{CerP} & \textbf{Static} & \textbf{OOD} & \textbf{A3FL} & \textbf{Static} & \textbf{OOD} & \textbf{DBA} & \textbf{Static} & \textbf{OOD} & \textbf{A3FL} & \textbf{CerP} & \textbf{Static} & \textbf{OOD} & \textbf{A3FL} & \textbf{Static} & \textbf{OOD} & \textbf{DBA} \\
\midrule
\multirow{2}{*}{\textbf{FedAvg}} & ASR & \textcolor{red}{73.6} & 30.3 & \textcolor{red}{93.2} & \textcolor{red}{86.7} & \textcolor{red}{76.3} & 22.9 & \textcolor{red}{93.8} & \textcolor{red}{96.1} & \textcolor{red}{80.1} & \textcolor{red}{94.6} & \textcolor{red}{86.8} & \textcolor{red}{91.2} & \textcolor{red}{88.1} & \textcolor{red}{82.9} & \textcolor{red}{70.7} & \textcolor{red}{82.7} & \textcolor{red}{74.8} & \textcolor{red}{97.9} & \textcolor{red}{98.4} & \textcolor{red}{97.1} \\
 & ACC & 90.1 & 89.9 & 90.2 & 90.1 & 89.8 & 90.0 & 90.0 & 83.9 & 80.3 & 82.2 & 83.3 & 83.3 & 83.3 & 83.3 & 83.2 & 83.0 & 83.3 & 83.1 & 83.1 & 82.9 \\
\midrule
\multirow{2}{*}{\textbf{TM (1)}} & ASR & \textcolor{red}{75.3} & 37.1 & \textcolor{red}{93.1} & \textcolor{red}{87.4} & \textcolor{red}{79.0} & 30.5 & \textcolor{red}{93.6} & \textcolor{red}{88.0} & \textcolor{red}{65.0} & \textcolor{red}{85.3} & 12.3 & \textcolor{red}{74.1} & 38.6 & 27.8 & 2.6 & \textcolor{red}{69.7} & 28.8 & 45.8 & \textcolor{red}{78.5} & 6.2 \\
 & ACC & 89.4 & 89.4 & 89.7 & 89.6 & 89.1 & 89.4 & 89.5 & 88.8 & 87.8 & 88.7 & 84.3 & 84.2 & 84.2 & 84.3 & 84.2 & 84.2 & 84.2 & 84.3 & 84.2 & 84.3 \\
\midrule
\multirow{2}{*}{\textbf{CMed (1)}} & ASR & 24.9 & 12.4 & 48.4 & 24.2 & 40.3 & 14.1 & \textcolor{red}{69.7} & 36.9 & 10.5 & 29.0 & 0.4 & \textcolor{red}{56.8} & 3.2 & 1.0 & 0.5 & \textcolor{red}{54.2} & 2.5 & 0.4 & \textcolor{red}{55.7} & 0.4 \\
 & ACC & \textbf{55.7} & \textbf{58.5} & \textbf{56.3} & \textbf{54.1} & \textbf{51.6} & \textbf{56.7} & \textbf{54.3} & \textbf{54.0} & \textbf{55.0} & \textbf{52.4} & 82.3 & 82.2 & 82.2 & 82.3 & 82.2 & 82.2 & 82.2 & 82.2 & 82.2 & 82.3 \\
\midrule
\multirow{2}{*}{\textbf{GMed (1)}} & ASR & \textcolor{red}{83.2} & 44.8 & \textcolor{red}{95.0} & \textcolor{red}{91.3} & \textcolor{red}{85.3} & 40.0 & \textcolor{red}{96.0} & \textcolor{red}{87.1} & \textcolor{red}{57.9} & \textcolor{red}{89.3} & \textcolor{red}{79.0} & \textcolor{red}{92.0} & \textcolor{red}{89.2} & \textcolor{red}{87.8} & \textcolor{red}{84.5} & \textcolor{red}{92.6} & \textcolor{red}{91.4} & \textcolor{red}{80.3} & \textcolor{red}{91.4} & \textcolor{red}{58.6} \\
 & ACC & 89.2 & 88.5 & 89.4 & 89.0 & 89.0 & 88.2 & 89.2 & 88.1 & 87.7 & 88.4 & 85.7 & 85.6 & 85.7 & 85.7 & 85.7 & 85.6 & 85.7 & 85.6 & 85.6 & 85.6 \\
\midrule
\multirow{2}{*}{\textbf{FoolsGold (2)}} & ASR & \textcolor{red}{72.7} & 25.8 & \textcolor{red}{94.5} & \textcolor{red}{86.3} & \textcolor{red}{74.7} & 20.8 & \textcolor{red}{94.4} & \textcolor{red}{95.9} & \textcolor{red}{78.6} & \textcolor{red}{94.5} & \textcolor{red}{86.7} & \textcolor{red}{91.4} & \textcolor{red}{92.3} & \textcolor{red}{84.5} & \textcolor{red}{70.0} & \textcolor{red}{84.9} & \textcolor{red}{82.5} & \textcolor{red}{98.0} & \textcolor{red}{97.6} & \textcolor{red}{97.6} \\
 & ACC & 89.8 & 90.1 & 89.9 & 89.8 & 89.8 & 90.0 & 89.7 & 83.5 & 80.2 & 82.3 & 86.2 & 86.2 & 86.2 & 86.2 & 86.2 & 86.2 & 86.2 & 85.6 & 85.6 & 85.6 \\
\midrule
\multirow{2}{*}{\textbf{FLTrust (2)}} & ASR & 10.9 & 11.6 & \textcolor{red}{61.3} & \textcolor{red}{56.9} & 6.4 & 8.3 & \textcolor{red}{53.4} & 7.3 & 23.2 & 2.4 & 0.5 & 0.7 & 1.4 & 5.2 & 38.3 & 1.4 & 44.8 & 0.5 & 1.6 & 0.3 \\
 & ACC & \textbf{67.3} & \textbf{67.6} & \textbf{67.2} & \textbf{67.7} & \textbf{68.2} & \textbf{68.1} & \textbf{66.2} & \textbf{68.2} & \textbf{67.3} & \textbf{67.4} & 82.2 & 82.2 & 82.1 & 82.2 & 82.2 & 82.1 & 82.2 & 82.2 & 82.1 & 82.2 \\
\midrule
\multirow{2}{*}{\textbf{Flare (2)}} & ASR & \textcolor{red}{81.2} & \textcolor{red}{58.4} & \textcolor{red}{94.5} & \textcolor{red}{89.8} & 13.2 & 48.1 & 25.1 & 2.0 & 4.9 & 2.7 & 0.6 & \textcolor{red}{94.8} & \textcolor{red}{72.8} & \textcolor{red}{93.7} & \textcolor{red}{93.5} & \textcolor{red}{96.0} & \textcolor{red}{96.1} & 0.2 & 5.6 & 0.3 \\
 & ACC & 87.0 & 86.0 & 86.9 & 87.0 & 86.8 & 85.9 & 86.6 & 86.4 & 86.7 & 86.6 & 79.8 & 80.2 & 80.0 & 81.0 & 80.9 & 80.7 & 80.6 & 80.4 & 80.7 & 80.1 \\
\midrule
\multirow{2}{*}{\textbf{NormClip (3)}} & ASR & \textcolor{red}{51.0} & 26.7 & \textcolor{red}{86.6} & \textcolor{red}{80.6} & \textcolor{red}{52.4} & 25.5 & \textcolor{red}{86.9} & \textcolor{red}{51.3} & 26.3 & \textcolor{red}{70.3} & 6.1 & \textcolor{red}{65.5} & 23.4 & 39.9 & 25.3 & \textcolor{red}{68.5} & 39.6 & 6.0 & \textcolor{red}{61.7} & 0.7 \\
 & ACC & 84.4 & 83.9 & 84.6 & 84.7 & 83.9 & 83.7 & 84.1 & 84.4 & 84.3 & 84.2 & 81.0 & 81.0 & 81.0 & 81.0 & 80.9 & 80.9 & 81.0 & 81.0 & 81.0 & 81.0 \\
\midrule
\multirow{2}{*}{\textbf{WeakDP (3)}} & ASR & 25.2 & 18.3 & \textcolor{red}{85.6} & \textcolor{red}{61.8} & 27.3 & 17.8 & \textcolor{red}{89.6} & 43.5 & 24.0 & \textcolor{red}{58.9} & \textcolor{red}{60.1} & \textcolor{red}{89.4} & \textcolor{red}{75.1} & \textcolor{red}{68.7} & 27.8 & \textcolor{red}{76.8} & \textcolor{red}{51.9} & \textcolor{red}{64.3} & \textcolor{red}{88.7} & \textcolor{red}{74.0} \\
 & ACC & \textbf{73.7} & \textbf{73.7} & \textbf{73.8} & \textbf{73.8} & \textbf{73.4} & \textbf{73.4} & \textbf{73.4} & \textbf{73.6} & \textbf{73.6} & \textbf{73.5} & \textbf{71.1} & \textbf{71.2} & \textbf{71.2} & \textbf{71.0} & \textbf{70.8} & \textbf{70.9} & \textbf{71.0} & \textbf{71.0} & \textbf{71.4} & \textbf{71.2} \\
\midrule
\multirow{2}{*}{\textbf{CRFL* (3)}} & ASR & 48.7 & 20.4 & \textcolor{red}{99.0} & \textcolor{red}{92.7} & \textcolor{red}{65.5} & 20.9 & \textcolor{red}{100.0} & \textcolor{red}{76.1} & 23.5 & \textcolor{red}{90.0} & \textcolor{red}{93.5} & \textcolor{red}{95.0} & \textcolor{red}{99.5} & \textcolor{red}{99.1} & \textcolor{red}{80.9} & \textcolor{red}{95.0} & \textcolor{red}{99.1} & \textcolor{red}{95.4} & \textcolor{red}{95.0} & \textcolor{red}{99.7} \\
 & ACC & \textbf{72.2} & \textbf{71.9} & \textbf{72.8} & \textbf{71.5} & \textbf{71.6} & \textbf{71.8} & \textbf{71.4} & \textbf{71.6} & \textbf{73.1} & \textbf{70.8} & \textbf{70.8} & \textbf{70.9} & \textbf{71.7} & \textbf{70.7} & \textbf{70.5} & \textbf{70.4} & \textbf{71.2} & \textbf{71.3} & \textbf{71.4} & \textbf{71.5} \\
\midrule
\multirow{2}{*}{\textbf{Krum (4)}} & ASR & 4.0 & 6.5 & 27.1 & 22.2 & 3.7 & 5.5 & 14.7 & 3.8 & 4.8 & 3.8 & 0.3 & 4.0 & 0.3 & \textcolor{red}{93.9} & \textcolor{red}{96.2} & \textcolor{red}{98.1} & \textcolor{red}{97.2} & 0.4 & 4.0 & 0.5 \\
 & ACC & \textbf{74.7} & \textbf{74.7} & \textbf{74.5} & \textbf{75.0} & \textbf{75.0} & \textbf{74.7} & \textbf{74.9} & 75.3 & 75.1 & \textbf{74.8} & \textbf{70.3} & \textbf{70.3} & \textbf{70.5} & \textbf{74.1} & \textbf{74.8} & \textbf{74.3} & \textbf{74.9} & \textbf{71.2} & \textbf{71.0} & \textbf{71.8} \\
\midrule
\multirow{2}{*}{\textbf{MKrum (4)}} & ASR & 8.1 & 7.9 & \textcolor{red}{64.8} & 37.4 & 6.5 & 6.0 & \textcolor{red}{70.0} & 4.2 & 5.0 & 3.2 & \textcolor{red}{83.2} & \textcolor{red}{91.6} & \textcolor{red}{91.9} & \textcolor{red}{86.2} & \textcolor{red}{71.3} & \textcolor{red}{86.1} & \textcolor{red}{84.3} & \textcolor{red}{95.9} & \textcolor{red}{95.3} & \textcolor{red}{93.7} \\
 & ACC & 88.5 & 88.7 & 88.6 & 88.6 & 88.7 & 88.7 & 88.2 & 88.1 & 88.6 & 88.4 & 86.1 & 86.1 & 86.1 & 86.1 & 86.1 & 86.1 & 86.1 & 85.9 & 85.8 & 85.8 \\
\midrule
\multirow{2}{*}{\textbf{Bulyan (4)}} & ASR & \textcolor{red}{84.7} & \textcolor{red}{51.4} & \textcolor{red}{95.0} & \textcolor{red}{91.8} & \textcolor{red}{86.8} & 42.9 & \textcolor{red}{95.3} & 1.4 & 2.5 & 1.5 & 0.5 & \textcolor{red}{72.8} & 3.0 & 34.1 & 30.4 & \textcolor{red}{85.6} & 47.6 & 0.3 & 6.3 & 0.4 \\
 & ACC & 87.9 & 87.4 & 88.1 & 88.2 & 87.5 & 87.6 & 87.8 & 87.8 & 87.9 & 87.8 & 79.8 & 80.0 & 79.7 & 79.7 & 79.7 & 79.7 & 79.9 & 80.6 & 80.7 & 80.5 \\
\midrule
\multirow{2}{*}{\textbf{RLR (5)}} & ASR & \textcolor{red}{67.7} & 44.5 & \textcolor{red}{91.0} & \textcolor{red}{89.4} & \textcolor{red}{72.4} & 40.4 & \textcolor{red}{90.2} & 1.1 & 1.8 & 1.5 & \textcolor{red}{69.4} & \textcolor{red}{88.3} & \textcolor{red}{81.3} & \textcolor{red}{70.1} & 47.2 & \textcolor{red}{79.7} & \textcolor{red}{66.2} & \textcolor{red}{97.0} & \textcolor{red}{97.5} & 42.1 \\
 & ACC & 84.2 & 83.7 & 84.7 & 84.4 & 83.6 & 83.5 & 84.1 & 82.5 & 80.7 & 82.3 & 84.4 & 84.4 & 84.4 & 84.4 & 84.4 & 84.3 & 84.4 & 84.2 & 84.1 & 77.9 \\
\bottomrule
\end{tabular}%
}
\end{table*}

\begin{table*}[h]
\centering
\caption{Full results with detection metrics of AD defenses on CIFAR10 + ResNet18 (left) and FEMNIST + SimpleCNN (right). Highest ASR values are highlighted in \textcolor{red}{red}, lowest ACC values are \textbf{bolded}.}
\label{tab:full_ad_results}
\resizebox{0.9\textwidth}{!}{%
\begin{tabular}{@{}lc|cccccccccc|cccccccccc@{}}
\toprule
\makebox[0pt][l]{Dataset+Model $\rightarrow$} & &
\multicolumn{10}{c|}{CIFAR10 + ResNet18} &
\multicolumn{10}{c}{FEMNIST + SimpleCNN} \\
\midrule
\multirow{2}{*}{Defense} &
\multirow{2}{*}{Metric} & \multicolumn{4}{c}{Baseline} & \multicolumn{3}{c}{PGD} & \multicolumn{3}{c|}{Model-Replacement} &
\multicolumn{4}{c}{Baseline} & \multicolumn{3}{c}{PGD} & \multicolumn{3}{c}{Model-Replacement} \\
\cmidrule(lr){3-6} \cmidrule(lr){7-9} \cmidrule(lr){10-12}
\cmidrule(lr){13-16} \cmidrule(lr){17-19} \cmidrule(lr){20-22}
 &  & \textbf{Static} & \textbf{OOD} & \textbf{A3FL} & \textbf{CerP} & \textbf{Static} & \textbf{OOD} & \textbf{A3FL} & \textbf{Static} & \textbf{OOD} & \textbf{DBA} & \textbf{Static} & \textbf{OOD} & \textbf{A3FL} & \textbf{CerP} & \textbf{Static} & \textbf{OOD} & \textbf{A3FL} & \textbf{Static} & \textbf{OOD} & \textbf{DBA} \\
\midrule
\multirow{2}{*}{\textbf{FedAvg}} & ASR & \textcolor{red}{73.2} & 39.3 & \textcolor{red}{93.1} & \textcolor{red}{85.3} & \textcolor{red}{75.0} & 29.8 & \textcolor{red}{93.2} & \textcolor{red}{95.7} & \textcolor{red}{82.0} & \textcolor{red}{75.9} & \textcolor{red}{86.6} & \textcolor{red}{91.1} & \textcolor{red}{92.2} & \textcolor{red}{84.3} & \textcolor{red}{69.8} & \textcolor{red}{84.4} & \textcolor{red}{83.3} & \textcolor{red}{97.6} & \textcolor{red}{97.9} & 45.9 \\
 & ACC & 91.0 & 90.9 & 91.0 & 91.0 & 90.9 & 90.9 & 91.1 & 85.8 & 82.6 & 84.1 & 86.8 & 86.8 & 86.9 & 86.9 & 86.9 & 86.8 & 86.9 & 86.2 & 86.2 & 86.2 \\
\midrule
\multirow{4}{*}{\textbf{MKrum}} & ASR & \textcolor{red}{76.6} & 46.3 & \textcolor{red}{94.1} & \textcolor{red}{87.3} & \textcolor{red}{79.3} & 36.8 & \textcolor{red}{94.1} & 0.9 & 5.8 & 0.8 & \textcolor{red}{77.7} & \textcolor{red}{91.1} & \textcolor{red}{88.6} & \textcolor{red}{86.1} & \textcolor{red}{74.5} & \textcolor{red}{86.8} & \textcolor{red}{85.7} & 0.1 & 4.6 & 0.1 \\
 & ACC & 90.8 & 90.8 & 90.9 & 91.1 & 91.0 & 90.4 & 90.9 & 91.0 & 90.3 & 91.0 & 86.8 & 86.8 & 86.8 & 86.8 & 86.8 & 86.8 & 86.8 & 86.8 & 86.8 & 86.8 \\
 & TPR & 0.0 & 0.0 & 0.0 & 0.0 & 0.0 & 0.0 & 0.0 & 100.0 & 99.2 & 100.0 & 7.2 & 2.0 & 3.6 & 0.5 & 0.0 & 0.0 & 0.0 & 99.5 & 99.9 & 99.7 \\
 & TNR & 81.6 & 81.6 & 81.6 & 81.6 & 81.6 & 81.6 & 81.6 & 100.0 & 99.9 & 100.0 & 89.7 & 89.0 & 89.2 & 88.8 & 88.7 & 88.7 & 88.7 & 99.9 & 100.0 & 100.0 \\
\midrule
\multirow{4}{*}{\textbf{AligIns}} & ASR & 1.0 & 4.1 & \textcolor{red}{88.8} & 4.6 & 1.0 & 4.1 & 3.6 & 1.1 & 9.1 & 0.9 & 0.1 & \textcolor{red}{66.4} & \textcolor{red}{51.4} & 0.1 & 0.1 & 48.1 & 0.1 & \textcolor{red}{89.5} & 46.3 & \textcolor{red}{88.7} \\
 & ACC & 90.2 & 89.8 & 89.7 & 90.0 & 90.4 & 90.2 & 90.1 & 89.8 & 90.0 & 89.8 & 86.5 & 86.3 & 86.4 & 86.5 & 86.6 & 86.3 & 86.5 & \textbf{6.5} & \textbf{7.9} & \textbf{6.5} \\
 & TPR & 100.0 & 99.2 & 71.7 & 96.7 & 100.0 & 100.0 & 100.0 & 98.8 & 92.3 & 97.6 & 99.6 & 62.4 & 70.1 & 100.0 & 100.0 & 69.8 & 100.0 & 13.6 & 25.3 & 13.3 \\
 & TNR & 65.1 & 64.6 & 50.1 & 58.5 & 77.6 & 68.7 & 70.2 & 64.6 & 57.8 & 60.4 & 64.5 & 48.4 & 50.1 & 63.6 & 71.5 & 49.9 & 66.4 & 71.9 & 71.7 & 71.2 \\
\midrule
\multirow{4}{*}{\textbf{DeepSight}} & ASR & 0.9 & 4.5 & 2.3 & 1.6 & 0.8 & 4.9 & 2.9 & 0.8 & 18.2 & 0.7 & \textcolor{red}{76.6} & \textcolor{red}{90.0} & \textcolor{red}{89.5} & \textcolor{red}{81.1} & \textcolor{red}{67.2} & \textcolor{red}{83.5} & \textcolor{red}{80.5} & \textcolor{red}{78.5} & \textcolor{red}{91.2} & 21.2 \\
 & ACC & 89.3 & 90.6 & 90.5 & 90.7 & 90.7 & 90.3 & 91.1 & 91.2 & 90.0 & 90.3 & 86.9 & 86.8 & 86.9 & 86.9 & 86.9 & 86.8 & 86.9 & 86.8 & 86.8 & 86.8 \\
 & TPR & 100.0 & 99.1 & 100.0 & 100.0 & 100.0 & 98.2 & 100.0 & 100.0 & 81.6 & 100.0 & 2.0 & 1.0 & 0.3 & 1.8 & 2.6 & 2.4 & 2.5 & 2.1 & 0.5 & 3.9 \\
 & TNR & 90.1 & 90.9 & 90.8 & 90.9 & 90.8 & 90.5 & 90.7 & 90.1 & 90.6 & 90.4 & 99.8 & 99.8 & 99.8 & 99.8 & 99.7 & 99.8 & 99.9 & 99.9 & 99.7 & 99.8 \\
\midrule
\multirow{4}{*}{\textbf{FLAME}} & ASR & \textcolor{red}{56.2} & 29.1 & \textcolor{red}{88.8} & \textcolor{red}{82.6} & 49.5 & 20.7 & \textcolor{red}{90.3} & \textcolor{red}{78.4} & \textcolor{red}{55.0} & 28.9 & \textcolor{red}{72.8} & \textcolor{red}{85.6} & \textcolor{red}{82.5} & \textcolor{red}{81.3} & 45.2 & \textcolor{red}{83.5} & \textcolor{red}{70.4} & \textcolor{red}{80.2} & \textcolor{red}{90.7} & 23.2 \\
 & ACC & 86.0 & 86.0 & 86.3 & 86.1 & 85.9 & 86.1 & 85.5 & 83.5 & 82.8 & 84.0 & 86.3 & 86.2 & 86.3 & 86.3 & 86.3 & 86.2 & 86.3 & 86.2 & 86.1 & 86.1 \\
 & TPR & 38.2 & 40.7 & 24.5 & 20.3 & 49.6 & 44.7 & 34.1 & 32.5 & 34.7 & 41.6 & 39.7 & 33.9 & 38.1 & 35.5 & 56.1 & 32.4 & 42.6 & 35.3 & 32.6 & 42.4 \\
 & TNR & 63.2 & 66.0 & 62.3 & 62.9 & 67.6 & 65.9 & 65.8 & 62.8 & 64.2 & 64.3 & 64.0 & 65.5 & 63.9 & 63.0 & 65.5 & 64.3 & 66.1 & 64.4 & 62.6 & 64.0 \\
\midrule
\multirow{4}{*}{\textbf{MultiM}} & ASR & \textcolor{red}{59.4} & 45.8 & \textcolor{red}{91.1} & \textcolor{red}{76.8} & \textcolor{red}{55.2} & 11.3 & \textcolor{red}{86.6} & \textcolor{red}{92.4} & \textcolor{red}{68.8} & 48.8 & \textcolor{red}{77.1} & \textcolor{red}{93.2} & \textcolor{red}{92.0} & \textcolor{red}{81.1} & \textcolor{red}{68.6} & \textcolor{red}{92.0} & \textcolor{red}{83.4} & 0.2 & 23.3 & 0.1 \\
 & ACC & 89.3 & 87.2 & 89.6 & 89.8 & 89.1 & 89.6 & 89.2 & 80.7 & 80.5 & 80.8 & 83.0 & 83.2 & 83.2 & 83.2 & 82.9 & 83.5 & 83.4 & 84.1 & 84.6 & 84.1 \\
 & TPR & 78.0 & 70.1 & 79.4 & 75.7 & 74.3 & 86.6 & 75.7 & 63.0 & 75.3 & 76.7 & 38.3 & 27.3 & 32.0 & 30.2 & 38.5 & 21.3 & 33.2 & 99.9 & 99.7 & 100.0 \\
 & TNR & 53.7 & 27.4 & 54.2 & 54.1 & 52.8 & 55.1 & 53.6 & 52.6 & 54.7 & 54.9 & 48.6 & 47.5 & 48.0 & 47.6 & 48.4 & 46.8 & 47.8 & 55.6 & 55.6 & 55.7 \\
\midrule
\multirow{4}{*}{\textbf{Indicator}} & ASR & 1.2 & 7.7 & \textcolor{red}{86.9} & 2.7 & 1.1 & 32.3 & 23.5 & 1.1 & 4.3 & 1.1 & 5.1 & 39.1 & \textcolor{red}{50.2} & 0.0 & 3.9 & \textcolor{red}{96.5} & 5.8 & \textcolor{red}{58.9} & 44.3 & 34.5 \\
 & ACC & 89.0 & 88.8 & 89.3 & 88.7 & 88.4 & 89.3 & 88.7 & 89.0 & 88.4 & 89.2 & \textbf{37.1} & \textbf{45.7} & \textbf{38.7} & \textbf{5.4} & \textbf{14.6} & \textbf{5.5} & \textbf{34.5} & \textbf{47.3} & \textbf{45.3} & \textbf{5.3} \\
 & TPR & 98.4 & 91.7 & 33.1 & 97.0 & 98.8 & 6.4 & 88.6 & 100.0 & 98.4 & 100.0 & 83.3 & 83.2 & 44.1 & 99.4 & 97.2 & 100.0 & 99.5 & 42.7 & 91.3 & 100.0 \\
 & TNR & 80.8 & 79.7 & 79.9 & 81.0 & 79.1 & 76.5 & 77.9 & 80.7 & 79.3 & 81.0 & 53.0 & 66.9 & 69.2 & 1.0 & 68.1 & 4.0 & 56.8 & 61.8 & 49.9 & 4.4 \\
\midrule
\multirow{4}{*}{\textbf{FedDLAD}} & ASR & 1.5 & 9.6 & 3.3 & 2.6 & 3.0 & 9.6 & 8.6 & \textcolor{red}{52.7} & 45.9 & 9.9 & 0.1 & 10.9 & 0.2 & 0.1 & 0.0 & 0.0 & 0.0 & 0.2 & 46.9 & 0.3 \\
 & ACC & 89.2 & 88.3 & 89.6 & 89.3 & 89.3 & 89.0 & 89.6 & 90.8 & 90.4 & 90.9 & 80.8 & 81.3 & 81.4 & 81.7 & 85.6 & 85.3 & 83.2 & 80.9 & 81.3 & 80.7 \\
 & TPR & 44.7 & 34.7 & 37.0 & 13.8 & 0.0 & 0.0 & 0.0 & 76.3 & 72.9 & 80.8 & 97.8 & 97.7 & 95.5 & 86.0 & 6.4 & 10.3 & 4.4 & 99.1 & 98.5 & 99.7 \\
 & TNR & 47.7 & 45.6 & 46.3 & 42.7 & 40.8 & 40.8 & 40.8 & 56.1 & 55.7 & 56.6 & 18.3 & 18.2 & 18.0 & 16.9 & 8.4 & 8.0 & 8.6 & 18.5 & 18.4 & 18.5 \\
\bottomrule
\end{tabular}%
}
\end{table*}

\subsection{Sensitivity Analysis of Attacks.}
\noindent\textbf{Impact of server's learning rate.} As can be seen in \cref{tab:ra_defense_eta}, a low $\eta=0.1$ improves the robustness of all RA defenses.

\begin{figure*}[t]
    \centering
    \begin{subfigure}{\linewidth}
        \centering
        \includegraphics[width=\linewidth]{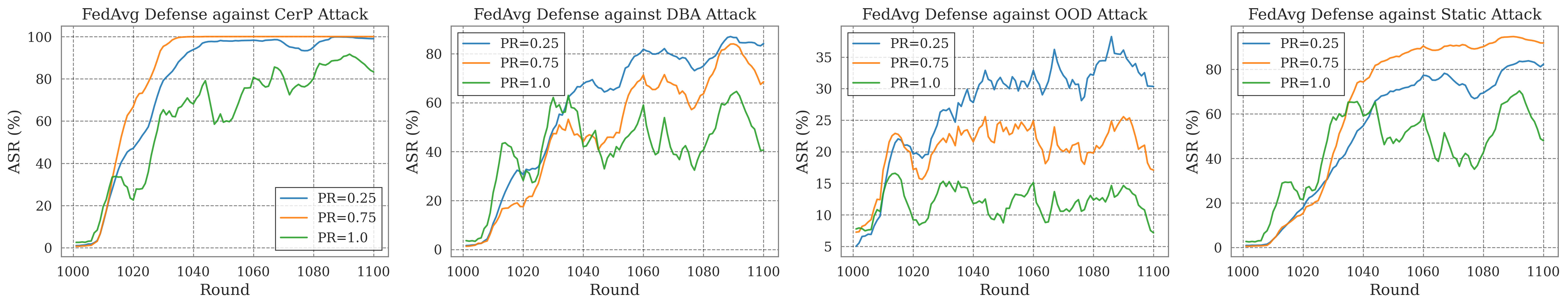}
        \caption{FedAvg}
    \end{subfigure}
    \caption{Poisoning ratio breakdown for FedAvg across all attacks.}
    \label{fig:poison_fedavg}
\end{figure*}

\noindent\textbf{Impact of poison ratios.} We vary the poison ratio from $25\%$ to $75\%$ and $100\%$. Interestingly, for the baseline FedAvg in \cref{fig:poison_fedavg}, the maximum poison ratio of $100\%$ actually has the lowest overall ASR while a low poison ratio of $25\%$ has the highest ASR for DBA and OOD attack. This suggests that a small poison ratio, which helps the model update of malicious clients to align better with the training dynamics of global model, is crucial for an effective attack even in the absence of defenses. 

\begin{figure*}[t]
    \centering

    \begin{subfigure}{\linewidth}
        \centering
        \includegraphics[width=\linewidth]{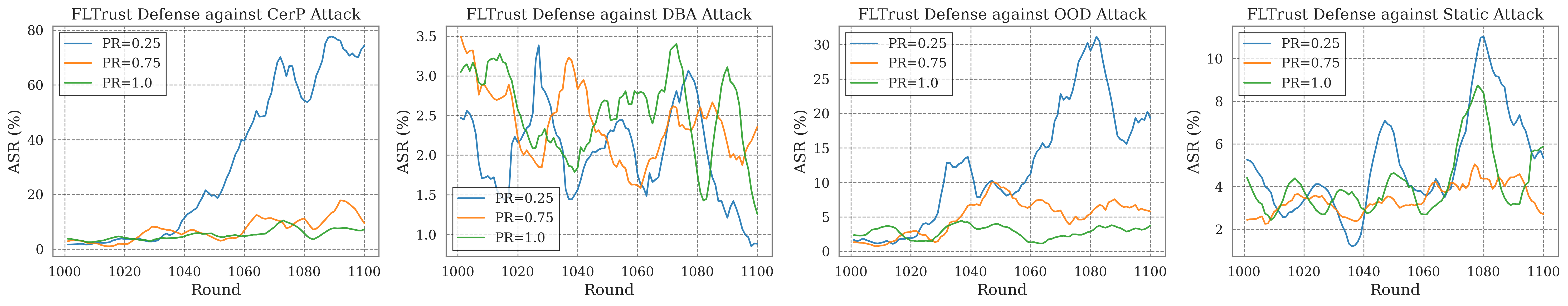}
        \caption{FLTrust}
    \end{subfigure}

    \begin{subfigure}{\linewidth}
        \centering
        \includegraphics[width=\linewidth]{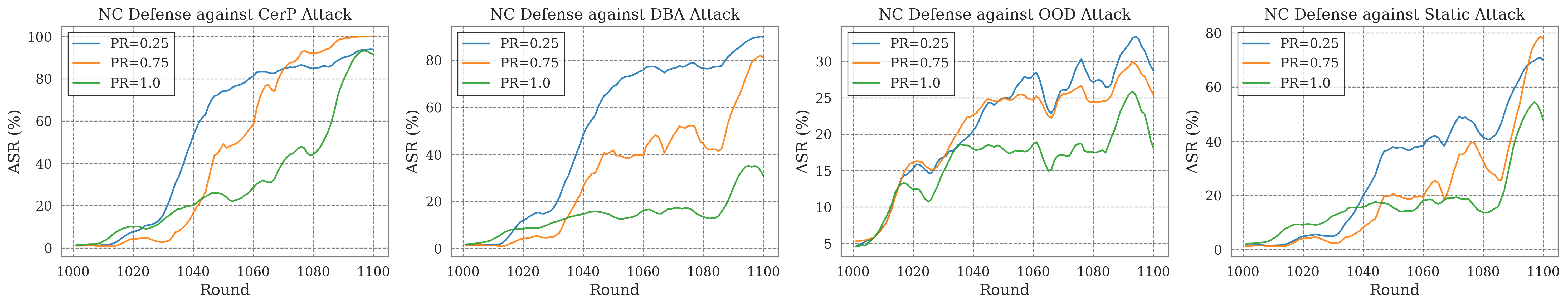}
        \caption{NormClipping}
    \end{subfigure}

    \begin{subfigure}{\linewidth}
        \centering
        \includegraphics[width=\linewidth]{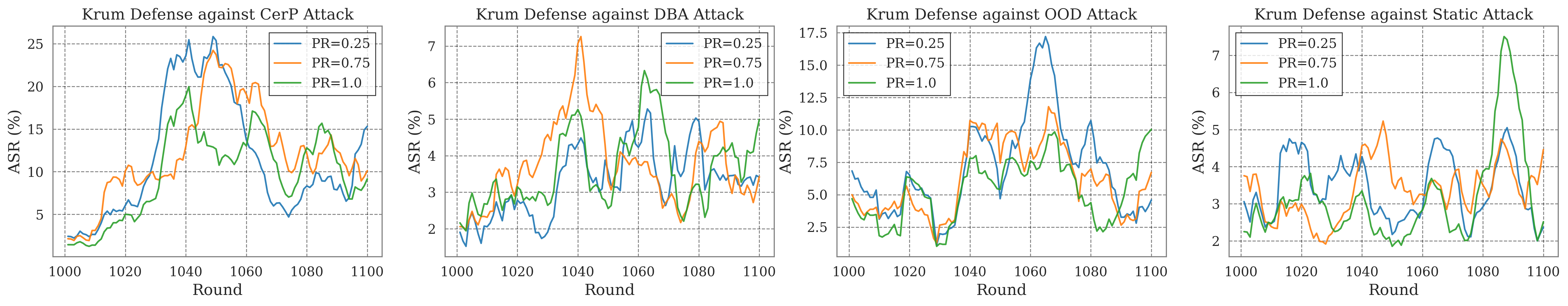}
        \caption{Krum}
    \end{subfigure}

    \caption{Poisoning ratio breakdown for Robust Aggregation (RA) defenses: FLTrust, NormClipping, and Krum.}
    \label{fig:poison_ra_group}
\end{figure*}

\begin{figure*}[t]
    \centering

    \begin{subfigure}{\linewidth}
        \centering
        \includegraphics[width=\linewidth]{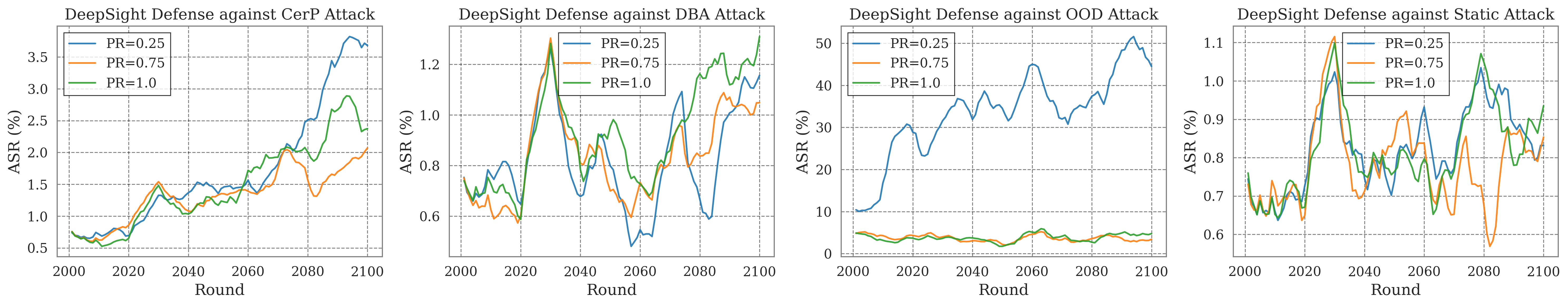}
        \caption{DeepSight}
    \end{subfigure}

    \begin{subfigure}{\linewidth}
        \centering
        \includegraphics[width=\linewidth]{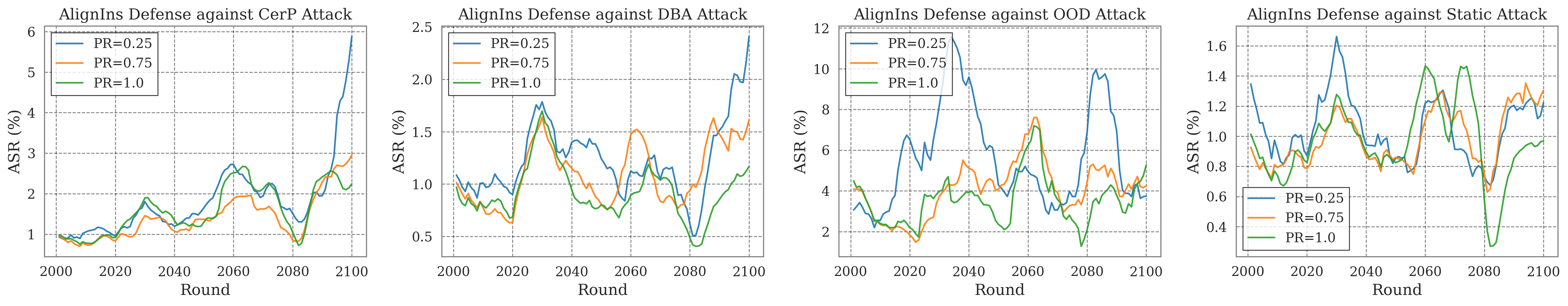}
        \caption{AlignIns}
    \end{subfigure}

    \begin{subfigure}{\linewidth}
        \centering
        \includegraphics[width=0.5\linewidth]{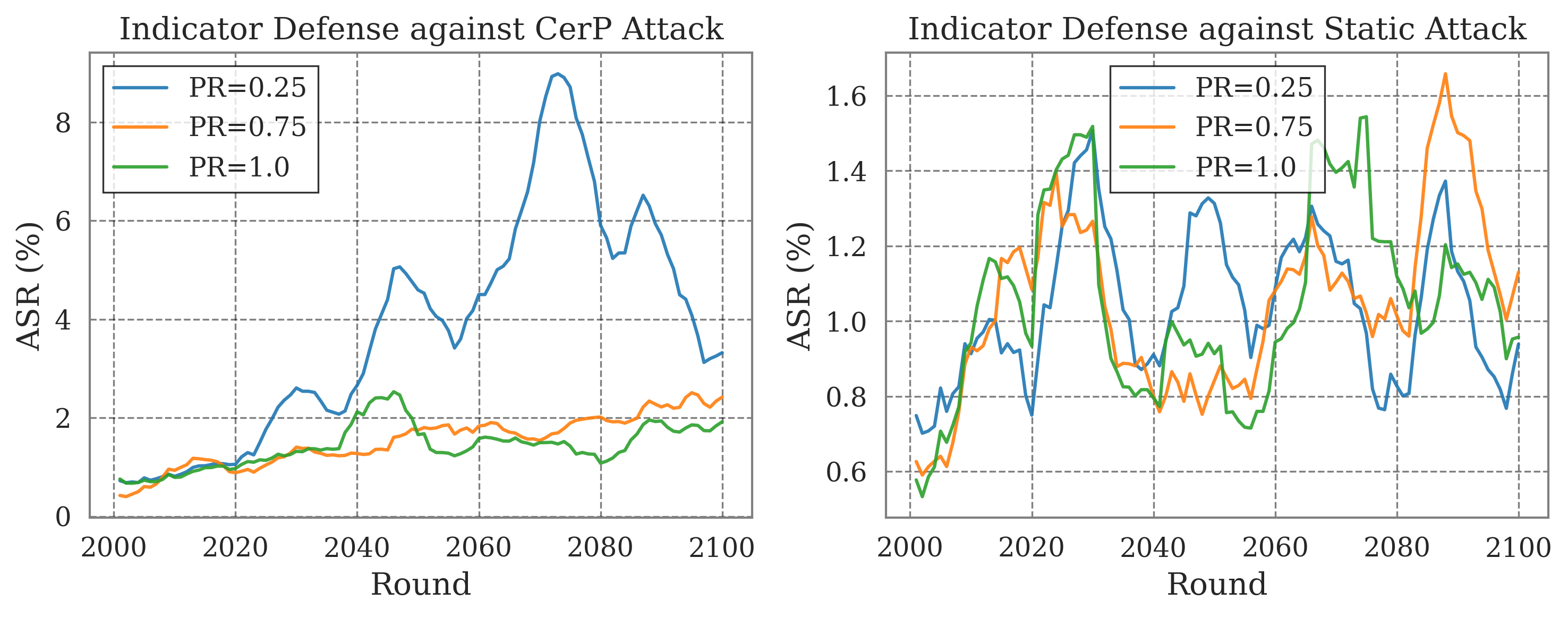}
        \caption{Indicator}
    \end{subfigure}

    \begin{subfigure}{\linewidth}
        \centering
        \includegraphics[width=\linewidth]{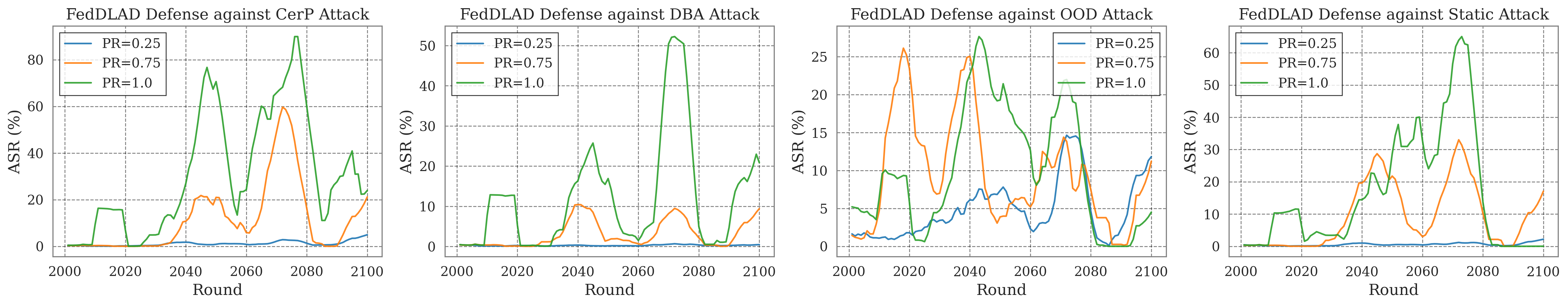}
        \caption{FedDLAD}
    \end{subfigure}

    \caption{Poisoning ratio breakdown for Anomaly Detection (AD) defenses: DeepSight, AlignIns, Indicator, and FedDLAD.}
    \label{fig:poison_ad_group}
\end{figure*}

We observe the same trend for RA defenses in \cref{fig:poison_ra_group}, where the low poison ratio of 0.25\% tends to yield the best ASR. For AD defenses in \cref{fig:poison_ad_group}, we find an interesting exception to this trend in FedDLAD, where the maximum poison ratio of 100\% actually results in the best ASR.

\noindent\textbf{Impact of increasing number of malicious clients.} We present additional defense performance when malicious clients occupy 5\% of total clients in \cref{tab:ra_mal_clients_005} and \cref{tab:ad_mal_clients_005}, and when malicious clients occupy 20\% of total clients in \cref{tab:ra_mal_clients_02} and \cref{tab:ad_mal_clients_02}. Overall, we find that none of the defenses are robust to an increasing number of malicious clients.

\begin{table}[t]
\centering
\caption{Performance of RA defenses on CIFAR10 and FEMNIST with $\eta=0.1$. We highlight \textcolor{red}{$>50\%$} and \textbf{ACC $<$ 70\%}.}
\label{tab:ra_defense_eta}
\resizebox{\columnwidth}{!}{%
\begin{tabular}{@{}lc|cccc|cccc@{}}
\toprule
\multirow{2}{*}{Defense} & \multirow{2}{*}{Metric} & \multicolumn{4}{c|}{CIFAR10 + ResNet18} & \multicolumn{4}{c}{FEMNIST + SimpleCNN} \\
\cmidrule(lr){3-6} \cmidrule(lr){7-10}
 &  & \textbf{Static} & \textbf{DBA} & \textbf{OOD} & \textbf{CerP} & \textbf{Static} & \textbf{DBA} & \textbf{OOD} & \textbf{CerP} \\
\midrule
\multirow{2}{*}{\textbf{FedAvg}} & ASR & 28.0 & 36.9 & 21.0 & \textcolor{red}{66.3} & 38.3 & 8.0 & \textcolor{red}{63.5} & \textcolor{red}{55.5} \\
 & ACC & 90.9 & 90.7 & 91.3 & 91.2 & 86.4 & 86.4 & 86.3 & 86.4 \\
\midrule
\multirow{2}{*}{\textbf{CMed}} & ASR & 16.5 & 7.9 & 6.1 & 17.3 & 0.3 & 0.3 & 14.2 & 0.2 \\
 & ACC & 72.3 & 72.0 & 72.9 & 72.2 & 82.4 & 82.4 & 82.3 & 82.3 \\
\midrule
\multirow{2}{*}{\textbf{Krum}} & ASR & 2.1 & 2.4 & 4.3 & 8.4 & 0.3 & 0.3 & 46.1 & 0.2 \\
 & ACC & 84.8 & 84.8 & 84.8 & 84.7 & 72.4 & 72.5 & 72.6 & 72.7 \\
\midrule
\multirow{2}{*}{\textbf{FLTrust}} & ASR & 3.5 & 2.3 & 4.9 & 4.9 & 0.3 & 0.3 & 0.6 & 0.2 \\
 & ACC & \textbf{67.2} & \textbf{67.9} & \textbf{68.3} & \textbf{66.5} & 82.0 & 82.0 & 82.0 & 82.0 \\
\midrule
\multirow{2}{*}{\textbf{WDP}} & ASR & 7.2 & 5.1 & 12.6 & 49.6 & 5.4 & 0.5 & \textcolor{red}{63.9} & 0.2 \\
 & ACC & 76.9 & 77.0 & 76.9 & 76.8 & 73.3 & \textbf{26.4} & 73.6 & \textbf{26.4} \\
\bottomrule
\end{tabular}%
}
\end{table}

\begin{table}[t]
\centering
\caption{Performance of RA defenses on CIFAR10 and FEMNIST when malicious clients occupy 5\% of total clients. We highlight \textcolor{red}{ASR $>50\%$} and \textbf{ACC $<$ 70\%}.}
\label{tab:ra_mal_clients_005}
\resizebox{\columnwidth}{!}{%
\begin{tabular}{@{}lc|cccc|cccc@{}}
\toprule
\multirow{2}{*}{Defense} & \multirow{2}{*}{Metric} & \multicolumn{4}{c|}{CIFAR10 + ResNet18} & \multicolumn{4}{c}{FEMNIST + SimpleCNN} \\
\cmidrule(lr){3-6} \cmidrule(lr){7-10}
 &  & \textbf{Static} & \textbf{DBA} & \textbf{OOD} & \textbf{CerP} & \textbf{Static} & \textbf{DBA} & \textbf{OOD} & \textbf{CerP} \\
\midrule
\multirow{2}{*}{\textbf{FedAvg}} & ASR & \textcolor{red}{53.4} & \textcolor{red}{51.0} & 20.8 & \textcolor{red}{83.5} & \textcolor{red}{71.0} & 46.3 & \textcolor{red}{79.0} & \textcolor{red}{80.0} \\
 & ACC & 89.8 & 89.7 & 90.3 & 90.2 & 86.2 & 86.2 & 86.2 & 86.2 \\
\midrule
\multirow{2}{*}{\textbf{CMed}} & ASR & 13.6 & 3.1 & 5.8 & 16.4 & 0.3 & 0.3 & 25.2 & 0.3 \\
 & ACC & \textbf{58.3} & \textbf{59.7} & \textbf{58.4} & \textbf{57.7} & 82.3 & 82.3 & 82.2 & 82.3 \\
\midrule
\multirow{2}{*}{\textbf{Krum}} & ASR & 3.0 & 3.7 & 5.3 & 9.5 & 0.3 & 0.3 & \textcolor{red}{90.1} & 0.2 \\
 & ACC & 75.5 & 75.7 & 75.5 & 75.8 & 70.3 & 70.5 & 71.0 & 70.3 \\
\midrule
\multirow{2}{*}{\textbf{FLTrust}} & ASR & 7.5 & 2.0 & 11.0 & 39.3 & 0.4 & 0.3 & 0.8 & 0.2 \\
 & ACC & \textbf{68.8} & \textbf{69.5} & \textbf{69.0} & \textbf{67.7} & 82.1 & 82.1 & 82.1 & 82.0 \\
\midrule
\multirow{2}{*}{\textbf{WDP}} & ASR & 11.1 & 11.4 & 10.2 & \textcolor{red}{59.2} & 0.0 & 0.2 & 4.7 & 10.0 \\
 & ACC & 73.9 & 74.0 & 74.0 & 73.8 & \textbf{11.0} & \textbf{13.3} & \textbf{14.7} & \textbf{27.4} \\
\bottomrule
\end{tabular}%
}
\end{table}

\begin{table}[t]
\centering
\caption{Performance of AD defenses on CIFAR10 and FEMNIST when malicious clients occupy 5\% of total clients. We highlight \textcolor{red}{ASR $>50\%$} and \textbf{ACC $<$ 70\%}.}
\label{tab:ad_mal_clients_005}
\resizebox{\columnwidth}{!}{%
\begin{tabular}{@{}lc|cccc|cccc@{}}
\toprule
\multirow{2}{*}{Defense} & \multirow{2}{*}{Metric} & \multicolumn{4}{c|}{CIFAR10 + ResNet18} & \multicolumn{4}{c}{FEMNIST + SimpleCNN} \\
\cmidrule(lr){3-6} \cmidrule(lr){7-10}
 &  & \textbf{Static} & \textbf{DBA} & \textbf{OOD} & \textbf{CerP} & \textbf{Static} & \textbf{DBA} & \textbf{OOD} & \textbf{CerP} \\
\midrule
\multirow{4}{*}{\textbf{AlignIns}} & ASR & 1.1 & 1.4 & 6.6 & \textcolor{red}{54.1} & 0.1 & 0.1 & \textcolor{red}{71.3} & 0.2 \\
 & ACC & 90.4 & 90.0 & 90.4 & 90.2 & 86.5 & 86.5 & 86.4 & 86.4 \\
 & TPR & 100.0 & 100.0 & 89.5 & 72.7 & 99.5 & 99.5 & 50.5 & 96.2 \\
 & TNR & 58.1 & 57.9 & 54.9 & 51.2 & 60.5 & 60.6 & 45.4 & 54.6 \\
\midrule
\multirow{4}{*}{\textbf{ADMkrum}} & ASR & \textcolor{red}{64.3} & \textcolor{red}{60.4} & 28.0 & \textcolor{red}{85.3} & \textcolor{red}{57.8} & 24.7 & \textcolor{red}{80.1} & \textcolor{red}{78.2} \\
 & ACC & 91.3 & 91.0 & 91.4 & 91.3 & 86.8 & 86.8 & 86.8 & 86.9 \\
 & TPR & 1.7 & 0.6 & 1.7 & 2.3 & 12.6 & 23.9 & 3.0 & 4.6 \\
 & TNR & 87.0 & 86.9 & 87.0 & 87.2 & 94.2 & 95.1 & 93.5 & 93.7 \\
\midrule
\multirow{4}{*}{\textbf{DeepSight}} & ASR & 0.9 & 0.9 & 10.7 & 2.3 & \textcolor{red}{57.1} & 20.2 & \textcolor{red}{79.0} & \textcolor{red}{74.8} \\
 & ACC & 91.6 & 91.4 & 91.6 & 91.4 & 86.8 & 86.9 & 86.8 & 86.9 \\
 & TPR & 98.8 & 100.0 & 77.7 & 98.4 & 0.2 & 1.5 & 0.7 & 1.9 \\
 & TNR & 89.9 & 91.2 & 90.6 & 90.3 & 99.9 & 99.8 & 99.9 & 99.7 \\
\midrule
\multirow{4}{*}{\textbf{Indicator}} & ASR & 1.0 & 0.9 & 6.5 & 2.2 & 2.6 & 7.1 & 20.6 & 0.0 \\
 & ACC & 89.6 & 89.8 & 89.9 & 89.8 & \textbf{37.9} & \textbf{41.2} & \textbf{52.1} & \textbf{5.5} \\
 & TPR & 96.5 & 97.7 & 80.2 & 97.9 & 67.2 & 65.1 & 92.0 & 100.0 \\
 & TNR & 84.2 & 84.6 & 82.1 & 82.3 & 70.6 & 59.6 & 47.2 & 0.0 \\
\midrule
\multirow{4}{*}{\textbf{FedDLAD}} & ASR & 1.0 & 0.6 & 5.2 & 1.4 & 0.1 & 0.2 & 4.9 & 0.1 \\
 & ACC & 90.7 & 90.3 & 90.7 & 90.5 & 81.6 & 81.5 & 81.9 & 81.4 \\
 & TPR & 58.7 & 58.5 & 39.5 & 53.5 & 97.3 & 96.9 & 97.3 & 97.8 \\
 & TNR & 51.3 & 50.6 & 48.4 & 50.0 & 17.6 & 17.6 & 17.6 & 17.6 \\
\bottomrule
\end{tabular}%
}
\end{table}

\begin{table}[t]
\centering
\caption{Performance of RA defenses on CIFAR10 and FEMNIST when malicious clients occupy 20\% of total clients. We highlight \textcolor{red}{ASR $>50\%$} and \textbf{ACC $<$ 70\%}.}
\label{tab:ra_mal_clients_02}
\resizebox{\columnwidth}{!}{%
\begin{tabular}{@{}lc|cccc|cccc@{}}
\toprule
\multirow{2}{*}{Defense} & \multirow{2}{*}{Metric} & \multicolumn{4}{c|}{CIFAR10 + ResNet18} & \multicolumn{4}{c}{FEMNIST + SimpleCNN} \\
\cmidrule(lr){3-6} \cmidrule(lr){7-10}
 &  & \textbf{Static} & \textbf{DBA} & \textbf{OOD} & \textbf{CerP} & \textbf{Static} & \textbf{DBA} & \textbf{OOD} & \textbf{CerP} \\
\midrule
\multirow{2}{*}{\textbf{FedAvg}} & ASR & \textcolor{red}{86.2} & \textcolor{red}{88.2} & \textcolor{red}{52.8} & \textcolor{red}{94.8} & \textcolor{red}{93.1} & \textcolor{red}{88.8} & \textcolor{red}{94.4} & \textcolor{red}{95.5} \\
 & ACC & 90.4 & 90.3 & 90.5 & 90.6 & 86.2 & 86.2 & 86.2 & 86.3 \\
\midrule
\multirow{2}{*}{\textbf{CMed}} & ASR & \textcolor{red}{67.8} & 11.7 & 23.0 & \textcolor{red}{71.7} & 10.7 & 2.2 & \textcolor{red}{83.5} & 31.5 \\
 & ACC & \textbf{62.0} & \textbf{61.4} & \textbf{61.8} & \textbf{61.9} & 82.2 & 82.2 & 82.2 & 82.1 \\
\midrule
\multirow{2}{*}{\textbf{Krum}} & ASR & 10.4 & 12.1 & 10.0 & 33.2 & \textcolor{red}{81.6} & 27.7 & \textcolor{red}{89.6} & \textcolor{red}{83.6} \\
 & ACC & 76.1 & 76.2 & 75.7 & 76.1 & 71.9 & 71.0 & 71.6 & 71.8 \\
\midrule
\multirow{2}{*}{\textbf{FLTrust}} & ASR & 21.1 & 9.4 & 18.7 & \textcolor{red}{67.2} & 11.3 & 0.4 & 14.2 & 1.4 \\
 & ACC & \textbf{68.7} & \textbf{68.9} & \textbf{68.4} & \textbf{69.2} & 82.2 & 82.1 & 82.1 & 82.1 \\
\midrule
\multirow{2}{*}{\textbf{WDP}} & ASR & \textcolor{red}{54.4} & \textcolor{red}{76.1} & 38.7 & \textcolor{red}{90.4} & \textcolor{red}{85.5} & \textcolor{red}{89.2} & \textcolor{red}{93.6} & \textcolor{red}{91.3} \\
 & ACC & 73.3 & 73.5 & 73.0 & 73.5 & 71.0 & 71.2 & 71.3 & 71.2 \\
\bottomrule
\end{tabular}%
}
\end{table}

\begin{table}[t]
\centering
\caption{Performance of RA defenses on CIFAR10 and FEMNIST when malicious clients occupy 20\% of total clients. We highlight \textcolor{red}{ASR $>50\%$} and \textbf{ACC $<$ 70\%}.}
\label{tab:ad_mal_clients_02}
\resizebox{\columnwidth}{!}{%
\begin{tabular}{@{}lc|cccc|cccc@{}}
\toprule
\multirow{2}{*}{Defense} & \multirow{2}{*}{Metric} & \multicolumn{4}{c|}{CIFAR10 + ResNet18} & \multicolumn{4}{c}{FEMNIST + SimpleCNN} \\
\cmidrule(lr){3-6} \cmidrule(lr){7-10}
 &  & \textbf{Static} & \textbf{DBA} & \textbf{OOD} & \textbf{CerP} & \textbf{Static} & \textbf{DBA} & \textbf{OOD} & \textbf{CerP} \\
\midrule
\multirow{4}{*}{\textbf{AlignIns}} & ASR & 45.5 & 45.7 & 18.3 & \textcolor{red}{80.3} & 0.1 & 12.6 & \textcolor{red}{74.3} & \textcolor{red}{65.0} \\
 & ACC & 90.6 & 90.1 & 90.2 & 90.5 & 86.6 & 86.6 & 86.5 & 86.4 \\
 & TPR & 82.3 & 86.1 & 86.6 & 77.8 & 99.6 & 88.2 & 61.5 & 65.9 \\
 & TNR & 55.0 & 59.5 & 58.1 & 52.9 & 76.6 & 69.3 & 49.9 & 52.5 \\
\midrule
\multirow{4}{*}{\textbf{ADMkrum}} & ASR & \textcolor{red}{89.2} & \textcolor{red}{90.2} & \textcolor{red}{71.1} & \textcolor{red}{95.8} & \textcolor{red}{87.7} & \textcolor{red}{75.3} & \textcolor{red}{95.3} & \textcolor{red}{92.0} \\
 & ACC & 91.5 & 91.4 & 91.4 & 91.7 & 86.6 & 86.7 & 86.6 & 86.7 \\
 & TPR & 1.9 & 2.1 & 2.6 & 2.0 & 7.7 & 24.7 & 2.5 & 5.6 \\
 & TNR & 70.0 & 70.0 & 70.2 & 70.1 & 77.8 & 82.1 & 76.5 & 77.2 \\
\midrule
\multirow{4}{*}{\textbf{DeepSight}} & ASR & 6.3 & 3.9 & \textcolor{red}{63.6} & \textcolor{red}{63.0} & \textcolor{red}{81.6} & \textcolor{red}{62.4} & \textcolor{red}{92.1} & \textcolor{red}{90.2} \\
 & ACC & 91.5 & 91.5 & 91.7 & 91.6 & 86.9 & 86.9 & 86.9 & 86.9 \\
 & TPR & 95.7 & 95.8 & 17.8 & 61.9 & 7.1 & 16.6 & 2.2 & 3.3 \\
 & TNR & 94.1 & 93.6 & 90.6 & 91.9 & 99.6 & 99.6 & 99.7 & 99.7 \\
\midrule
\multirow{4}{*}{\textbf{Indicator}} & ASR & 5.6 & 2.9 & 25.9 & \textcolor{red}{78.4} & 8.3 & 7.3 & \textcolor{red}{51.9} & \textcolor{red}{67.2} \\
 & ACC & 89.8 & 89.9 & 89.8 & 90.5 & \textbf{12.6} & \textbf{28.8} & \textbf{23.9} & \textbf{31.5} \\
 & TPR & 95.5 & 97.1 & 70.9 & 34.8 & 60.2 & 55.3 & 84.0 & 68.5 \\
 & TNR & 84.0 & 84.9 & 83.6 & 81.1 & 60.4 & 76.9 & 56.6 & 73.3 \\
\midrule
\multirow{4}{*}{\textbf{FedDLAD}} & ASR & 42.1 & 1.5 & 39.6 & \textcolor{red}{80.2} & 0.1 & 0.2 & 11.7 & 0.4 \\
 & ACC & 90.2 & 88.0 & 90.9 & 91.1 & 79.4 & 80.6 & 81.6 & 80.5 \\
 & TPR & 11.1 & 17.4 & 14.2 & 15.4 & 96.3 & 96.8 & 88.1 & 95.5 \\
 & TNR & 37.5 & 38.5 & 38.5 & 38.9 & 19.4 & 19.9 & 16.9 & 19.1 \\
\bottomrule
\end{tabular}%
}
\end{table}


\begin{thebibliography}{66}
\providecommand{\natexlab}[1]{#1}
\providecommand{\url}[1]{\texttt{#1}}
\expandafter\ifx\csname urlstyle\endcsname\relax
  \providecommand{\doi}[1]{doi: #1}\else
  \providecommand{\doi}{doi: \begingroup \urlstyle{rm}\Url}\fi

\bibitem[Bagdasaryan and Shmatikov(2021)]{backdoors101}
Eugene Bagdasaryan and Vitaly Shmatikov.
\newblock Blind backdoors in deep learning models.
\newblock In \emph{30th {USENIX} Security Symposium ({USENIX} Security 21)}, pages 1505--1521. {USENIX} Association, 2021.

\bibitem[Bagdasaryan et~al.(2020)Bagdasaryan, Veit, Hua, Estrin, and Shmatikov]{howto}
Eugene Bagdasaryan, Andreas Veit, Yiqing Hua, Deborah Estrin, and Vitaly Shmatikov.
\newblock How to backdoor federated learning.
\newblock In \emph{Proceedings of the Twenty Third International Conference on Artificial Intelligence and Statistics}, pages 2938--2948. PMLR, 2020.

\bibitem[Barni et~al.(2019)Barni, Kallas, and Tondi]{sig}
M. Barni, K. Kallas, and B. Tondi.
\newblock A new backdoor attack in cnns by training set corruption without label poisoning.
\newblock In \emph{2019 IEEE International Conference on Image Processing (ICIP)}, pages 101--105, 2019.

\bibitem[Blanchard et~al.(2017{\natexlab{a}})Blanchard, El~Mhamdi, Guerraoui, and Stainer]{blanchard2017krum}
Peva Blanchard, El~Mahdi El~Mhamdi, Rachid Guerraoui, and Julien Stainer.
\newblock Machine learning with adversaries: Byzantine tolerant gradient descent.
\newblock \emph{Advances in neural information processing systems}, 30, 2017{\natexlab{a}}.

\bibitem[Blanchard et~al.(2017{\natexlab{b}})Blanchard, El~Mhamdi, Guerraoui, and Stainer]{blanchard2017machine}
Peva Blanchard, El~Mahdi El~Mhamdi, Rachid Guerraoui, and Julien Stainer.
\newblock Machine learning with adversaries: Byzantine tolerant gradient descent.
\newblock \emph{Advances in neural information processing systems}, 30, 2017{\natexlab{b}}.

\bibitem[Caldas et~al.(2019)Caldas, Duddu, Wu, Li, Konečný, McMahan, Smith, and Talwalkar]{leaf}
Sebastian Caldas, Sai Meher~Karthik Duddu, Peter Wu, Tian Li, Jakub Konečný, H.~Brendan McMahan, Virginia Smith, and Ameet Talwalkar.
\newblock Leaf: A benchmark for federated settings, 2019.

\bibitem[Cao et~al.(2020)Cao, Fang, Liu, and Gong]{cao2020fltrust}
Xiaoyu Cao, Minghong Fang, Jia Liu, and Neil~Zhenqiang Gong.
\newblock Fltrust: Byzantine-robust federated learning via trust bootstrapping.
\newblock \emph{arXiv preprint arXiv:2012.13995}, 2020.

\bibitem[Chen et~al.(2017{\natexlab{a}})Chen, Liu, Li, Lu, and Song]{blended}
Xinyun Chen, Chang Liu, Bo Li, Kimberly Lu, and Dawn Song.
\newblock Targeted backdoor attacks on deep learning systems using data poisoning.
\newblock \emph{arXiv preprint arXiv:1712.05526}, 2017{\natexlab{a}}.

\bibitem[Chen et~al.(2017{\natexlab{b}})Chen, Liu, Li, Lu, and Song]{chen2017targeted}
Xinyun Chen, Chang Liu, Bo Li, Kui Lu, and Dawn Song.
\newblock Targeted backdoor attacks on deep learning systems using data poisoning.
\newblock In \emph{Proceedings of the 2017 ACM SIGSAC Conference on Computer and Communications Security}, pages 603--618. ACM, 2017{\natexlab{b}}.

\bibitem[Dai and Li(2023)]{dai2023chameleon}
Yanbo Dai and Songze Li.
\newblock Chameleon: Adapting to peer images for planting durable backdoors in federated learning.
\newblock In \emph{International Conference on Machine Learning}, pages 6712--6725. PMLR, 2023.

\bibitem[Dao et~al.(2025)Dao, Doan, and Wong]{anonymous2025cleanlabel}
Thinh Dao, Khoa~D Doan, and Kok-Seng Wong.
\newblock Clean-label physical backdoor attacks with data distillation.
\newblock In \emph{NeurIPS 2025 Workshop: Reliable ML from Unreliable Data}, 2025.

\bibitem[Ding et~al.(2025)Ding, Yang, and Huang]{feddlad}
Binbin Ding, Penghui Yang, and Sheng-Jun Huang.
\newblock Feddlad: A federated learning dual-layer anomaly detection framework for enhancing resilience against backdoor attacks.
\newblock In \emph{Proceedings of the Thirty-Fourth International Joint Conference on Artificial Intelligence, {IJCAI-25}}, pages 5021--5029. International Joint Conferences on Artificial Intelligence Organization, 2025.
\newblock Main Track.

\bibitem[El~Mhamdi et~al.(2018)El~Mhamdi, Guerraoui, and Rouault]{bulyan}
El~Mahdi El~Mhamdi, Rachid Guerraoui, and S{\'e}bastien Rouault.
\newblock The hidden vulnerability of distributed learning in {B}yzantium.
\newblock In \emph{Proceedings of the 35th International Conference on Machine Learning}, pages 3521--3530. PMLR, 2018.

\bibitem[Fang and Chen(2023)]{f3ba}
Pei Fang and Jinghui Chen.
\newblock On the vulnerability of backdoor defenses for federated learning.
\newblock In \emph{Proceedings of the 37th AAAI Conference on Artificial Intelligence (AAAI)}, pages 11800--11808. AAAI Press, 2023.

\bibitem[Fung et~al.(2018)Fung, Yoon, and Beschastnikh]{fung2018foolsgold}
Clement Fung, Chris~JM Yoon, and Ivan Beschastnikh.
\newblock Mitigating sybils in federated learning poisoning.
\newblock \emph{arXiv}, abs/1808.04866, 2018.

\bibitem[Gu et~al.(2019)Gu, Liu, Dolan-Gavitt, and Garg]{gu2019badnets}
Tianyu Gu, Kang Liu, Brendan Dolan-Gavitt, and Siddharth Garg.
\newblock Badnets: Evaluating backdooring attacks on deep neural networks.
\newblock \emph{IEEE Access}, 7:\penalty0 47230--47244, 2019.

\bibitem[Huang et~al.(2023{\natexlab{a}})Huang, Li, Chen, Shi, and Gao]{huang2023multi}
Siquan Huang, Yijiang Li, Chong Chen, Leyu Shi, and Ying Gao.
\newblock Multi-metrics adaptively identifies backdoors in federated learning.
\newblock In \emph{Proceedings of the IEEE/CVF International Conference on Computer Vision}, pages 4652--4662, 2023{\natexlab{a}}.

\bibitem[Huang et~al.(2023{\natexlab{b}})Huang, Hu, Chow, Ilhan, Tekin, and Liu]{huang2023lockdown}
Tiansheng Huang, Sihao Hu, Ka-Ho Chow, Fatih Ilhan, Selim~Furkan Tekin, and Ling Liu.
\newblock Lockdown: Backdoor defense for federated learning with isolated subspace training.
\newblock In \emph{Advances in Neural Information Processing Systems (NeurIPS)}, 2023{\natexlab{b}}.

\bibitem[Karra et~al.(2020)Karra, Ashcraft, and Fendley]{karra2020trojai}
Kiran Karra, Chace Ashcraft, and Neil Fendley.
\newblock The trojai software framework: An opensource tool for embedding trojans into deep learning models, 2020.

\bibitem[Kone{\v{c}}n{\`y}(2016)]{konevcny2016federated}
Jakub Kone{\v{c}}n{\`y}.
\newblock Federated learning: Strategies for improving communication efficiency.
\newblock \emph{arXiv preprint arXiv:1610.05492}, 2016.

\bibitem[Li et~al.(2023)Li, Ye, Hu, Li, Wang, Fang, and Shi]{3dfed}
Haoyang Li, Qingqing Ye, Haibo Hu, Jin Li, Leixia Wang, Chengfang Fang, and Jie Shi.
\newblock 3dfed: Adaptive and extensible framework for covert backdoor attack in federated learning.
\newblock In \emph{2023 IEEE Symposium on Security and Privacy (SP)}, pages 1893--1907, 2023.

\bibitem[Li and Dai(2024)]{li2024backdoorindicator}
Songze Li and Yanbo Dai.
\newblock Backdoorindicator: Leveraging ood data for proactive backdoor detection in federated learning.
\newblock \emph{arXiv preprint arXiv:2405.20862}, 2024.

\bibitem[Lin et~al.(2020)Lin, Zhang, He, Zhang, and Xu]{lin2020dual}
Zhun Lin, Hongyang Zhang, Zhiqiang He, Pin-Yu Zhang, and Kaizhao Xu.
\newblock Dual adversarial training: Balancing robustness and accuracy with two adversaries.
\newblock In \emph{NeurIPS}, 2020.

\bibitem[Liu et~al.(2024)Liu, Zhang, Feng, Yang, Xu, Man, and Yang]{fcba}
Tao Liu, Yuhang Zhang, Zhu Feng, Zhiqin Yang, Chen Xu, Dapeng Man, and Wu Yang.
\newblock Beyond traditional threats: A persistent backdoor attack on federated learning.
\newblock In \emph{Proceedings of the AAAI Conference on Artificial Intelligence}, pages 21359--21367, 2024.

\bibitem[Lyu et~al.(2023)Lyu, Han, Wang, Liu, Wang, Liu, and Zhang]{ceberus}
Xiaoting Lyu, Yufei Han, Wei Wang, Jingkai Liu, Bin Wang, Jiqiang Liu, and Xiangliang Zhang.
\newblock Poisoning with cerberus: Stealthy and colluded backdoor attack against federated learning.
\newblock \emph{Proceedings of the AAAI Conference on Artificial Intelligence}, 37\penalty0 (7):\penalty0 9020--9028, 2023.

\bibitem[Ma et~al.(2021)Ma, Chen, Zhang, and Guo]{ma2021effective}
Xiaojun Ma, Yunfei Chen, Shuo Zhang, and Yinzhi Guo.
\newblock Effective and efficient adversarial examples detection with dynamic thresholding.
\newblock In \emph{AAAI}, 2021.

\bibitem[Madry et~al.(2018)Madry, Makelov, Schmidt, Tsipras, and Vladu]{madry2018towards}
Aleksander Madry, Aleksandar Makelov, Ludwig Schmidt, Dimitris Tsipras, and Adrian Vladu.
\newblock Towards deep learning models resistant to adversarial attacks.
\newblock In \emph{International Conference on Learning Representations (ICLR)}, 2018.

\bibitem[McMahan et~al.(2017)McMahan, Moore, Ramage, Hampson, and y~Arcas]{mcmahan2017communication}
Brendan McMahan, Eider Moore, Daniel Ramage, Seth Hampson, and Blaise~Aguera y Arcas.
\newblock Communication-efficient learning of deep networks from decentralized data.
\newblock In \emph{Artificial intelligence and statistics}, pages 1273--1282. PMLR, 2017.

\bibitem[Moritz et~al.(2018)Moritz, Nishihara, Wang, Tumanov, Liaw, Liang, Elibol, Yang, Paul, Jordan, and Stoica]{ray}
Philipp Moritz, Robert Nishihara, Stephanie Wang, Alexey Tumanov, Richard Liaw, Eric Liang, Melih Elibol, Zongheng Yang, William Paul, Michael~I. Jordan, and Ion Stoica.
\newblock Ray: A distributed framework for emerging {AI} applications.
\newblock In \emph{13th USENIX Symposium on Operating Systems Design and Implementation (OSDI 18)}, pages 561--577, Carlsbad, CA, 2018. USENIX Association.

\bibitem[Naseri et~al.(2020)Naseri, Hayes, and De~Cristofaro]{localdp}
Mohammad Naseri, Jamie Hayes, and Emiliano De~Cristofaro.
\newblock Local and central differential privacy for robustness and privacy in federated learning.
\newblock \emph{arXiv preprint arXiv:2009.03561}, 2020.

\bibitem[Nguyen et~al.(2025)Nguyen, Doan, and Wong]{flat}
Tuan Nguyen, Khoa~D Doan, and Kok-Seng Wong.
\newblock Flat: Latent-driven arbitrary-target backdoor attacks in federated learning, 2025.

\bibitem[Nguyen et~al.(2022)Nguyen, Rieger, De~Viti, Chen, Brandenburg, Yalame, M{\"o}llering, Fereidooni, Marchal, Miettinen, et~al.]{nguyen2022flame}
Thien~Duc Nguyen, Phillip Rieger, Roberta De~Viti, Huili Chen, Bj{\"o}rn~B Brandenburg, Hossein Yalame, Helen M{\"o}llering, Hossein Fereidooni, Samuel Marchal, Markus Miettinen, et~al.
\newblock $\{$FLAME$\}$: Taming backdoors in federated learning.
\newblock In \emph{31st USENIX Security Symposium (USENIX Security 22)}, pages 1415--1432, 2022.

\bibitem[Nguyen et~al.(2024{\natexlab{a}})Nguyen, Nguyen, Nguyen, Pham, Doan, and Wong]{backdoor_survey}
Thuy~Dung Nguyen, Tuan Nguyen, Phi~Le Nguyen, Hieu~H. Pham, Khoa~D. Doan, and Kok-Seng Wong.
\newblock Backdoor attacks and defenses in federated learning: Survey, challenges and future research directions.
\newblock \emph{Engineering Applications of Artificial Intelligence}, 127:\penalty0 107166, 2024{\natexlab{a}}.

\bibitem[Nguyen et~al.(2024{\natexlab{b}})Nguyen, Nguyen, Tran, Doan, and Wong]{nguyen2024iba}
Thuy~Dung Nguyen, Tuan~A Nguyen, Anh Tran, Khoa~D Doan, and Kok-Seng Wong.
\newblock Iba: Towards irreversible backdoor attacks in federated learning.
\newblock \emph{Advances in Neural Information Processing Systems}, 36, 2024{\natexlab{b}}.

\bibitem[Nguyen et~al.(2024{\natexlab{c}})Nguyen, Vu, Nguyen, Nguyen, Doan, and Wong]{fedunlearn}
Thai-Hung Nguyen, Hong-Phuc Vu, Dung~Thuy Nguyen, Tuan~Minh Nguyen, Khoa~D Doan, and Kok-Seng Wong.
\newblock Empirical study of federated unlearning: {E}fficiency and effectiveness.
\newblock In \emph{Proceedings of the 15th Asian Conference on Machine Learning}, pages 959--974. PMLR, 2024{\natexlab{c}}.

\bibitem[Nie et~al.(2022)Nie, Gupta, et~al.]{nie2022diffpure}
Weili Nie, Arash Vahdat Anima~Anandkumar Gupta, et~al.
\newblock Diffusion models for adversarial purification.
\newblock In \emph{International Conference on Machine Learning (ICML)}, 2022.

\bibitem[Ozdayi et~al.(2021{\natexlab{a}})Ozdayi, Kantarcioglu, and Gel]{ozdayi2021defending}
Mustafa~Safa Ozdayi, Murat Kantarcioglu, and Yulia~R Gel.
\newblock Defending against backdoors in federated learning with robust learning rate.
\newblock In \emph{Proceedings of the AAAI Conference on Artificial Intelligence}, pages 9268--9276, 2021{\natexlab{a}}.

\bibitem[Ozdayi et~al.(2021{\natexlab{b}})Ozdayi, Kantarcioglu, and Gel]{rlr}
Mustafa~Safa Ozdayi, Murat Kantarcioglu, and Yulia~R Gel.
\newblock Defending against backdoors in federated learning with robust learning rate.
\newblock In \emph{Proceedings of the AAAI conference on artificial intelligence}, pages 9268--9276, 2021{\natexlab{b}}.

\bibitem[Pang et~al.(2022{\natexlab{a}})Pang, Zhang, Gao, Xi, Ji, Cheng, Luo, and Wang]{trojanzoo}
Ren Pang, Zheng Zhang, Xiangshan Gao, Zhaohan Xi, Shouling Ji, Peng Cheng, Xiapu Luo, and Ting Wang.
\newblock Trojanzoo: Towards unified, holistic, and practical evaluation of neural backdoors.
\newblock In \emph{2022 IEEE 7th European Symposium on Security and Privacy (EuroS\&P)}, page 684–702. IEEE, 2022{\natexlab{a}}.

\bibitem[Pang et~al.(2022{\natexlab{b}})Pang, Yang, Dong, and Zhu]{pang2022defense}
Tianyu Pang, Xiao Yang, Yinpeng Dong, and Jun Zhu.
\newblock Defense-prompted adversarial detection: Detecting adversarial examples via robust feature alignment.
\newblock In \emph{CVPR}, 2022{\natexlab{b}}.

\bibitem[Pillutla et~al.(2022)Pillutla, Kakade, and Harchaoui]{pillutla2022robust}
Krishna Pillutla, Sham~M Kakade, and Zaid Harchaoui.
\newblock Robust aggregation for federated learning.
\newblock \emph{IEEE Transactions on Signal Processing}, 70:\penalty0 1142--1154, 2022.

\bibitem[Qin et~al.(2024)Qin, Chen, Zhi, Yan, and Deng]{snowball}
Zhen Qin, Feiyi Chen, Chen Zhi, Xueqiang Yan, and Shuiguang Deng.
\newblock Resisting backdoor attacks in federated learning via bidirectional elections and individual perspective.
\newblock In \emph{Proceedings of the AAAI Conference on Artificial Intelligence}, pages 14677--14685, 2024.

\bibitem[Rieger et~al.(2022)Rieger, Nguyen, Miettinen, and Sadeghi]{rieger2022deepsight}
Phillip Rieger, Thien~Duc Nguyen, Markus Miettinen, and Ahmad-Reza Sadeghi.
\newblock Deepsight: Mitigating backdoor attacks in federated learning through deep model inspection.
\newblock \emph{Annual Network and Distributed System Security Symposium}, 2022.

\bibitem[Song et~al.(2017)Song, Shu, Kushman, and Ermon]{song2017pixeldefend}
Yang Song, Rui Shu, Nate Kushman, and Stefano Ermon.
\newblock Pixeldefend: Leveraging generative models to understand and defend against adversarial examples.
\newblock In \emph{NeurIPS}, 2017.

\bibitem[Sun et~al.(2021)Sun, Li, DiValentin, Hassanzadeh, Chen, and Li]{flwbc}
Jingwei Sun, Ang Li, Louis DiValentin, Amin Hassanzadeh, Yiran Chen, and Hai Li.
\newblock Fl-wbc: Enhancing robustness against model poisoning attacks in federated learning from a client perspective.
\newblock \emph{Advances in neural information processing systems}, 34:\penalty0 12613--12624, 2021.

\bibitem[Wan and Chen(2021)]{rfa}
Ching~Pui Wan and Qifeng Chen.
\newblock Robust federated learning with attack-adaptive aggregation.
\newblock \emph{arXiv preprint arXiv:2102.05257}, 2021.

\bibitem[Wang et~al.(2020{\natexlab{a}})Wang, Sreenivasan, Rajput, Vishwakarma, Agarwal, Sohn, Lee, and Papailiopoulos]{edge_case}
Hongyi Wang, Kartik Sreenivasan, Shashank Rajput, Harit Vishwakarma, Saurabh Agarwal, Jy-yong Sohn, Kangwook Lee, and Dimitris Papailiopoulos.
\newblock Attack of the tails: Yes, you really can backdoor federated learning.
\newblock \emph{Advances in Neural Information Processing Systems}, 33:\penalty0 16070--16084, 2020{\natexlab{a}}.

\bibitem[Wang et~al.(2020{\natexlab{b}})Wang, Sreenivasan, Rajput, Vishwakarma, Agarwal, Sohn, Lee, and Papailiopoulos]{sun2019can}
Hongyi Wang, Kartik Sreenivasan, Shashank Rajput, Harit Vishwakarma, Saurabh Agarwal, Jy-yong Sohn, Kangwook Lee, and Dimitris Papailiopoulos.
\newblock Attack of the tails: yes, you really can backdoor federated learning.
\newblock In \emph{Proceedings of the 34th International Conference on Neural Information Processing Systems}, Red Hook, NY, USA, 2020{\natexlab{b}}. Curran Associates Inc.

\bibitem[Wang et~al.(2024)Wang, Zhang, Xiao, Chen, Lou, and Hou]{flare}
Ning Wang, Chaoyu Zhang, Yang Xiao, Yimin Chen, Wenjing Lou, and Y.~Thomas Hou.
\newblock Flare: Defending federated learning against model poisoning attacks via latent space representations.
\newblock \emph{IEEE Transactions on Dependable and Secure Computing}, pages 1--17, 2024.

\bibitem[Wang et~al.(2022)Wang, Zhai, Zhan, and Xia]{wang2022rflbat}
Yongkang Wang, Dihua Zhai, Yufeng Zhan, and Yuanqing Xia.
\newblock Rflbat: A robust federated learning algorithm against backdoor attack.
\newblock \emph{arXiv preprint arXiv:2201.03772}, 2022.

\bibitem[Wen et~al.(2022)Wen, Geiping, Fowl, Souri, Chellappa, Goldblum, and Goldstein]{anticipate}
Yuxin Wen, Jonas Geiping, Liam Fowl, Hossein Souri, Rama Chellappa, Micah Goldblum, and Tom Goldstein.
\newblock Thinking two moves ahead: Anticipating other users improves backdoor attacks in federated learning.
\newblock In \emph{Advances in Neural Information Processing Systems (NeurIPS) 35}, 2022.
\newblock Preprint available on arXiv.

\bibitem[Wu et~al.(2022)Wu, Chen, Zhang, Zhu, Wei, Yuan, and Shen]{wu2022backdoorbench}
Baoyuan Wu, Hongrui Chen, Mingda Zhang, Zihao Zhu, Shaokui Wei, Danni Yuan, and Chao Shen.
\newblock Backdoorbench: A comprehensive benchmark of backdoor learning.
\newblock \emph{Advances in Neural Information Processing Systems}, 35:\penalty0 10546--10559, 2022.

\bibitem[Xie et~al.(2020)Xie, Huang, Chen, and Li]{xie2020dba}
Chulin Xie, Keli Huang, Pin~Yu Chen, and Bo Li.
\newblock Dba: Distributed backdoor attacks against federated learning.
\newblock In \emph{8th International Conference on Learning Representations, ICLR 2020}, 2020.

\bibitem[Xie et~al.(2021)Xie, Chen, Chen, and Li]{xie2021crfl}
Chulin Xie, Minghao Chen, Pin-Yu Chen, and Bo Li.
\newblock Crfl: Certifiably robust federated learning against backdoor attacks.
\newblock In \emph{International Conference on Machine Learning}, pages 11372--11382. PMLR, 2021.

\bibitem[XIE et~al.(2024)XIE, Fang, and Gong]{xie2024fedredefense}
Yueqi XIE, Minghong Fang, and Neil~Zhenqiang Gong.
\newblock Fed{RED}efense: Defending against model poisoning attacks for federated learning using model update reconstruction error.
\newblock In \emph{Forty-first International Conference on Machine Learning}, 2024.

\bibitem[Xu et~al.(2025)Xu, Zhang, and Hu]{aligninns}
Jiahao Xu, Zikai Zhang, and Rui Hu.
\newblock Detecting backdoor attacks in federated learning via direction alignment inspection.
\newblock In \emph{Proceedings of the Computer Vision and Pattern Recognition Conference}, pages 20654--20664, 2025.

\bibitem[Yin et~al.(2018{\natexlab{a}})Yin, Chen, Kannan, and Bartlett]{pmlr-v80-yin18a}
Dong Yin, Yudong Chen, Ramchandran Kannan, and Peter Bartlett.
\newblock {B}yzantine-robust distributed learning: Towards optimal statistical rates.
\newblock In \emph{Proceedings of the 35th International Conference on Machine Learning}, pages 5650--5659. PMLR, 2018{\natexlab{a}}.

\bibitem[Yin et~al.(2018{\natexlab{b}})Yin, Chen, Kannan, and Bartlett]{yin2018byzantine}
Dong Yin, Yudong Chen, Ramchandran Kannan, and Peter Bartlett.
\newblock Byzantine-robust distributed learning: Towards optimal statistical rates.
\newblock In \emph{International conference on machine learning}, pages 5650--5659. Pmlr, 2018{\natexlab{b}}.

\bibitem[Zhang et~al.(2019)Zhang, Yu, Jiao, Xing, Ghaoui, and Jordan]{zhang2019theoretically}
Hongyang Zhang, Yaodong Yu, Jiantao Jiao, Eric Xing, Laurent~El Ghaoui, and Michael Jordan.
\newblock Theoretically principled trade-off between robustness and accuracy.
\newblock In \emph{International Conference on Machine Learning (ICML)}, 2019.

\bibitem[Zhang et~al.(2024)Zhang, Jia, Chen, Lin, and Wu]{zhang2024a3fl}
Hangfan Zhang, Jinyuan Jia, Jinghui Chen, Lu Lin, and Dinghao Wu.
\newblock A3fl: Adversarially adaptive backdoor attacks to federated learning.
\newblock \emph{Advances in Neural Information Processing Systems}, 36, 2024.

\bibitem[Zhang et~al.(2025)Zhang, Liu, He, Wu, Cong, and Huang]{zhang2025sok}
Heyi Zhang, Yule Liu, Xinlei He, Jun Wu, Tianshuo Cong, and Xinyi Huang.
\newblock Sok: Benchmarking poisoning attacks and defenses in federated learning.
\newblock \emph{arXiv preprint arXiv:2502.03801}, 2025.

\bibitem[Zhang et~al.(2023)Zhang, Tao, Xu, Cheng, An, Liu, Feng, Shen, Chen, Ma, and Zhang]{zhang2023flip}
Kaiyuan Zhang, Guanhong Tao, Qiuling Xu, Siyuan Cheng, Shengwei An, Yingqi Liu, Shiwei Feng, Guangyu Shen, Pin-Yu Chen, Shiqing Ma, and Xiangyu Zhang.
\newblock {FLIP}: A provable defense framework for backdoor mitigation in federated learning.
\newblock In \emph{The Eleventh International Conference on Learning Representations}, 2023.

\bibitem[Zhang et~al.(2022{\natexlab{a}})Zhang, Cao, Jia, and Gong]{zhang2022fldetector}
Zaixi Zhang, Xiaoyu Cao, Jinyuan Jia, and Neil~Zhenqiang Gong.
\newblock Fldetector: Defending federated learning against model poisoning attacks via detecting malicious clients.
\newblock In \emph{Proceedings of the 28th ACM SIGKDD Conference on Knowledge Discovery and Data Mining}, pages 2545--2555, 2022{\natexlab{a}}.

\bibitem[Zhang et~al.(2022{\natexlab{b}})Zhang, Panda, Song, Yang, Mahoney, Mittal, Kannan, and Gonzalez]{zhang2022neurotoxin}
Zhengming Zhang, Ashwinee Panda, Linyue Song, Yaoqing Yang, Michael Mahoney, Prateek Mittal, Ramchandran Kannan, and Joseph Gonzalez.
\newblock Neurotoxin: Durable backdoors in federated learning.
\newblock In \emph{International Conference on Machine Learning}, pages 26429--26446. PMLR, 2022{\natexlab{b}}.

\bibitem[Zhuang et~al.(2024{\natexlab{a}})Zhuang, Yu, Wang, Hua, Li, and Yuan]{backdoor_critical}
Haomin Zhuang, Mingxian Yu, Hao Wang, Yang Hua, Jian Li, and Xu Yuan.
\newblock Backdoor federated learning by poisoning backdoor-critical layers.
\newblock In \emph{The Twelfth International Conference on Learning Representations}, 2024{\natexlab{a}}.

\bibitem[Zhuang et~al.(2024{\natexlab{b}})Zhuang, Yu, Wang, Hua, Li, and Yuan]{layer-aware}
Haomin Zhuang, Mingxian Yu, Hao Wang, Yang Hua, Jian Li, and Xu Yuan.
\newblock Backdoor federated learning by poisoning backdoor-critical layers.
\newblock In \emph{The Twelfth International Conference on Learning Representations}, 2024{\natexlab{b}}.

\end{thebibliography}
\end{document}